\documentstyle[11pt,fancyheadings,twoside,doublespace,epsfig,calc,hangcaption,a4]{book}
\renewcommand{\chaptermark}[1]%
                  {\markboth{#1}{}}
\renewcommand{\sectionmark}[1]%
                  {\markright{\thesection\ #1}}
\newcommand{\be}{\begin{equation}}
\newcommand{\ee}{\end{equation}}

\newcommand{\eqref}[1]{Eq. (\ref{#1})}

\newcommand{\ai}[1]{#1 \index{#1}}

\newcommand{\captionm}[1]{\isucaption{
{\footnotesize #1}
}}

\def\beq{\begin{equation}}
\def\eeq{\end{equation}}
\def\bea{\begin{eqnarray}}
\def\eea{\end{eqnarray}}

\def\ddf{D_{2\rightarrow 4}}
\def\dtf{D_{3\rightarrow 4}}
\def\dff{D_{4\rightarrow 4}}  
\def\lb{\label}  
\def\kt{k_{\bot}}
\def\ku{k_1}
\def\kup{k'_1}
\def\qu{q_1}
\def\qup{q'_1}
\def\oq{\omega(q)}
\def\oa{\omega(q_{1})}
\def\ob{\omega(q_{2})}
\def\eq{\eta (q)}
\def\ea{\eta (q_{1})}
\def\eb{\eta (q_{2})}

\def\fq{f (q)}
\def\fa{f (q_{1})}
\def\fb{f (q_{2})}
\def\fc{f (q'_{1})}
\def\fd{f (q'_{2})}
\def\ql{(q \cdot l)}
\def\qlp{(q \cdot l')}                      
\def\kit{k^{2}_{i\bot}}
\def\mit{m^{2}_{i\bot}}
\def\mjt{m^{2}_{i+1,\bot}}

\def\qit{q^{2}_{i\bot}}
\def\qjt{q^{2}_{i+1,\bot}}
\def\kip{k_{i+}}

\def\sij{s_{i,i+1}}
\def\sji{s_{i-1,i}}
\def\ai{\alpha_{i}}
\def\aj{\alpha_{i+1}}
\def\bi{\beta_{i}}

\makeindex
\begin{document}
\begin{spacing}{1.1}

\title{\bf The hard QCD pomeron: some aspects of its phenomenology
           and interactions}
\date{}
\author{\\ G.P. Vacca \\ \\ \\
\centerline{{\it Department of Physics,
University of Bologna and INFN, Italy}}} 
%\vspace{5cm}
%\\ \\ \\ \\ \\ \\ \\ \\ 
%\hspace{5cm} \\ \\ \\ \\ \\ \\ \\ \\}

\maketitle

%\chapter*{}
\stepcounter{page}
\tableofcontents

% Introduction

\addcontentsline{toc}{chapter}{Introduction}
\chapter*{Introduction}
Strong interactions continue to be a challenge in theoretical physics.
They involve all the phenomena regarding hadrons and
actually the best suggestion for a model to describe them is given by
Quantum Chromodynamics (QCD), which is a gauge quantum field theory based
on a local $SU(3)$ symmetry, which satisfies the general requirements of
renormalizability and unitarity of the scattering $S$-matrix.
The elementary degrees of freedom of QCD are quarks and gluons and
its complexity has prevented the solution of the theory, i.e. finding the
spectrum and correlation functions.
On one side interesting work is devoted to analyzing numerically the 
euclidean version of the theory on a lattice, even if until now the available
computational power really restrict the dimension of the lattice involved and
thus the significance of obtainable results.

On the other hand analytical efforts are done in two directions.
Due to the asymptotic freedom property related to the decrease of the 
interaction strength with the increase of the energy of the particles 
involved, perturbation theory in the small coupling regime is applied. 
A perturbative approach has been also considered for other ``small'' 
parameters, such as $1/N_c$ \cite{venez,thooft,witten} where $N_c$ is the 
number of colours which defines the special unitary gauge group.
Typical non-perturbative methods concern condensates of quarks and gluons
and topological objects such as instantons; however we are still far from the
understanding the nature of composite bound objects in QCD.

One of these objects seems to be the pomeron, originating from the works
in Regge theory, before a consistent field theory approach to strong 
interactions was established with QCD. The pomeron is related, in a hadronic
scattering process, to the rightmost singularity (preferably a pole) of the 
partial wave amplitude in the complex angular momentum plane.
The mathematical structure 
of the amplitude is commonly associated to the presence of an exchanged object
with vacuum quantum numbers which can lead to a power-law-increase of the
total cross section with the center of mass energy, in such a case in contrast 
with the unitarity bounds.
In general the process is deeply related to non-perturbative phenomena and
the only possibility to understand as much as possible of it with perturbative
tools is to consider processes involving the scattering of highly virtual
photons or onia-particles where the perturbative analysis in terms of
elementary quarks and gluons make sense.

At each order of perturbation theory one finds a leading behaviour
proportional to a proper power of the logarithm of the center of mass 
energy ($\sqrt{s}$). The approach which resums these dominant contributions is
called leading log(s) approximation (LLA).
One of its principal limitations is related to the fact that the coupling
constant is considered fixed.
The works of Lipatov et al. \cite{lip1,bfkl} were devoted to these
calculations for a $SU(N_c)$ gauge theory. According to previous ideas
of Regge theory they use the gluon reggeization property, computed 
considering the dominant virtual corrections and which can be 
associated to a bootstrap condition related to unitarity, to construct the BFKL
equation which describes the interactions in LLA between two reggeized gluons;
the contributions to these interactions may be represented by ladder-type
diagrams, resummed in a Bethe-Salpeter-like equation.
The vacuum quantum number in the $t$-channel corresponds to the BFKL hard
pomeron. Its singularity is not a pole but a branch-cut in the complex
angular momentum plane.

In order to go beyond the fixed coupling case a next-to-leading order (NLO)
analysis has been carried on in the last years \cite{lipfa89,nlomix,fali98}.
A quite complicated generalized BFKL equation was found but still it is
not clear if such a model is really applicable to the HERA kinematical region
or one may need to go beyond the NLO approximation to give satisfactory
predictions and interpretations of the experimental data.

All the theoretical predictions may be tested in various experiments in the
kinematical Regge region. In the last years particular interest was devoted
to the deep inelastic scattering (DIS) processes at HERA where the small $x$
region is probed. Recent obtained measures \cite{HERA,ZEUS2} have shown
a sharp rise of the $F_2$ proton structure function at low $x$ and this fact 
may be interpreted as a manifestation of the hard pomeron. Much attention was
also given to the observed large rate of rapidity-gap events at small $x$,
another possible signature of the pomeron interaction.

The basic ingredients in the LLA analysis and a brief derivation of the BFKL
equation are reported in the first chapter of this thesis. A well known
problem related to the consistency of the approach is evidenced by noting
that the diffusion-like process described by the BFKL equation may spoil
the applicability of perturbation theory, since a huge contribution at large
energies may involve small transverse momenta, deeply related to a
non-perturbative regime in QCD. The other main problem in the BFKL theory
is the violation of unitarity due to the high value of its intercept.
The remarkable property of global conformal symmetry of the BFKL equation in
the transverse two dimensional space is also shown and this permits one to 
also have a general explicit solution for the system.

An alternative approach in perturbative QCD to describe the scattering of two
onia-particles is given by the colour dipole picture \cite{mueller,muelpat}, 
which is based on the computation of the corrections to the light-cone wave 
functions of onia or virtual photons due to the emission of soft gluons 
strongly ordered in their scaling variables. This implies a leading logarithmic
approach which turns out to be equivalent to the BFKL one.
In the large $N_c$ limit it is possible to write an equation for the
generating functional of the $n$-dipole distributions \cite{mueller} which
should be considered in general to have unitarity corrections. 
A brief introduction to the basic equations of this model is given in the
last section of the first chapter, since they will be needed in the
analysis performed in the third chapter.

The problems related to the infrared sector one finds in the BFKL theory,
which is constructed on a small fixed coupling analysis thus far away from 
the non-perturbative region which instead can be reached in the diffusion
precess, has motivated many phenomenological works, beyond the NLO analysis,
devoted to the modification of the infrared sector of the BFKL equation.

In the second chapter we try to analyze some properties of a phenomenological
model which have the merit of preserving the gluon reggeization property
\cite{braun1,braun2}
with the insertion of the running coupling in the generalized BFKL equation.
The bootstrap condition is the crucial point since in this way unitarity is
satisfied to leading order for the production amplitude with one gluon 
exchange in the $t$-channel while in all the other phenomenological attempts
it is lost.
The second input for the model is the compatibility with the predictions
of the renormalization group equations for the gluon density (DGLAP) in the
double leading log approximation (DLLA) 
The model depends at least on a mass parameter, related to infrared sector, 
on which the trajectories of all the singularities depend.

The insertion of the running coupling provides a deep change in the
properties of the pomeron equation \cite{bvv}. 
A part of the spectrum is found to be discrete. The values are found 
numerically while the corresponding asymptotic behaviour of the eigenfunction
can be derived analytically. The parameter dependence can be fixed, for 
example, by forcing the value of the slope (computed perturbatively) of the
leading singularity. For some reasonable values one finds two supercritical
pomeron states (corresponding to negative ``energy'' for the two reggeized
gluons bound states in the Schr\"odinger-like picture). 
A rough attempt to estimate some effects of unitarity corrections
has been made by considering the multipomeron exchange in the scattering
of two virtual photons in the large $N_c$ limit, a case for which a 
corresponding eikonal formulation can be given.
As a result, for the parameter choice considered, one finds that these
corrections occur at energies much higher than the ones now-a-days reached.
However unitarity requires one to include also the effects of the triple
pomeron interaction, as is also shown in the third chapter.
Thus unitarity effects can still be significant at lower energies.

In the framework of the standard BFKL approach the process of jet production
has been extensively studied showing the importance of mini-jet production
and the logarithmic rise of the multiplicity. The fixed coupling analysis
requires the introduction of an infrared parameter and the inclusions of
some unitarity and coherence effects to restore the correct asymptotic
behaviour of the differential cross section for large jet transverse momentum.
In the model with running coupling considered \cite{bv2} one does not have,
of course, any problem in the infrared sector and finds, due to the asymptotic
behaviour of the pomeron states, a correct tail for the cross section at
large jet transverse momenta. The calculations have been carried out on taking
into account the two supercritical pomeron states. 

In order to obtain a better description of the physical processes at present 
energies a first improvement consists in considering all the states present 
in the spectrum of the pomeron equation, and not only the leading 
contributions of the two supercritical states.
This program \cite{bv3} can be achieved by considering the associated evolution
equation in $1/x$ (with $x$ the Bjorken variable) for the pomeron wave 
function. To this end we are forced to address the problem of the coupling 
of a virtual photon to the gluon in the running coupling framework, since 
otherwise there is no link, for example, between the gluon distribution in
the proton and its structure function.
The bootstrap relation, which concerns only the gluon interaction, is of no
help. The idea is to find a corrected dipole density distribution for the
photon in order to take into account a minimal running coupling effect.
We adopt here a phenomenological approach and guess a possible form of the 
dipole density from the matching in the DLLA of the quark distribution
(which one has after coupling the gluons to the dipole) with the one obtained
by the DGLAP evolution.
Numerical calculations show a still steep increase of the gluon distribution
and the singlet proton structure function, but smaller than the one implied
by the standard BFKL theory.

In the second part of the thesis we deal with the unitarity problem
from a theoretical point of view.
There is a limitation to the validity of the BFKL (fixed coupling)
theory for large $s$ or equivalently large $1/x$ and to extend the kinematical
region of applicability of the model one has to go beyond the LLA.
This problem is shared also by the DGLAP formalism and to solve it one
needs to find a systematic way to include unitarity corrections.
Continuing the restriction to the onia-particles or virtual photons scattering
there is the hope to reach some new insights by means of perturbative tools
even if the complete understanding surely requires a true non-perturbative
approach (the idea of grasping non-perturbative effects with the help
of a perturbative approach can be seen for example in the papers about
renormalons).

In order to understand which kind of corrections one has to deal with, it is
very useful to consider the evolution of the parton densities in the rapidity.
For example it is well know for DIS processes that at
lower $x$ the photon probes in the proton a higher density of partons, of
typical size $1/Q$ at virtuality $Q^2$. The resulting cross section then
increases; the unitarity corrections correspond in this picture to the
fusion of partons which diminishes the rate of increase of the parton density.
A first quantitative model for such a process was given by Gribov, Levin and
Ryskin \cite{grlpr} by constructing a modified non linear BFKL equation
generating a fan diagrams structure (based on triple ladder vertices).
The key point is the presence of multi-parton states in the $t$-channel and
eventually also multi-scattering processes between the partons of the two
scattering onia have to be considered as suggested by the dipole picture.

To this end a systematic approach to take into account a minimal set of
corrections necessary to restore unitarity has been studied by Bartels 
\cite{bart2}.
Higher order amplitudes, which contain any number of (reggeized) gluons 
in the $t$-channel, are constructed using unitarity and dispersion relations
starting from the lower order ones and one has a coupled system of equations
for the $n$ reggeized gluon state amplitudes.
The main feature is the presence of transitions vertices which change the
number of reggeons in the $t$-channel and thus leads to a reggeon field theory.

These corrections have been studied by considering a system of up to four
reggeized gluons in the $t$-channel \cite{bartwue}.
Among the building blocks an effective vertex for the transition between $2$ 
and $4$ gluons states in the $t$-channel appears. In \cite{baliwu} its
conformal invariance property was proven and in the appropriate color 
subspace it is related to the triple pomeron vertex. 
Another element of some interest is the $4$-gluon interacting state (and its
generalizations to $n$ gluons) which, as a system with a fixed number of
reggeized gluons in the $t$-channel, can be studied by means of the so called
BKP equation \cite{bart1,bart3}.
The related amplitude has been studied in the large
$N_c$ limit and the system has been proven to be complete integrable and 
equivalent to an XXX Heisenberg zero spin model \cite{lip2,fadkorch,tatafa}.   
These elements have suggested an interpretations in terms of correlation 
functions of a $2$D conformal field theory (CFT) \cite{pertqcd,korch97} and
one hopes to have an effective formulation in terms of a (2+1) dimensional 
CFT for QCD in the Regge limit.

In the third chapter some efforts are addressed to obtaining a clearer picture
of the interactions present in the system of $4$ reggeized gluons in the 
$t$-channel in the large $N_c$ limit \cite{braun4, bv4}, in particular on
having some understanding of the relation between multipomeron exchange and
triple pomeron interactions and making a connections between this reggeon 
analysis based on the $s$-channel unitarity and the colour dipole picture
for QCD.

We shall study the formal general system of equations for the
$4$ gluon system at leading and next-to-leading order in $1/N_c$
The double pomeron amplitude at NLO is studied to understand the role
of the direct double pomeron exchange (DPE)  and the triple pomeron 
interaction (TPI). The TPI term is analyzed in terms of 3 point 
function building blocks (a generalization to the non-forward case of a
function previously introduced in \cite{bartwue}) to describe in simpler
terms the conformal invariance properties of the system.
It is later shown in detail the fact, already noted for $N_c=3$ in 
\cite{bartwue}, that the DPE contribution can be substituted by a completely
equivalent TPI term while the inverse is not true and there is a complete
agreement with the vertex form in \cite{bartwue} in the limit $N_c\to \infty$.
On coupling to pomerons and working in the coordinate representation one can
compare the results obtained from $s$-channel unitarity with the ones
found by R. Peschanski in \cite{pesh} starting from the colour dipole picture.
A substantial agreement is found, apart from an asymmetry factor in the 
triple pomeron vertex and some single pomeron contributions absent in the 
dipole approach.
A final section is devoted to studying some properties of higher order dipole
densities in the A.H. Mueller approach which are found to be represented by
a set of pomeron fan diagrams with only a triple pomeron coupling.
It is expected that the $s$-channel unitarity analysis for a larger number
of reggeized gluons in the $t$-channel might tell us the form of higher
order pomeron vertices and thus also if the dipole picture has to be
improved, as already suggested \cite{pesh}.\\ 

The thesis is structured in the following way.

\noindent
In the first chapter we review some important results of QCD in LLA which
are at the basis of the developments in the second and third chapters.
In section 1.1 the BFKL equation is briefly derived and in the following one
the pomeron solution with some properties presented.
Section 1.3 is devoted to show in some details the conformal invariance
property of the BFKL equation and finally in section 1.4 a minimal review
of the colour dipole approach in the A.H. Mueller formulation is given.
In the second chapter we analyze a phenomenological model for considering
the running coupling in the hard pomeron.
In section 2.1 the definition of the model is given and the properties of the
pomeron studied \cite{bvv}; in particular
in section 2.1.1 and 2.1.2 the problem is formulated to make possible a
numerical analysis, performed in section 2.1.3, while multi-scattering effects
are considered in section 2.1.4
In section 2.2 the inclusive jet production process is considered \cite{bv2}.
Section 2.2.1 is devoted to set up the formalism and to make asymptotic
estimates while section 2.2.2 collects some numerical results. 
The evolution of the gluon density \cite{bv3} is considered in section 2.3
and in particular in section 2.3.1 we address the problem of the coupling to
the virtual photon, in section 2.3.2 we look at the starting distribution
and in section 2.3.3 some numerical calculations are shown.
Finally in the third chapter the triple pomeron vertex is analyzed in the
large $N_c$ limit \cite{bv4}.
In section 3.1 we introduce some basic elements of the unitarization program
and in section 3.2 we present the analysis for the 4 gluons system in the
large $N_c$ limit, in particular at the leading order in 3.2.1 and at the NLO
in 3.2.2 where we establish the form of the DPE and TPI terms.
In section 3.3 the properties of conformal invariance are re-analyzed in
terms of a simple function and in section 3.4 the interplay between the DPE
and the TPI terms is analyzed in detail.
In section 3.5 the coupling to pomerons and the relation to the dipole
approach is studied while section 3.6 is devoted to study the solutions
for general higher order dipole distributions in the standard dipole picture
and analyze the triple pomeron interactions structure resulted.
In the appendix one can find some more technical parts.
In A.1, related to the phenomenological model considered in the third chapter,
the asymptotic form of the bound state solution of the pomeron equation is
derived. In A.2 the dipole density in the non-forward case for the coupling of
two gluons to a quark loop is given.
In A.3 the coordinate representation of the function $G$, needed in chapter 4,
is calculated and its infrared stability explicitly shown.
In A.4 some relations, used in the fourth chapter, are derived in a graphical
notation.

% cap1

\chapter{Pomeron and the Regge limit in perturbative QCD}

To understand the concept of Pomeron it is necessary to go back to the
old Regge theory and Pomeron field theory.
The Pomeron is the rightmost singularity in the complex angular momentum plane
of the partial wave signatured amplitude of a hadronic scattering process
characterized by the quantum number of the vacuum exchanged in the $t$-channel.
The singularity can be associated to a rise of the cross-section with
the center of mass energy, thus leading to a violation of unitarity which
can be restored only if more complicated processes than a single pomeron
exchange are taken into account. The understanding of the strong interactions
in the Regge limit is therefore deeply connected to unitarity from a general
point of view.

The study of strong interactions as described by QCD theory is well known
to be an incredibly difficult problem and due to the asymptotic freedom which
characterizes QCD many efforts have been directed in perturbative analysis,
hoping to not lose some important aspects, which could serve as a guideline
to understand the features of the complete real QCD.

In particular the concept of the Pomeron has been intensively studied
in perturbative QCD in the Regge limit for a long time \cite{lip1}.
At the basis of the approach is the property of Reggeization of the gluon.
In general the poles in the complex angular momentum plane of the partial
wave amplitude depend on the exchanged $t$ squared momenta so that they may 
describe a family of exchanged particles of integer even or odd angular 
momenta, thus rising with spin and mass. 
An elementary particle, when exchanged in the $t$ channel, gives a non-analytic
contribution to the partial amplitude dependence in the angular momentum, 
and the modern idea of reggeization is related to the fact that summing all the
contributions to the amplitude not containing poles in the $t$-channel leads
to a partial wave amplitude with a pole in the $t$-channel which is located at
the spin of the elementary particles if $t$ is taken to be the mass squared
of the particle itself.
This was verified for the gluon (massive after considering the Higgs 
mechanism) up to the 8-th order (and partially up to 
10-th) in perturbation theory by Cheng and Lo \cite{chelo}, after the work of 
Grisaru \cite{grisaru} where a first confirmation that the gluon reggeizes 
was given.

Another method \cite{lip1} which has been used to check gluon 
reggeization consists in taking a gluon ladder diagram, constructed with 
reggeized gluons on the basis of elastic unitarity (the on shell condition
turns the reggeized gluons into elementary gluons).
On summing these ladder diagrams one
obtains a partial wave amplitude which, in the colour channel, has the same
reggeized gluon Regge pole. This is called a ``bootstrap'' condition and
it is at the basis of the BFKL theory and all its extensions.

The analysis of the processes in the Regge limit in perturbation theory
reduces to summing the leading log terms. In fact for a small fixed
coupling $g$ and high c.m.s. energy $s$ such that $s \to \infty$,
 $g \to 0$ and $g^2 \ln s \sim 1$ the Regge pole contribution has a behaviour
$s^{\alpha(t)}$ where $\alpha(t)=1+\omega(t)$ is the trajectory and 
$\omega(t)$ is of second order in the coupling constant $g$. 
Therefore a perturbative expansion leads one to consider amplitudes of the
order $g^2 s (g^2 \ln s)^n$.

Apart from considering a fixed coupling, its smallness is generally ensured
in the Lipatov perturbative approach by giving to the gluon a mass $M>>\Lambda$
using the Higgs-Brout-Englert mechanism \cite{higgs,broeng}.
Thus one has also an infrared 
regularization but we should remember that in this way one is not really 
working with true QCD since it is not possible to safely perform the 
$m\to 0$ limit without avoiding the non-perturbative region.

The LLA analysis of the amplitude for gluon-gluon scattering in the vacuum 
channel was started in \cite{lip1,bfkl}.
Multiparticle production and unitarity were the ingredients used to
construct a Bethe-Salpeter equation for ladder contributions to the partial
wave amplitude with reggeized gluons in the $t$-channel.
The resulting singularity was found to be a branch cut and not a pole in the
complex angular momentum and the BFKL pomeron to be a ``bound'' state of two 
reggeized gluons. In the colour octet channel instead the Regge pole
ansatz for the gluon applied to the BFKL equation gave directly the 
self-consistency bootstrap condition.

The BFKL equation can be seen as a Schr\"odinger-like equation in the
transverse space with the evolution flow in the rapidity variable.
We just remember, beyond the violation of unitarity, an intrinsic problem in 
the BFKL perturbative approach due to the diffusion of the gluon
momenta in the ladders.  For a highly virtual projectile ($Q^2$) the 
scattering is characterized by a distribution which tends to be Gaussian in 
$\ln {\bf k}^2$ with an average value $\ln Q^2$ and a width proportional to 
$\sqrt{\ln s}$, thus with a support which can be strongly present in the
non-perturbative region.

The opportunity of applying the results found in this perturbative approach
to analyze some real processes is based on the possibility of separating in
them the contribution of short distance interactions (hard scales), where 
perturbation theory can be applied, from the long distance ones (soft scales)
where perturbation theory is meaningless.

This is the meaning of the factorization theorems. A first kind used to
study many hard processes characterized by a hard scale $Q^2$ at which 
the coupling is small is the collinear factorization theorem \cite{collith}.
An object such as, for example, a structure function in DIS can be written as
a convolution between a cross section for an hard parton subprocess 
(perturbatively calculable and depending on the process virtuality) and a 
distribution for the parton probed in the hadron involved.
The latter distribution depends on the virtuality chosen to probe the hadron 
by means of a renormalization group equations (DGLAP \cite{dglap}) so that a 
non-perturbative input at a lower scale $Q^2_0$ is necessary to provide
the initial conditions for these equations (which are constructed on all the
collinear logarithmic contributions).
The DGLAP equations can be associated, in the physical gauge, with summing all
the contributions of ladder graphs with strongly ordered parton momenta
but, in the small $x$ (high energy) limit, higher order contributions, beyond 
the leading logs, are found to be significant. This fact has led to consider
the perturbative resummation of contributions of the order 
$(\alpha_s \ln 1/x)^n$ which correspond to the ladder graphs with reggeized
gluons in the $t$-channel and no ordering in the transverse momenta.
In this framework a new ${\bf k}$-factorization theorem has been given
\cite{kfacth}.

We shall introduce in the next section the BFKL equation and show some of its 
properties and the solutions in the subsequent one.
In the third section the proof of its conformal invariance is given, a feature 
which turns out to be very interesting and also characterizes the symmetry of
the three pomeron vertex. We shall return to these properties in the last
chapter.
The Colour dipole approach in the A. H. Mueller formulation 
\cite{mueller,muelpat} is briefly reviewed
in the last section since its use is pertinent in the high energy (and
a large number of colours) limit and it has been shown to give an alternative
formulation to the hard pomeron phenomena.
In chapter 3 we shall try to compare the pomeron interactions described in
the two formulations to see the degree of equivalence of the two approaches.

%%%%%%%%%%%%%%%%%%%%%%%%%%%%%%%%%%%%%%%%%%%%%%%%%%%%%%%%%%%%%%%%%%%%%%%
\section{The BFKL equation}
The main results about the LLA are briefly derived in the following
\cite{bfkl,braun94}.
We consider the absorptive part 
$A_s(s,t)={\mbox Im}A(s,t)$
of the dominant non-helicity-flip amplitude for the scattering of two gluons 
in the limit $s \to \infty$, $t$ finite and $\alpha_s =g^2/(4\pi)<<1$ in the
$t$-channel colour state $R$. 
Using unitarity one can compute this quantity by means of the amplitude for 
the production of $(n+2)$ gluons from two:
\beq
{\mbox Im}A^{(R)}(s,t)=\frac{P^{(R)}}{2}\sum_{n}\int d\tau_{n+2}
A^{\ast}(n+2,p'_{1},p'_{2})A(n+2,p_{1},p_{2})
\label{absoamp}
\eeq   
where $=P^{(R)}$ is the colour projector, $p_1=(p_+,0,0_\perp)$ and 
$p_2=(0,p_-,0_\perp)$ are the momenta of the
incoming gluons in the c.m.s. in the light cone representation, so that 
$s=2p_1\cdot p_2=2p_+^2$ and $d\tau_{n+2}$ is the $(n+2)$ particle phase 
measure given by
\beq
d\tau_{n+2}=\prod_{i=0}^{n+1}(d^{3}k_{i}/(2\pi)^{3}2\kip)
(2\pi)^{4}\delta^{4}(\sum k_{i}-p_{1}-p_{2})
\eeq   
The momenta $k_i$ of the intermediate particles satisfy an on-shell condition.
It is convenient to use the Sudakov variables $\alpha_i$ and $\beta_i$ such 
that $k_i=\alpha_i p_{1}+\beta_i p_{2}+k_{i\perp}$, 
$\ai\bi=m_{i\perp}^2/s$ with $m_{i\perp}^2=m^2-k_{i\perp}^2$.
Other useful variables are the transferred momenta 
$q_i=p_1 - \sum_{j=0}^{i-1}k_{j}$ ($k_{i}=q_{i}-q_{i+1}$) and the partial
energies carried by two adiacent intermediate real particles 
$\sij=(k_{i}+k_{i+1})^{2}$.

In the Regge limit the dominant leading log(s) contribution comes from
the region in the phase space characterized by
\beq
\alpha_i>>\alpha_{i+1} \quad , \quad \beta_i << \beta_{i+1}
\eeq
and thus one has approximatively the equalities $\sij=(\ai /\aj )\mjt$ and
$q_i^2 =q^2_{i\perp}$.
Changing the variables which represent the measure in phase space to $q_i$
and $s_{i,i+1}$ and performing an integration in the tranverse space and
the integrations w.r.t. $\alpha_0$ and $\alpha_{n+1}$ one finally may write
\beq
d\tau_{n+2}=\frac{\pi}{s}\prod_{i=1}^{n+1}\frac{d^{2}q_{i}d\sji}
{(2\pi)^{3}}\delta(\prod_{i=0}^{n}\sij-s\prod_{i=1}^{n}\mit)
\eeq

Let us now consider the amplitude for the production of $(n+2)$ gluons.
In particular we shall denote by $\lambda_i$, $a_i$ and  $\lambda_i'$, $a_i'$
the helicities and colours of the two incoming gluons and of the two outgoing
gluons with nearly the same momenta as the initial ones and by $e_i$ and
$d_i$ the polarizations and colours of the remaining $n$ gluons.

In the Regge limit the virtual corrections to the tree diagrams may
be analyzed by looking at the one-loop corrections of the 
scattering process $gg \to gg$ with both the diagrams related to the 
$s$-channel and $u$-channel physical region.
One finds a correction of the type $\alpha(t) \ln s$ and it is natural
to make the reggeization ansatz for a correction $(s/s_0)^{\alpha(t)}$ in
all orders in $\alpha_s$ in the LLA.
The trajectory found for the gluon is $\alpha(t)=1+\omega(t)$ with
\beq
\omega(t)=N_c\alpha_{s}t\int\frac{d^{2}l}{4\pi^{2}l^{2}(q-l)^{2}} \quad ;\quad
t=-q^{2}
\eeq    
Alternatively one may use a general reggeization ansatz for the gluons in the
$t$-channel and find from the boostrap condition the trajectory, for a case of
the two gluons amplitude in the colour adjoint representation; this confirms
the fact that the gluon reggeizes.

The real corrections in the LLA leads one to consider a three gluons
reggeon-physical-reggeon vertex (the reggeons in the $t$-channel) whose 
explicit expression is
\beq
F(q_{i+1},q_{i})=-q_{i\bot}-q_{i+1,\bot}+
p_{1}(\frac{k_{i}p_{2}}{p_{1}p_{2}}+\frac{\qit}{k_{i}p_{1}})
-p_{2}(\frac{k_{i}p_{1}}{p_{1}p_{2}}+\frac{\qjt}{k_{i}p_{2}})
\eeq    
It is orthogonal ($F(q_{i+1},q_{i})k_{i}=0$) to the emitted gluon momentum.
From the analysis of the two physical-reggeon-physical gluons vertices
(external in the ladder) the dominant contribution is helicity conserving
and proportional to $s$ so that the amplitude $A_{2\to 2+n}$ can be written as
\beq
A_{2\to 2+n}=2sg\delta_{\lambda_{1}\lambda'_{1}}T_{a'_{1}a_{1}}
^{i_{1}}s_{01}^{\omega(t_{1})}t_{1}^{-1}gT_{i_{2}i_{1}}^{d_{1}}
(e_{1}F(q_{2},q_{1}))s_{12}^{\omega(t_{2})}t_{2}^{-1}
gT_{i_{3}i_{2}}^{d_{2}}(e_{2}F(q_{3},q_{2}))...\]
\[...s_{n,n+1}^{\omega(t_{n+1})} t_{n+1}^{-1}g\delta_
{\lambda_{2}\lambda'_{2}}T_{a'_{2}a_{2}}^{i_{n+1}}
\eeq    
where $t_i=-q_i^2$
In order to obtaining an expression for (\ref{absoamp}) it is necessary to sum
over the polarizations and the colours of the intermediate real gluons and
consider the projection onto the desired colour subspace (we shall be mainly
interested in the vacuum state).

The sum over polarizations leads to an effective interaction vertex between
the two reggeized gluons. Starting from $\sum_{\lambda_i} e^{\mu_i}_{\lambda_i}
e^{\nu_i}_{\lambda_i} F_{\mu_i}(q_i,q_{i+1}) F_{\nu_i}(q_i',q_{i+1}')$
and performing the sum over the polarizations (actually using the replacement
$\sum_{\lambda_i} e^{\mu_i}_{\lambda_i} e^{\nu_i} \to - g^{\mu_i\nu_i}$)
one obtains
\beq
F \cdot F'= -2 K_q(q_i,q_i',q_{i+1},q_{i+1}') \quad ; \quad
K_q(q_i,q_i',q_{i+1},q_{i+1}')=
q^{2}-(q_{1\bot}^{2}{q'}_{2\bot}^{2}+
q_{2\bot}^{2}{q'}_{1\bot}^{2})/k_{\bot}^{2}   
\eeq  

Regarding the colour structure in the $t$-channel, the gluon-gluon amplitude
has to be decomposed into irreducible representations of the 
$(N_c^2-1) \otimes (N_c^2-1)$; by means of a projection operator
one has
\beq 
A_{a_{2}a'_{2},a_{1}a'_{1}}(s,t)=
\sum_{(R)}P^{(R)}_{a_{2}a'_{2},a_{1}a'_{1}}A^{(R)}(s,t)
\eeq                                                       
For each real gluon emission one has a colour factor which depends on the
representation $R$ according to the decomposition of the product of the two
corresponding colour matrices
\beq
\Bigl[ \sum_{d_{i}}
T^{d_{i}}_{b_{i+1},b_{i}}(T^{d_{i}}_{b'_{i+1},b'_{i}})^{\ast}\Bigr]^{(R)}=
\Bigl[-(TT')_{b_{i+1}b'_{i+1},b_{i}b'_{i}}\Bigr]^{(R)}=
\lambda_{R}\delta_{b_{i+1}b_{i}}\delta_{b'_{i+1}b'_{i}}  
\eeq    
so that the interation term is given by
\beq
\tilde{K}^{(R)}=\lambda_{R}K_{q}
\eeq   
Also for the first and last intermediate gluons one has the same colour factor.
In particular one has for the vacuum channel $\lambda_1 \equiv \lambda_0=N_c$
and for the adjoint representation $\lambda_{N_c^2-1}=N_c/2$.

The absorptive amplitude (\ref{absoamp}) can thus be written in the form
\[ {\mbox
Im}A^{(R)(n)}(s,t)=\pi sg^{2(n+2)}\lambda_{R}^{2}
\int\prod_{i=1}^{n+1}\frac{d^{2}q_{i}d\sji}{2(2\pi)^
{3}}\Bigl(\frac{\sji}{s_0}\Bigr)^{\omega(t_{i})+\omega(t'_{i})}
\frac{1}{t_{i}{t'}_{i}}\]
\beq
\prod_{i=1}^{n}2\tilde{K}^{(R)}_{q}(q_{i},q_{i+1})
\delta(\prod_{i=0}^{n}\sij-s\prod_{i=1}^{n}|\kit|)
\eeq                                            
and has the signature fixed by the $R$ representation (positive for symmetric
and negative for antisymmetric representations).
Taking the Mellin transform one obtains for it an asymptotic contribution
(which thus corresponds to the Mellin transform nearby the rightmost 
singularity)
\[ a_{j}^{(R)(n)}(t)=2
g^{2(n+2)}\lambda_{R}^{2}
\int\prod_{i=1}^{n+1}\frac{d^{2}q_{i}}{16\pi^{3}t_{i}{t'}_{i}
(j-1-\omega(t_{i})-\omega(t'_{i}))}\]
\beq
\prod_{i=1}^{n}|\kit|^{j-1}
2K^{(R)}_{q}(q_{i},q_{i+1})
\eeq       
We note that in the validity region ($j$ close to $1$) the $|\kit|^{j-1}$ 
factor may be omitted.
Changing the notation to the 2-dimensional transverse vectors
$l_{1i}=q_{i\perp}$, $l_{2i}=-q'_{i\perp}$ such that $l_{1i}+l_{2i}=q$,
$t_i=-l_i^2$ and $t=-q^2$ one may write in an operatorial form
\beq
a_{j}^{(R)(n)}(t)=2 g^{2}\lambda_{R}^{2}
\int \frac{d^{2}l_{1,1}}{16\pi^{3}t_{11}t_{21}}
\langle l_{11}|D(V^{(R)}D)^{n}|l_{1,n+1} \rangle
\frac{d^{2}l_{1,n+1}}{16\pi^{3}}
\eeq   
where we have defined the operators
\beq
D(l_{1})=\frac{1}{j-1-\omega(t_{1})-\omega(t_{2})}
\quad ; \quad
V^{(R)}(l_{1},l'_{1})=2\frac{\tilde{K}_{q}^{(R)}(l_{1},l'_{1})}
{{l'_{1}}^{2}{l'_{2}}^{2}}
\eeq
It is now very easy to sum the contributions of all the ladder diagrams and
one has
\beq
a_{j}^{(R)}(t)=2 g^{2}\lambda_R^2
\int \frac{d^{2}l_{1,1}} {16\pi^3} \frac {\phi^{(R)}_{jq}}{t_{11}t_{21}}
\eeq
with
\beq
\phi_{jq}^{(R)}(l_{1})=\int \frac{d^{2}l'_{1}}{16\pi^{3}}
\langle l_{1}|D(1-V^{(R)}D)^{-1}|l'_{1} \rangle
\eeq     
The Lipatov equation is a Bethe-Salpeter equation for $\phi_{jq}^{(R)}(l_{1})$
which can be obtained by taking the matrix element of the operatorial identity 
$D^{-1} T^{(R)}=1+V^{(R)} T^{(R)}$ w.r.t. the states $\langle l_1 |$ and 
$| l'_1 \rangle$ where $T^{(R)}=D(1-V^{(R)}D)^{-1}$ and integrating w.r.t. 
$l'_1$. It reads
\beq
[j-1-\omega(t_{1})-\omega (t_{2})]\phi_{jq}^{(R)}(l_{1})=
1+2\alpha_{s}\lambda_{R}\int
\frac{d^{2}l'_{1}}{4\pi^{2}}
V_{q}(l_{1},l'_{1})\phi_{jq}^{(R)}(l'_{1})
\eeq                       
with the interaction potential given by
\beq
V_{q}(l_{1},l'_{1})=(\frac{l_{1}^{2}}{{l'_{1}}^{2}}+
\frac{l_{2}^{2}}{{l'_{2}}^{2}})\frac{1}{
(l_{1}-l'_{1})^{2}}-\frac{q^{2}}{{l'_{1}}^{2}{l'_{1}}^{2}}
\label{intpot}
\eeq
This equation in the gluon channel gives the bootstrap condition if one
requires that the solution of this equation must have a Regge pole of
a reggeized gluon, i.e. at $j=1+\omega(t)$. Denoting by $c_q(l_1)$ the residue
of the solution at this pole, the bootstrap condition can be written for 
$N_c=3$ as
\beq [\omega(t)-\omega(t_{1})-\omega (t_{2})]c_{q}(l_{1})=
3\alpha_{s}\int
\frac{d^{2}l'_{1}}{4\pi^{2}} V_q(l_{1},l'_{1}) c_{q}(l'_{1})
\eeq    
The gluon trajectory found by perturbative calculations corresponds to
a residue $c_q$ independent of $l_1$. Hence in the gluon channel the whole 
solution for $\phi_{jq}^{(8_A)}(l_{1})$ is determined by this pole.

%%%%%%%%%%%%%%%%%%%%%%%%%%%%%%%%%%%%%%%%%%%%%%%%%%%%%%%%%%%%%%%%%%%%%%%
\section{Solution for the pomeron amplitude}
The Lipatov equation can be written as an inhomogeneous Schr\"odinger equation
$(H-E)\phi=f$ with an ``energy'' $E=1-j$ and an Hamiltonian
\beq
H=-\omega(l_1)-\omega(l_2)-2\alpha_s \lambda_R V_q
\eeq
In the vacuum case one has $\lambda_R=N_c$ and one may check that for
$q \ne 0$, as will be shown also in the next section, the infrared 
infrared divergences cancels between the gluon trajectories and the 
interaction potential term. For $q=0$ the divergences at $l'_1=l'_2=0$
remain but actually disappear for the coupling of the pomeron to
a colourless physical object, since the function $\phi_{jq}(l\to0,q)$
vanishes.

Let us consider the forward scattering case ($q=0$). It is convenient to
consider a new function $\psi(l_1)=\phi(l_1)/l_1^2$ for which one can write
the equation
\beq
[j-1-2\omega (t_{1})]\psi(l_{1})=\frac{1}{l_{1}^{2}}+4\alpha_{s}\lambda_{R}
\int\frac{d^{2}l'_{1}}{4\pi^{2}}\frac{\psi(l'_{1})}{(l_{1}-l'_{1})^2}
\eeq
The interaction term is of the Coulomb type and both the trajectory and the
potential can be studied in the limit of zero regularizing mass.
Using the expressions given in the next section (\ref{intappr}) and
(\ref{trajappr}) one may write the Hamiltonian for the vacuum colour channel
($\lambda_0=N_c=3$)
\beq
H=(3\alpha_{s}/\pi)\,h \quad ; \quad 
h=\ln l^{2}+\ln \rho^{2}-2(\ln 2+\psi(1)) \quad ; \quad 
E=1-j=(3\alpha_{s}/\pi)\epsilon   
\eeq
where the dependence on the gluon mass has disappeared and $\psi(1)$ is
referred to the logarithmic derivative of the $\Gamma$ function.
To find the eigenvalues and eigenfunction of this Hamiltonian one can take
advantage of the scaling symmetry $\rho \to c \rho \ , \ l \to l /c$ which is
manifest. The symmetry between the coordinate and the momentum space is 
complete and one may write easily the proper functions for the system
both in coordinate and momentum space
\beq
\psi_{\nu n}(\rho)=a_{\nu n}\rho^{-1+i\nu} e^{i n \phi}
\quad ; \quad
\psi_{\nu n}(l)=b_{\nu n}l^{-1-i\nu}e^{i n \phi}  
\eeq 
for any real $\nu$ and integer $n$.
It is also easy to show that these functions are orthonormalized if
$|a_{\nu n}|^{2}=|b_{\nu n}|^{2}=1/4\pi^{2}$.
Defining $\mu=(1/2)(1+n+i \nu)$ the relation between the two normalizing
factors is for $n\ge0$
\beq
b_{\nu n}=a_{\nu n}i^{n}2^{-i\nu}\frac{\Gamma (\mu^{\ast})}{\Gamma(\mu)}
\eeq
while the $n<0$ case is obtained by complex conjugation.
The eigenvalues are computed through the definition $h \, \psi_{\nu n}=
\epsilon_{\nu n} \,\psi_{\nu n}$; in particular one has
\beq
\epsilon_{\nu n}=2{\mbox Re}\, \psi\Bigl(\frac{1+|n|+i\nu}{2}\Bigr)-2\psi(1)
\eeq
The minimal value $\epsilon_0=-4 \, \ln 2$ is reached at $\nu=0$ and $n=0$ 
and gives for the BFKL pomeron the intercept
\beq
j=1+\frac{3\alpha_{s}}{\pi}\, 4 \ln 2
\eeq    
From this value the violation of the Froissart bound
\cite{froiss,butchm,ayala} derived from unitarity is evident.
The pomeron Green function in the forward direction can be easily constructed
considering the inhomogeneous term of the equation for $\psi$, 
$f=(2\pi)^{2}\delta (l-l')l^{-2}{l'}^{-2}$, and in momentum space is
\beq
G_{\epsilon}(l,l')=
\frac{1}{4\pi^{2}l^3l^{'3}}\sum_{n=-\infty}^{\infty}
e^{i n (\phi-\phi')}\int d\nu\,\Bigl(\frac{l'}{l}\Bigr)^{i\nu}
\frac{1}{\epsilon_{\nu n}-\epsilon}
\eeq    

To see which behaviour the solution has in the high momentum (short range)
limit one should study the singularities of the Green function in the $\nu$
plane. Since we are interested in the dominant contribution, which one has for
$\epsilon$ close to $\epsilon_0$, we consider the expansion up to the
second order
\beq
\epsilon_{\nu 0}=\epsilon_{0}+a\nu^{2}
\eeq
where $a=(7/2)\zeta(3)$. Defining $w=\epsilon_{0}-\epsilon=
(\pi/3\alpha_s)(j-1)-|\epsilon_{0}|$ the singularity is at $w=0$ and
the asymptotic behaviour for large $l$ is calculated by taking the residue at
$\nu=i\sqrt{w/a}$. Thus one obtains
\beq
\psi_\epsilon(l)=\sum_{n=-\infty}^{\infty}e^{i n \phi}\int_{-\infty}^{+\infty}
d\nu l^{-1+i\nu}b_{\nu n} \frac{\langle \psi_{\nu n} | f \rangle}
{\epsilon_{\nu n}-\epsilon} \simeq
\frac{\langle \psi_{00} | f \rangle}{2 l \sqrt{a w}}
e^{-\sqrt{\frac{w}{a}}\ln l}
\eeq
and it is evident that the pomeron singularity is branch cut of type
$1/\sqrt{j-1-|\epsilon_0|}$.
The leading behavior at high energies $s$ can be found by performing an inverse
Mellin transform for the positive signatured amplitude
\beq
\psi(s,l)=\frac{i}{4}\int dj\psi_{j}(l)s^{j}\frac{e^{-i\pi j}+1}{\sin\pi j}
\eeq   
giving the dominant contribution
\beq
\psi(s,l)\simeq
\frac{i\pi}{4l} s^{1+\frac{3\alpha_{s}}{\pi}|\epsilon_{0}|}
\frac{3\alpha_s}{\pi}\frac{\langle \psi_{00} | f \rangle}
{\sqrt{3\alpha_{s}a\ln s}}
e^{-\pi\frac{\ln^{2} l}{12\alpha_{s}a\ln s}}
\eeq    
Thus the average value of $l$ increases with $s$ since the distribution is
Gaussian in $\ln l$ with a width proportional to $\sqrt{\ln s}$.
This fact is characteristic of a diffusion process in the transverse momenta;
the Lipatov equation can in fact be seen as an diffusion equation in the
rapidity flow.
From this fact we address an important problem in the BFKL theory.
Such a diffusion process may spoil the applicability of perturbation
theory since at sufficient large energy the distribution in the transverse
momenta can have support in the small values region corresponding to large
distances and typically regarding a non-perturbative regime.

%%%%%%%%%%%%%%%%%%%%%%%%%%%%%%%%%%%%%%%%%%%%%%%%%%%%%%%%%%%%%%%%%%%%%%%
\section{Conformal Invariance}
The BFKL equation manifests two important properties: the holomorphic 
separability and conformal invariance.
An elegant way to prove them \cite{lip3,pertqcd,braun94} can be achieved by 
rewriting the Hamiltonian of the two interacting reggeized gluons in the 
complex coordinate space representation for the transverse plane.
In the following we will fix the notation: given two vectors 
${\bf a}=(a_x,a_y)$ and ${\bf b}=(b_x,b_y)$ we have
\beq
a=a_x+i a_y; \quad, 
{\bf a}\cdot {\bf b}=\frac{1}{2}(a b^*+a^*b)
\eeq
The complex vectors for the impact parameter $r_i$ will be conjugate to the
momenta $l_i=-i\frac{\partial}{\partial r_i}$. It is also convenient to 
introduce the relative momentum $l=\frac{1}{2}(l_1-l_2)$ and relative distance
${\bf \rho}={\bf r_1}-{\bf r_2}$ (vector) and 
$\rho=\frac{1}{2}(r^*_1-r^*_2)$ (complex number) so that $\rho$,$l$ 
and $\rho^*$,$l^*$ forms, each one, an independent pair of canonical 
variables and $|{\bf \rho}|^2=4 \rho \rho^*$. 
It should be clear from the context which is, each time, used.
Further if in the configuration space we are referring to a vector or to a
complex number should be clear from the context.
Moreover we define $k=l_1-l'_1$.

The interaction potential (\ref{intpot}) can be rewritten as
\beq
V_q(l_1,l'_1)=l_1 l^*_2\frac{1}{k k^*} (l'_1 {l'}^*_2)^{-1} + c.c.=
l_1 l^*_2 W(\rho) (l'_1 {l'}^*_2)^{-1} + c.c.
\eeq
where $W(\rho)$ is a Coulombian potential in the coordinate representation.
The explicit form of $W$, after having introduced an infrared regulator $m$, is
\beq
W(\rho)=\frac{K_0(m\rho)}{2\pi} \mathop{\approx}_{m \to 0}
\frac{1}{2\pi}\Bigl[\psi(1)-\ln\frac{m \rho}{2}\Bigr]
\label{intappr}
\eeq 
and consequently one obtains
\beq
V_q=-\frac{1}{4\pi}\Bigl[ l_1 \ln\rho \, l_1^{-1} + l_2 \ln\rho \, l_2^{-1}
+\ln m^2-2\psi(1)\Bigr] + c.c.
\eeq
The regularized form of the gluon Regge trajectory reads
\beq
\omega(l_1)\mathop{\approx}_{m \to 0} -\frac{N_c \alpha_s}{2\pi}
(\ln l_1 - \ln m +c.c.)
\label{trajappr}
\eeq
so that the Hamiltonian is holomorphic separable, i.e. 
$H=\tilde{H}(\rho,l_i)+ c.c.$, and, for the R colour channel,
\beq
\tilde{H}^R(\rho,l_i)=\frac{N_c \alpha_s}{2\pi}
\Bigl[ \ln l_1 + \ln l_2 +  \frac{\lambda_R}{N_c}
(l_1 \ln\rho \, l_1^{-1} + l_2 \ln\rho \, l_2^{-1})
+(\frac{\lambda_R}{N_c}-1) \ln m^2 -2 \frac{\lambda_R}{N_c}\psi(1) \Bigr]
\eeq
In the vacuum channel the infrared divergences (regularized by $m$) 
in the interaction and the trajectory terms cancel and on rescaling the wave
function $\phi$ in the form $\phi=l_1^2 l_2^2 \psi$, the new Hamiltonian for
$\psi$ will be given by $H'=\frac{N \alpha_s}{2\pi} (h_{12}+\bar{h}_{12})$ with
\beq
h_{12}=  \ln l_1 + \ln l_2 +  l_1^{-1} \ln\rho \, l_1 + l_2^{-1} \ln\rho \, l_2
  -2 \psi(1)
\label{holoinv}
\eeq
The holomorphic separability of the Hamiltonian is an important property and
as a consequence the eigenvalues will be given by the sum of two contributions,
one the complex conjugate of the other, and the corresponding eigenfunctions
will be factorized in two parts, the first depending only on $r_1$,$r_2$
and the second part on the conjugate quantities.

Let us now consider a global conformal transformation in two dimension, 
represented by the M\"obius mapping
\beq
r \to r'=\frac{a r + b}{c r + d}
\eeq
where $a,b,c,d$ are complex number such that $a d -b c=1$. The generators
of the M\"obius group are $M^0=r l$, $M^-=l$, $M^+=r^2 l$ where 
$l=-i \frac{\partial}{\partial r}$; $M_0$ generates the rotation and scale
transformations and $M^-$ the translations. The Casimir operator is given by
$M^2={M^0}^2-\frac{1}{2}\{M^+,M^-\}$. For a system of $n$ gluons the
generators are given by the direct sum of the generators acting on the
different gluon spaces.
  
The Hamiltonian , which describes the two gluon system, is evidently invariant
under the action of the $M_0$ and $M^-$ generators 
(also $h_{12}$ given in (\ref{holoinv}) is invariant). To prove the conformal
invariance it is convenient \cite{lip3,pertqcd,braun94} to consider the finite 
transformation defined by $r \to -\frac{1}{r}$ which leads to
\beq
\rho \to \frac{\rho}{r_1 r_2} \quad ; \quad l_i \to r_i^2 l_i
\label{somefinite}
\eeq 
and rewrite the Hamiltonian in a different form.
Let us therefore consider the following operator identities
\bea
\partial_1=\Gamma^{-1}(1+z)\rho^{-1} \, \Gamma(1+z) \quad &\to&
\ln \partial_1=-\Gamma^{-1}(1+z) \ln\rho \, \Gamma(1+z) \nonumber \\
\rho^2\partial_1=\Gamma(z)\rho \, \Gamma^{-1}(z) \quad &\to&
\ln \rho^2\partial_1=\Gamma(z)\ln \rho \, \Gamma^{-1}(z)
\label{pseudoide1}
\eea
where $\partial_1=\frac{\partial}{\partial r_1}$ and $z=\rho \partial_1$.
They are similarity transformations between $\rho^{-1}$ and $\partial_1$
and between $\rho$ and $\rho^2\partial_1$.
By noting that $[z,\ln \rho]=1$ one can use the representation $\ln \rho=
- \frac{\partial}{\partial z}$ and derive from the relations 
(\ref{pseudoide1})
\bea
\ln \partial_1&=& -\ln \rho +\psi(z)+\frac{1}{z} \nonumber \\
\ln \rho^2\partial_1&=&\ln \rho + \psi(z)
\label{pseudoide2}
\eea
Introducing these relations in (\ref{holoinv}) and also substituting the 
operator identity $l_1^{-1}\ln \rho \, l_1=\ln \rho -\frac{1}{z}$
the holomorphic part $h_{12}$ can be written as
\beq
h_{12}=  \ln \rho^2\partial_1 + \ln \rho^2\partial_2 -2 \ln \rho -2 \psi(1)
\label{holoinv2}
\eeq
In this last expression we made use of the cancellation of imaginary terms
between the holomorphic and antiholomorphic part.

It is now easy to prove the invariance of the Hamiltonian with respect to the
transformation $r_i \to -\frac{1}{r_i} \, ,i=1,2$.
In fact applying the transformations (\ref{somefinite}) one has
\beq
\rho^2\partial_1 \to \frac{\rho^2\partial_1}{r_2^2} \quad ; \quad
\rho^2\partial_2 \to \frac{\rho^2\partial_2}{r_1^2} 
\eeq
and the conformal invariance is verified by direct substitution.

Once the conformal invariance is known to be a property of the system, one
can take advantage of it by expanding the general solution of the BFKL
equation in terms of the conformal basis functions, which are eigenfunctions
of the Casimir operator of the M\"obius group and are then related to
its irreducible representations.

For a system of two gluon the Casimir operator for the holomorphic part is 
given by $M^2 =-\rho l_1 l_2$. Using the notation $\rho=r_{12}=r_1-r_2$
the conformal basis is defined by
\beq
M^2 \bar{M}^2 E_\mu(r_1,r_2)=\frac{1}{16} r_{12}^4 \partial_1^2 \partial_2^2
E_\mu(r_1,r_2)=\frac{1}{16} \mu(\mu-1)\bar{\mu}(\bar{\mu}-1)E_\mu(r_1,r_2)
\eeq
where $\mu=\frac{1-n}{2}+i \nu$ and $\bar{\mu}=\frac{1+n}{2}+i \nu$ are
the conformal weights which label the representation; $n$, the conformal spin, 
is an integer and $\nu$ is a real number.
The basis functions (in complex notation) are explicitly given by
\beq
E_\mu(r_1,r_2)=E_{n,\nu,r_0}(r_1,r_2)=
\Bigl( \frac{r_{12}}{r_{10} r_{20}} \Bigr)^{ \frac{1-n}{2}+i \nu}
\Bigl( \frac{r_{12}^*}{r_{10}^* r_{20}^*}\Bigr)^{\frac{1+n}{2}+i \nu}
\eeq
One notes the presence of an additional quantum number, the coordinate $r_0$,
in some way related to translational invariance.
The corresponding eigenvalues of the Casimir operator can be
written conveniently to obtain
\beq
M^2 \bar{M}^2 E_\mu(r_1,r_2)=\frac{\pi^8}{4 a_{n+1,\nu} a_{n-1,\nu}}
E_\mu(r_1,r_2)
\label{eqconfbas}
\eeq
where we have used the standard notation
\beq
a_{n,\nu} \equiv a_\mu = \frac{\pi^4}{2}\frac{1}{\nu^2+n^2/4}
\eeq
The conformal basis forms a complete system:
\beq
r_{12}^4\delta^2(r_{11'})\delta^2(r_{22'})=
\sum_{\mu}
E_{\mu}(r_1,r_2)E^*_{\mu}(r'_1,r'_2)
\label{comple}
\eeq       
The notation used regarding the sum over the quantum numbers is the following
\beq
\sum_{\mu}=\sum_{n=-\infty}^{\infty}\int d\nu\frac{1}{2a_{n,\nu}}
 \int d^2r_0
\eeq
Another important property is the orthogonality relation which contains
an additional term due to the equivalence of the representations which 
correspond to the $(n,\nu)$ and $(-n,-\nu)$ quantum numbers. It is given by
\bea
&&\int\frac{d^2r_1d^2r_2}{r_{12}^4}
E_{n,\nu,r_0}(r_1,r_2)E^*_{n',\nu',r'_0}(r_1,r_2) =\nonumber \\
&&a_{n,\nu}\delta_{nn'}\delta(\nu-\nu')\delta^2(r_{00'})+
b_{n,\nu}\delta_{n,-n'}\delta(\nu+\nu')|r_{00'}|^{-2-4i\nu}(\frac{r_{00'}}
{r^*_{00'}})^n
\label{orthorel}
\eea   
where
\beq
b_{n,\nu}=\pi^3 \frac{2^{4 i \nu}}{-i \nu +|n|/2}
\frac{\Gamma(-i\nu+(1+|n|)/2)}{\Gamma(i\nu+(1+|n|)/2)}
\frac{\Gamma(i\nu+|n|/2)}{\Gamma(-i\nu+|n|/2)}
\eeq
The proof of these relations can be found in\cite{lip3}.
The spectrum of the BFKL Hamiltonian for the general non-forward case 
($q \ne 0$) is the same as the forward case, as a consequence of the global
conformal invariance:
\beq
\omega_\mu=\omega_{n,\nu}=\frac{N_c \alpha_s}{\pi}
\Bigl[ 2 \psi(1)-\psi(\frac{1+|n|}{2}+i\nu)-\psi(\frac{1+|n|}{2}-i\nu) \Bigr]
\eeq
The solution of the BFKL equation with an inhomogeneous term $\Psi_0$ can then
be expanded in terms of the conformal basis functions: 
in the Mellin representation one has
\beq
\Psi_j(r_1,r_2)=\sum_{\mu} \frac{E_{\mu}(r_1,r_2)
\langle\mu|\Psi_{0}\rangle}{j-1-\omega_\mu}
\label{bfkljexp}
\eeq    
and moving to the rapidity space representations, where the BFKL equation
is an evolution equation in the rapidity variable, one finds
\beq
\Psi(r_1,r_2;y)=\sum_{\mu}e^{y\omega_{\mu}}E_{\mu}(r_1,r_2)
\langle\mu|\Psi_0\rangle
\label{bfklyexp}     
\eeq
Here we have defined    
\beq
\langle \mu|\Psi_0\rangle=
\int \frac{d^2r_1d^2r_2}{r_{12}^4}E^*_{\mu}(r_1,r_2)\Psi_0(r_1,r_2)
\eeq   

%%%%%%%%%%%%%%%%%%%%%%%%%%%%%%%%%%%%%%%%%%%%%%%%%%%%%%%%%%%%%%%%%%%%%%
\section{The colour dipole approach in the large $N_c$ limit}
The light-cone wave function of colourless objects, such as onia or virtual
photons, has been studied by many authors \cite{mueller,muelpat,nikzak1} to 
compute the corrections to the
wave function given by the emission of soft gluons with scaling variables
strongly ordered. The kinematical region $z_{k+1}>>z_k$ thus relates the
results found to the one obtained in the BFKL approach.
Incidentally the derivation is quite different and simpler and avoids some
suspicious assumptions taken in the construction of the BFKL theory.

The colour dipole approach has also the merit of giving, without further
effort, some insight in the multipomeron exchange and interactions as we
shall see in chapter 3.
In this section we shall sketch a brief derivation of the model in the
approach given by A.H. Mueller \cite{mueller,muelpat,braun94} and show its 
relation to the BFKL pomeron.

The first object introduced is the wave function of the colourless object
(onium or virtual photon) with momentum $(p_+,0,{\bf 0})$
in the lowest order in the coupling, with no soft gluons emitted, so that
we may think of it as split in a quark-antiquark pair.
Let $k_1$ and $-k_1$ denote the transverse momenta of the antiquark and of
the quark respectively; the scaling variable for the antiquark is defined by
$z_1=k_{1+}/p_+$. We shall mainly use the coordinate representation in which 
the wave function reads
\beq
\psi^{(0)}_{\alpha\beta}(r_1,z_1)=
\int\frac{d^{2}k_{1}}{(2\pi)^{2}}\psi^{(0)}_{\alpha\beta}(k_1,z_1)
e^{ik_1 r_1}
\eeq    
where $\alpha$ and $\beta$ are indices for degrees of freedom such as 
spin and colour. The incoherent sum over spin and colour of the square
moduli of the wave function is
\beq
\Phi^{(0)}(r_1,z_1)=\sum_{\alpha\beta} |\psi^{(0)}_{\alpha\beta}(r_1,z_1)|^2
\eeq
and the normalization will be given by
\beq
N=\int_{0}^{1}\frac{dz_{1}}{2z_{1}(1-z_{1})}
\int\frac{d^{2}r_{1}}{2\pi}\Phi^{(0)}(r_{1},z_{1})
\eeq  
For a hadron $N$ is finite and can be chosen to be $1$ while for a photon
it would diverge.  

Let us consider the contribution to order $\alpha_s$ due to the emission
of one soft gluon. One will search for a component of the wave function of the
form $\psi_{\alpha\beta}^{(1)a}(r_{1},z_{1};r_{2},z_{2})$ where $r_2$, $z_2$ 
and $a$ are the transverse coordinate, the scaling and the colour variables
of the soft gluon ($z_2 <<z_1,1-z_1$) respectively. 
Summing the contributions due to the emission of a soft gluon from
the quark and the antiquark one finds 
\beq
\psi_{\alpha\beta}^{(1)a}(r_{1},z_{1};r_{2},z_{2})=
-i(g/\pi)t^{a}\psi^{(0)}_{\alpha\beta}(r_{1},z_{1})
[(r_{2}e)/r_{2}^{2}-(r_{21}e)/r_{21}^{2}]
\eeq  
where $e$ is the polarization vector of the soft gluon and $r_{ij}=r_i-r_j$.
Introducing the quark coordinate $r_0=0$, taking the square modulus of the
wave function and the sum over colour and spin one finds the distribution
\beq
\Phi^{(1)}(r_{1},z_{1};r_{2},z_{2})=
\Phi^{(0)}(r_{1},z_{1})c\frac{r_{10}^{2}}{r_{20}^{2}r_{21}^{2}}
\eeq   
where $c=4 \alpha_s C_F/\pi$, $C_F=(N_c^2-1)/(2N_c)$.

In order to consider the contributions due to the multiple emissions of soft
gluons some simplifications have been taken into account so that
it is possible to find an integral equation which resums all the orders.
Precisely the large $N_c$ limit has been considered, under which the gluon
lines can be represented in terms of $q\bar{q}$ pairs and only
diagrams with a planar topology survive.

For a second soft gluon emitted (characterized by $r_3$,$z_3$) one finally
obtains
\bea
\Phi^{(2)}(r_{1},z_{1};r_{2},z_{2}; r_{3},z_{3})&=&
\Phi^{(1)}(r_{1},z_{1};r_{2},z_{2})c \Bigl(\frac
{r_{02}^{2}}{r_{03}^{2}r_{32}^{2}}+
\frac{r_{12}^{2}}{r_{13}^{2}r_{32}^{2}}\Bigr) \nonumber \\
&=&\Phi^{(0)}(r_{1},z_{1})c^{2}r_{01}^{2}
\Bigl(\frac{1}{r_{02}^{2}r_{23}^{2}r_{31}^{2}}+
\frac{1}{r_{03}^{2}r_{32}^{2}r_{21}^{2}}\Bigr)
\eea  
and the formula for an emission of $n$ soft gluons in the kinematical region
$z_{n+1}<<z_{n}<<...z_{2}$ is easily inferred
\beq
\Phi^{(n)}(r_{1},z_{1};r_{2},z_{2};... r_{n+1},z_{n+1})=
\Phi^{(0)}(r_{1},z_{1}) c^{n}r_{01}^{2}
\Bigl(\frac{1}{r_{12}^{2}r_{23}^{2}...r_{n+1,0}^{2}}+
perm. \  of \ 2,3,...n+1\Bigr)
\eeq     
In the strongly ordered region the integration in the scaling variables
of the emitted soft gluons can evidently be done and in the leading order
one obtains a trivial logarithmical factor $(\ln z_{10})^n/n!$ where 
$z_{10}=z_1/z_0$.
It is possible to introduce \cite{mueller} a generating functional for the 
distribution of the transverse coordinates of the emitted gluons:
\beq
\Phi^{(n)}(r_{1},z_{1};r_{2},... r_{n+1}; z_{0})=
\Phi^{(0)}(r_{1},z_{1})(1/n!)\prod_{2}^{n+1}\frac{\delta}{\delta u(r_{i})}
Z(r_{1},r_{0},z_{10})_{u=0}
\label{distriZ}
\eeq      
The generating functional satisfies the following nonlinear equation
\beq
Z(r_{1},r_{0},z_{10})=1+
cr_{10}^{2}\int_{z_{0}}^{z_{1}}\frac{dz_{2}}{z_{2}}
\int \frac{d^{2}r_{2}}{4\pi r_{02}^{2}r_{21}^{2}}u(r_{2})
Z(r_{2},r_{0},z_{20})Z(r_{2},r_{1},z_{20})
\label{genZreal}
\eeq    
One may be interested in a functional generating not the distribution in terms
of the transverse positions of the emitted gluons but instead in terms
of $r_{ik}$, already present in (\ref{genZreal}), which correspond to the
dimensions of the colour dipoles.
The proper generating functional in this case is
\beq
D(r_{10},z_{10};u)
=u(r_{10})+cr_{10}^{2}\int_{z_{0}}^{z_{1}}
\frac{dz_{2}}{z_{2}}\int \frac{d^{2}r_{2}}{4\pi r_{12}^{2}r_{20}^{2}}
D(r_{12},z_{20}; u)D(r_{21},z_{20}; u)
\label{genDreal}
\eeq   
The generating functionals given above are very useful because
using them one can easily
calculate the multiple inclusive distributions for gluons or dipoles.
In fact after taking the functional derivatives with respect to $u$ one
has just to set its value to $1$ instead of $0$ as in (\ref{distriZ}).
On setting $u=1$ without taking any derivative one has the probability
to find any number of gluons or dipoles emitted with a scaling variable $z$
in the range $z_0 <z <z_1$.
These quantities for both (\ref{genZreal}) and (\ref{genDreal}) are evidently
not normalized and divergent, the divergence due to the integration in the
region of small inter-gluon distances in the transverse plane ($r_{ij}=0$).

In order to cure these ultraviolet divergences one has to include also the 
virtual contributions given by loop diagrams; 
in fact the previous computation was performed only at the tree level.
One can even compute the contribution of the virtual corrections by imposing
that the divergent part of real and virtual gluon emissions compensate 
each other to give a normalized probability and checking that the 
contribution is correct at the lowest loop order. 
This procedure has been followed for example in \cite{mueller} and leads
to the following equations for the functionals $Z$ and $D$:
\beq
Z(r_{1},r_{0},z_{10}; u)=z_{10}^{2\omega (r_{10})}+
cr_{10}^{2}\int_{z_{0}}^{z_{1}}\frac{dz_{2}}{z_{2}}
\int \frac{d^{2}r_{2}}{4\pi r_{02}^{2}r_{21}^{2}}u(r_{2})
z_{12}^{2\omega
(r_{10})}Z(r_{2},r_{0},z_{20}; u)Z(r_{2},r_{1},z_{20}; u) 
\label{genZ}
\eeq       
\beq
D(r_{10},z_{10};u)
=u(r_{10})z_{10}^{2\omega (r_{10})}+
cr_{10}^{2}\int_{z_{0}}^{z_{1}}
\frac{dz_{2}}{z_{2}}\int \frac{d^{2}r_{2}}{4\pi r_{12}^{2}r_{20}^{2}}
z_{12}^{2\omega (r_{10})}
D(r_{12},z_{20}; u)D(r_{20},z_{20}; u)
\label{genD}
\eeq
where the quantity $\omega(r)$ has the same functional form of the gluon Regge
trajectory respect to its momentum argument, i.e.
\beq
\omega (r_{01})=
-(c/2)r_{01}^{2}\int \frac{d^{2}r_{2}}{4\pi r_{02}^{2}r_{21}^{2}}
=(2\alpha_{s}C_{F}/\pi)r_{01}^{2}
\int \frac{d^{2}r_{2}}{4\pi r_{02}^{2}r_{21}^{2}}
\eeq
The virtual corrections to the functional equations are related to the
property of gluon reggeization of the BFKL approach.
    
Let us now look at the connection between the colour dipole model and the
BFKL pomeron.
One first obtains an equation for the inclusive one dipole distribution
taking a functional derivative with respect to $u$ of relation (\ref{genD}) 
and setting $u=1$. The resulting equation in term of the rapidity
$y=\ln z_{10}$ reads 
\beq
n(r_{10},r,y)=e^{2 y \omega (r_{10})}\delta^{2}(r-r_{10})+
2cr_{10}^{2}\int_{0}^{y}
d y' e^{2(y-y') \omega (r_{10})} 
\int \frac{d^{2}r_{2}}{4\pi r_{12}^{2}r_{20}^{2}}
n(r_{21},r,y')
\eeq  
and taking a derivative with respect to $y$ one has the evolution equation
\beq
\frac{\partial}{\partial y} n(r_{10},r,y)= 2 \omega (r_{10}) n(r_{10},r,y)
+\frac{2\alpha_s C_F}{\pi^2}  \int d^{2}r_{2} 
\frac{r_{10}^2}{r_{12}^{2}r_{20}^{2}}
 n(r_{21},r,y)
\eeq
which describes a dual BFKL diffusion in the transverse coordinate space 
instead of in the transverse momentum space. This equation is also
infrared and ultraviolet finite but one may need to regularize it with
an ultraviolet cutoff; in it the role of the coordinates and momenta is
interchanged.

% cap2

\chapter{Hard pomeron with a running coupling constant: 
a phenomenological approach}

The basic elements in the BFKL approach for the multi-Regge kinematics of 
Yang-Mills theories are the effective vertices (real contribution) and
the Regge trajectory (virtual contribution) whose separate bad behaviour in
the infrared sector cancels in the total cross section.
In the LLA approximation all the contributions, which are present in the BFKL
equation for the pomeron, are based on calculations in the first non-trivial
order in perturbation theory.
The effective QCD coupling constant (taken to be small since 
$\alpha_s \ln s \sim 1$) is considered fixed and one does not know the region
of applicability of LLA and in particular the energies and momenta fixing
the scale for $\alpha_s$.
This is at the basis of the weakness of the LLA approximation, since the 
quantitative results can be strongly modified by changing the scale of
virtuality. 

For this reason a program to compute the next-to-leading order corrections
to the BFKL kernel of the Lipatov equation has been carried out in the last 
years \cite{lipfa89} with the hope to study soon the so called quasi-Regge 
kinematics for gluon production. 
The one loop corrections to the effective vertices, the
two loop corrections to the gluon trajectory and terms for the productions
of two gluons and quark-antiquark with fixed invariant mass have been 
computed \cite{nlomix}. 
The nice feature is that also at the NLO order the pomeron can be seen as a
compound state of two reggeized gluons. The generalized BFKL equation however
is quite complicated \cite{fali98} and there is no hope to obtain an
explicit analytical solution but the study of such a kernel should be very 
useful to understand by means of numerical calculations the validity of the 
predictions in the Regge kinematics and corrections in the quasi-Regge limit.
Moreover it is still not clear if such results are applicable in the
HERA kinematical region or one should go even beyond the NLO analysis 
\cite{fali98}.  

A part of this well established program some attempts have been made in order
to take into account the running of the QCD coupling with a minimal added
work. 
The first idea \cite{pertqcd} has been to introduce directly in the BFKL 
equation for the pomeron the running coupling with some assumptions about 
the momentum and distance scales at which the gluons interact. 
Also in other attempts \cite{lr89,nikzak2,hanross,kmsgb,vacven} the BFKL 
equation has been modified in the infrared
sector in order to model some non-perturbative effects and at least
try to avoid the strong inconsistency one has using the BFKL evolution which
does not separate perturbative and non-perturbative scales.
This fact is evident on considering the diffusion mechanism of the
gluon distribution in the transverse momentum space during the evolution
in the rapidity leading to a typical normal distribution in $\ln {\bf k}^2$.
In all these phenomenological models a better behaviour also due to some
decrease of the pomeron intercept to more reasonable values was found, but
one of the fundamental properties shown by the QCD in the Regge kinematical 
limit, that the gluon reggeizes, is lost.

The idea to introduce a running coupling in a manner consistent with the gluon
reggeization has been proposed by M.Braun \cite{braun1,braun2} and in this way 
one is able to guarantee that the production amplitude in the one reggeized 
gluon exchange approximation, which serves as an input in the BFKL theory, 
satisfy unitarity in the leading order \cite{bart1}. Thereby the whole scheme
becomes self-consistent: otherwise one should add to the input amplitudes
corrections following from the unitarity. Since the bootstrap condition,
as discussed in section 1.x, is a crucial element for the reggeization of the 
gluon and for the theory of the reggeized gluons as a whole the only way to 
introduce the running coupling constant in a manner compatible with the gluon 
reggeization is to preserve the bootstrap.
A second basic ingredient is the requirement that the outcoming gluon
distribution has the same asymptotic behavior as the one predicted by the
DGLAP equations (describing the renormalization group flow) in the DLLA.

In section 1 we present the basic equations in the form suitable for
numerical analysis \cite{bvv} for the cases $q=0$ (forward scattering) and 
$q\neq 0$.
After that we describe the method of the solution and present the
numerical results for the intercept, slope and the wave function at $q=0$.
The results for the intercept and slope, on the whole, agree with those
found in \cite{braun1,braun2} by the variational approach. 
An interesting new result is the existence of a second pomeron with the 
intercept minus one, roughly speaking, two time less than for the leading one, 
but still positive.
We end this section applying these results to the study of the asymptotical 
behaviour of the cross-section for the $\gamma^{\ast}\gamma^{\ast}$ scattering,
including some unitarization effects due to multipomeron scattering by mean
of an eikonal formulation valid in the large $N_c$ limit \cite{braun3},
and try also to extrapolate to the hadron case.

Therefore the results of our study show that with the running coupling 
included the pomeron equation possesses bound state solutions which have 
negative energy and thus intercepts greater than unity. 
These solutions correspond to supercritical pomerons in the old sense, 
that is, they represent simple poles in the complex angular momentum plane. 
However the sub-dominant pomeron does not seem to play an important 
role in describing the asymptotical behaviour of the amplitudes. The
intercept of the leading pomeron singularity depend weakly on the infrared 
regulator parameter and stay in the region 0.35--0.5 for its physically 
reasonable values. The introduction of the running coupling and thus a scale
provides for a nontrivial slope for the pomeron, which is responsible for the
physically reasonable behaviour of the cross-sections at very high energies.

For realistic photonic cross-sections and, with a rather crude approximation,
for the hadronic ones the estimated unitarization effects begin to be felt at
extraordinary high energies, of the order $100-1000\ TeV$  (or equivalently
$x<10^{-10}-10^{-12}$). Until these energies a single pomeron exchange
remains a very good approximation to the asymptotic amplitude.
By the way we remember that no effects due to triple pomeron interaction are
here taken into account and these are seen to be required as shown by the
unitarization program (see the third chapter). 
Assuming that the picture is not completely changed by the last consideration,
a comparison to the experimental cross-sections and structure functions at the
highest energy (lowest $x$) achieved seems to confirm the widespread opinion
that we are still rather far from the asymptotical regime and that other
states, different from the supercritical pomerons, give the dominant
contribution.

In section 2 we apply our model with the two found supercritical 
pomerons to jet production \cite{bv2}. This process has been extensively 
studied in the framework of the standard BFKL approach with a 
fixed coupling \cite{glr83,lr89}. As is well-known, this analysis has lead to 
some far-reaching conclusions as to the importance of 
mini-jet production at high energies and the logarithmic rise 
of the multiplicity. However in the fixed coupling approach,
various {\it ad hoc} modifications of the basic BFKL model had 
to be introduced  to cut off the spectrum at low $\kt$ and 
also to correctly reproduce the high $\kt$ tail of the 
spectrum.  Both these problems are naturally resolved by the 
introduction of a running coupling in our model, in which no 
new parameters appear in contrast to the fixed coupling approach.
We shall study the asymptotic behaviour of the jet production cross section
using the asymptotic form of the pomeron wave functions shown in 
appendix A.1 and present some numerical calculations.
 
Since we feel that to describe the present experimental data it is 
necessary to take into account all the states from the the spectrum of 
pomeron equation, this one has been converted into an evolution 
equation in $1/x$ and solved with an initial condition at some
(presumably small) value $x=x_{0}$. In such an approach \cite{bv3}, taking a 
non-perturbative input at $x=x_{0}$ adjusted to the experimental data,
also the problem of coupling the pomeron to the hadronic target is 
solved in an effective way.
This approach is shown in details in section 3 where we state the 
basic equations and we try to pass from the gluon density to the observable 
structure function.

%%%%%%%%%%%%%%%%%%%%%%%%%%%%%%%%%%%%%%%%%%%%%%%%%%%%%%%%%%%%%%%%%%%%
\section{Pomeron equation with running coupling}

Let us consider the bootstrap condition for the reggeization of the gluon:
\beq
[\omega(q)-\omega(q_1)-\omega(q_2)]\phi(q_1)=
\int \frac{d^2q'_1}{(2\pi)^2} K_q(q_1,q'_1)\phi(q'_1)
\eeq
This is in general a very complicated functional integral equation which 
relates $\omega$, $K$ and $\phi$.
Looking for a particular class of solutions of this equation (we look for
$\omega$ and $K$, limiting ourselves to constant $\phi$) we can start to 
write them making use of more than one arbitrary function.
In fact we can write the reggeized gluon trajectory
\beq
\oq=-\frac{N_c}{2}\int
\frac{d^2q_1}{(2\pi)^2}\frac{f_A(q)}{f_B(q_1) f_C(q_2)},\ \
q=q_{1}+q_{2}
\eeq
and the gluon pair interaction kernel in the gluon channel
\beq
K_q(q_{1},q'_{1})=\frac{N_c}{2} 
\left( (\frac{f_A(q_1)}{f_B(q'_1)}+\frac{f_A(q_2)}{f_B(q'_2)})
\frac{1}{f_C(q_{1}-q'_{1})}-\frac{f_A(q)}{f_B(q'_1)f_C(q'_2)}\right)
\eeq
One must note that there are some constrains on the kernel $K$ which must be 
satisfied.
The first one is that $K$ should be symmetric in the two gluons 
(indices 1 and 2) so we must require that $f_B=f_C$ from the last term in the
kernel.
The second requirement is that $K$ should be symmetric in the
initial and final gluons (quantities with momentum not primed and primed), 
after separating the two "gluon propagators" ( $\frac{1}{f_B(q'_1)f_B(q'_2)}$).
From the first term in the kernel we must thus require this symmetry for the 
expression
\beq 
\frac{f_A(q_1)f_B(q'_2)+f_A(q_2)f_B(q'_1)}{f_C(q_{1}-q'_{1})}
\eeq
Matching the expression with the corresponding one where the primed quantities 
are exchanged with the not primed we finally get $f_A=f_B$.
So we arrive at $f_A=f_B=f_C=\eta$.
The standard BFKL theory is characterized by $\eta=q^2/(2\alpha_s)$ and in
general the gluon trajectory and the interaction kernel are correlated in
terms of the $\eta$ function.

Let us write the equation for the amplitude in the vacuum channel in the
form from which the typical cancellation of the infrared divergences in the
pure BFKL theory is manifest
\beq
(j-1)\phi(q_1)=N_c \int\frac{d^2q'_1}{(2\pi)^2}
\Bigl[ \frac{\eta(q_1)}{\eta(q_1-q'_1)} 
\Bigl(\frac{\phi(q'_1)}{\eta(q'_1)}-\frac{\phi(q_1)}{\eta(q'_1)+\eta(q_1-q'_1)}
\Bigr) +(1 \leftrightarrow 2) -\frac{\eta(q)}{\eta(q'_1)\eta(q'_2)}\phi(q'_1)
\Bigr]
\eeq
The idea in \cite{braun1,braun2} was to change $\eq$ so that it corresponds 
to a running rather then to a fixed coupling. 
For the running coupling some conclusions about the
form of $\eq$ can be made considering the vacuum channel equation in the
limiting case of very large $q$. 
In this case one has, assuming that $\eq$ grows with $q$, the approximated 
equation in the forward direction ($q=0$)
\beq
(j-1)\phi(q_1)=2 N_c \int\frac{d^2q'_1}{(2\pi)^2} \theta(q_1^2-{q'_1}^{2})
\frac{\phi(q'_1)}{\eta(q'_1)}
\eeq
and in the leading approximation one gets, after an inverse Mellin 
transformation, an evolution equation for the fully amputated wave function
\beq
\frac{\partial}{\partial\ln q_1^2}
\frac{\partial}{\partial\ln 1/x}\phi(x,q_1)=\frac{N_c}{2\pi} 
\frac{q_1^2}{\eta(q_1)} \phi(x,q_1)
\eeq
which, on comparing with the DGLAP evolution equation in the leading order 
in $\ln 1/x$ (that is, in the double leading log approximation (DLLA)) for 
the gluon distribution,
\beq
\frac{\partial}{\partial\ln q_1^2}
\frac{\partial}{\partial\ln 1/x}x g(x,q_1)=N_c\frac{\alpha_s(q_1^2)}{\pi} 
x g(x,q_1)
\eeq
allows one to find the
asymptotic form of $\eq$:
\beq
\eq \simeq \frac{q^{2}}{2\alpha_{s}(q^{2})},\ \ q\rightarrow\infty
\label{etaasym}
\eeq
and a proportionality relation between the amputated pomeron wave function
$\phi(x,k^2)$ and the gluon distribution $x g(x,k^2)$.
It evidently differs from the fixed coupling case by changing the
fixed coupling constant $\alpha_{s}$ to a running one $\alpha_{s}(q^{2})$.
As a result, with a running coupling, both the gluon
trajectory and its interaction have to be changed simultaneously in an
interrelated manner, so that the resulting equation is different from the
BFKL one already in the leading order.

The behaviour of $\eq$ at small $q$, comparable or even smaller than the QCD
parameter $\Lambda$, cannot be established from any theoretical calculation,
since this domain is non-perturbative. To take into account these
confinement effect we choose $\eq$ at finite $q$ in a simple form:
\beq \eq=(b/2\pi)f(q) \quad ; \quad
f(q)=(q^{2}+m_{1}^{2})\ln ((q^{2}+m^{2})/\Lambda^{2})
\label{eta}
\eeq
with $b=(1/4)(11-(2/3)N_{F})$ and $m\geq\Lambda$, which agrees with 
(\ref{etaasym}) for large $q$ and remains finite up to $q=0$.
It allows for the freezing of the coupling and the confinement proper to occur
at somewhat different scales ($m$ and $m_{1}$ respectively). However, on
physical grounds, one feels that they should be of the same order.

A preliminary study of the properties of the pomeron with $\eq$ given by
(\ref{eta}) for $m=m_1$ was performed by the variational technique in 
\cite{braun1,braun2}. It was found that
the intercept depended on the ratio $m/\Lambda$ quite weakly: as $m/\Lambda$
changes from 1.5 to 5.0 the intercept (minus one)  $\Delta$ falls from 0.4
to 0.25. On the other hand, the slope depends on this ratio very strongly.
This allows to fix the ratio $m/\Lambda$ to values in the interval
3.0$\div$4.0.

This variational study, although very simple, cannot however give values for
the intercept and especially for the slope with some precision.
Still less can be found by this
method about the properties of the pomeron wave function essential for the
high-energy behaviour of the physical amplitudes, for some
aspects studied in \cite{braun3}. 
Finally, one does not
receive any knowledge about the existence of other solutions with a
positive intercept. All these reasons give us a motivation to undertake a
numerical study of the two-gluon vacuum channel equation with in general the
gluon trajectory and interaction given by
\bea
\oq&=&-\frac{N_c}{2}\int
\frac{d^2q_1}{(2\pi)^2}\frac{\eta(q)}{\eta(q_1) \eta(q_2)} \nonumber \\
K_q(q_{1},q'_{1})&=&-T_1 T_2 
\left( (\frac{\eta(q_1)}{\eta(q'_1)}+\frac{\eta(q_2)}{\eta(q'_2)})
\frac{1}{\eta(q_{1}-q'_{1})}-\frac{\eta(q)}{\eta(q'_1) \eta(q'_2)}\right)
\label{tra_int}
\eea
where $T$ is the gluon colour vector ($T_1T_2=-N_c$ in the 
vacuum channel) and the function $\eq$ satisfying (\ref{etaasym}).

The parametrization (\ref{eta}), with $b$ fixed, does not evidently take into
account the change in $N_{F}$ which occurs on flavour  thresholds. In
principle this can easily be remedied by assuming $N_{F}$ and thus $b$
depending on $q^{2}$ and varying accordingly. Our theory possesses a
scale (actually several scales, $m$,$m_{1}$ and $\Lambda$), so that the
asymptotic behaviour (5) used to determine $\eta$ refers to
$q^{2}>>m^{2},m^{2}_{1},\Lambda^{2}$. Since $m\sim m_{1}$ is of the order of
$1\ GeV$ (the actual choice we make in following is $m=m_{1}=0.82\ GeV$),
the only thresholds to be accounted for are those of the heavier quarks, 
$b$ and $t$. However
one should take into account that our theory is much better behaved than the
scale-less BFKL model. In particular, in our case the pomeron possesses a
normalizable wave function. So, in contrast to the BFKL case, the
characteristic  momenta which form the pomeron are determined by the same
scale $m\sim m_{1}$ and are therefore of the order of several $GeV/c$.
For this reason we expect to find the pomeron properties with sufficient
precision fixing $N_{F}=4$ and forgetting about $b$ and $t$ thresholds.
Accordingly, the calculations presented are performed with a
constant $N_{F}=4$. Clearly the found momentum wave function is strictly
speaking valid up to momenta squared of the order $100\ (GeV/c)^{2}$, when
the beauty threshold should be taken into account to slightly lower the
value of our parameter $b$ (by 8\%). In applications to the structure
functions this restricts the domain of applicability of the calculated
pomeron wave function to $Q^{2}\leq 100\ (GeV/c)^{2}$.

%%%%%%%%%%%%%%%%%%%%%%%%%%%%%%%%%%%%%%%%%%%%%%%%%%%%%%%%%%%%%%%%%%%%%%%%%%
\subsection{Basic equations. Pomeron at $q=0$}
We consider the physical case $N_{c}=3$. The units are chosen to have
$\Lambda=1$. In relating to observable quantities we take $\Lambda=0.2\ GeV$.

The pomeron equation is the eigenvalue equation
\beq
\Bigl(-\oa-\ob\Bigr)\phi(q_{1})+\int \frac{d^{2}q'_{1}}{(2\pi)^{2}}
K_{q}^{vac}(q_{1},q'_{1})\phi(q'_{1})=
E(q)\phi(q_{1})
\label{eigeneq1}
\eeq
where the "energy" eigenvalue  $E(q)$ is related to the pomeron trajectory via
\beq
\alpha(q)=1-E(q) \simeq 1+\Delta-\alpha' q^{2}
\label{slope}
\eeq
The last equation, well satisfied for small $q$, defines the intercept 
$\Delta$ and the slope $\alpha'$. 
In (\ref{eigeneq1}) the trajectories $\omega$ and the kernel $K^{vac}$ are
given by the eqs. (\ref{tra_int}) with $T_{1}T_{2}=-3$ and
the function $\eta$ given by (\ref{eta}).
To symmetrize the kernel we pass to the function
\beq
\psi(q_{1})=\phi(q_{1})/\sqrt{\ea\eb}
\label{semiamppsi}
\eeq
We also take out the common numerical factor
$6/((11-2/3N_{F})\pi)$ and express all terms via the
function $\fq$ also defined in (\ref{eta}). 
Then the  equation for $\psi$ takes the form
\beq
A_{q}(q_{1})\psi(q_{1})+\int d^{2}q'_{1}L_{q}(q_{1},q'_{1})\psi(q'_{1})=
\epsilon(q)\psi(q_{1})
\label{eigeneq2}
\eeq
Here the "kinetic energy" is
\beq
A_{q}(q_{1})=(1/2) \int\frac{d^{2}q'_{1}\fa}{\fc f(q_{1}-q'_{1})}+
(1/2)\int\frac{d^{2}q'_{2}\fb}{\fd f(q_{2}-q'_{2})}
\eeq
The interaction kernel consists of two parts, a quasi-local  and a
separable one:
$
L=L^{(ql)}+L^{(sep)}
$
They are given by
\beq
L^{(ql)}_{q}(q_{1},q'_{1})=
-\sqrt{\frac{\fa}{\fb}}\frac{1}{f(q_{1}-q'_{1})}\sqrt{\frac{\fd}{\fc}}-
\sqrt{\frac{\fb}{\fa}}\frac{1}{f(q_{2}-q'_{2})}\sqrt{\frac{\fc}{\fd}})
\label{quasilocal}
\eeq
and
\beq
L^{(sep)}_{q}(q_{1},q'_{1})=\frac{\fq}{\sqrt{\fa\fb\fc\fd}}
\eeq
Both parts are evidently symmetric in $q_{1}$ and $q'_{1}$.
The scaled energy $\epsilon$ is related to the initial one by
\beq
E=\frac{6}{\pi (11-(2/3)N_{F})}\epsilon
\eeq
eq. (\ref{eigeneq2}) simplifies in the case when the total momentum of the two
gluons is equal to zero. With $q=0$ the two parts of the kinetic energies
become equal and the square roots in (\ref{quasilocal}) turn to unity. 
So at $q=0$ the equation retains its form (\ref{eigeneq2}) with
\beq
A_{0}(q_{1})=\int\frac{d^{2}q'_{1}\fa}{\fc f(q_{1}-q'_{1})}
\eeq
and the interaction  where now
\beq
L^{(ql)}_{0}(q_{1},q'_{1})=
-\frac{2}{f(q_{1}-q'_{1})}
\eeq
has really become local and
\beq
L^{(sep)}_{0}(q_{1},q'_{1})=\frac{f(0)}{\fa\fc}
\eeq
This is the equation which we shall solve numerically.

To reduce to a one-dimensional problem we introduce the angular momentum of 
the gluons $n$ and choose the solution in the form
\beq
\psi(q)=\psi_{n}(q^{2})\exp in\phi
\eeq
where $\phi$ is the azimuthal angle. Integrating over it in the
equation, we obtain an one-dimensional integral equation for the radial
function $\psi_{n}(q^{2})$:
\beq
A_{0}(q)\psi_{n}(q^{2})+\int dq_{1}^{2}
 L_{n}(q^{2},q_{1}^{2})\psi(q_{1}^{2})=
\epsilon\psi_{n}(q^{2})
\label{eigeneq3}
\eeq
with the kernel  now given by
\beq
L_{n}(q^{2},q_{1}^{2})=-B_{n}(q^{2},q_{1}^{2})
+\delta_{n0}\pi\frac{f(0)}{\fq\fa}
\label{aveeq1}
\eeq
where
\beq
B_{n}(q^{2},q_{1}^{2})=\int_{0}^{2\pi} d\phi
\frac{\cos n\phi}{f(q^{2}+q_{1}^{2}-2qq_{1}\cos
\phi)}
\label{aveeq2}
\eeq
Note that $A_{0}$ can be expressed via $B_{0}$:
\beq
A_{0}(q)=(1/2)\int dq_{1}^{2}B_{0}(q^{2},q_{1}^{2})\frac{\fq}{\fa}
\label{aveeq3}
\eeq

Evidently eq. (\ref{eigeneq3}) is very similar to a
Schr\"odinger equation with an attractive interaction provided by the local
term and a positive kinetic energy described by $A$, which however grows
very slowly at high momenta (as $\ln\ln q$). Therefore the attraction becomes 
smaller with growing $n$ and we expect to
find negative energies, corresponding to intercepts larger than unity, only
for small $n$. Remember that for the BFKL pomeron only the isotropic state
with $n=0$ has a negative energy.  Our calculations reveal that the
introduction of the running coupling following does not change this situation:
states with $|n|>0$ all have positive energies. So in the following
we consider the case $n=0$.

%%%%%%%%%%%%%%%%%%%%%%%%%%%%%%%%%%%%%%%%%%%%%%%%%%%%%%%%%%%%%%%%%%%%%
\subsection{Pomeron at $q\neq 0$: the slope}
With $q\neq 0$ the pomeron equation becomes essentially two dimensional.
Rather than attempt to solve it numerically at all $q$ we limit ourselves
to small values of $q$ and determine not the whole trajectory $\alpha(q)$
but only the slope $\alpha'$ defined by (\ref{slope}). 
This can be done in a much simpler manner using a perturbative approach. 
We present "the Hamiltonian" in (\ref{eigeneq2})
$
H_{q}=A_{q}+L_{q}
$
in the form
$
H_{q}=H_{0}+W(q)
$
and calculate analytically $W(q)$ up to terms of the second order in $q$.
Then for small $q$ the value of the energy $\epsilon(q)$ will be given
 by the standard perturbation formula
\beq
\epsilon(q)=\epsilon (0)+<W(q)>
\eeq
where $<\ >$ means taking the average with the wave function at $q=0$,
determined from the numerical solution of the equation discussed in the
previous section. Thus we avoid solving the two-dimensional problem, but, of
course, cannot determine more than the slope. Fortunately it is practically
all we need to study the high-energy asymptotics (although, of course, the
knowledge of the trajectory as a whole might be of some interest).

In order to derive an expression for $W(q)$ we pass to the relative momenta
$l$ and $l'$
$
q_{1(2)}=(1/2)q+(-)l$
and similarly for the primed momenta.
 Up to the second order in $q$ we have
\beq
\fa = f(l) [ 1 + a_1 \ql + \frac{a_1}{4}q^2 + \frac{a_2}{2}
	\ql^2 ]
\eeq
where
\beq
a_1 = a_1(l) = \bigl[1+\ln(l^2+m^2)-\frac{m^2-m_1^2}{l^2+m^2}\bigr]
\frac{1}{f(l)} \eeq
\beq a_2 = a_2(l) =
\bigl[\frac{1}{l^2+m^2}+\frac{m^2-m_1^2}{(l^2+m^2)^2}\bigr]\frac{1}{f(l)}
\eeq
 The expansion for $\fb$ differs by changing the sign of $l$ (or of $q$);
for $\fc$ and $\fd$ it suffices to replace $l$ by $l'$ in the expressions for
$\fa$ and $\fb$. We use  the notation $a'_1=a_1(l')$ and $a'_2=a_2(l')$.
We also need  the expansion for $f(q)$:
\beq
f(q)=f(0)(1+ a_3 q^2 ) + O(q^4) \quad ; \quad
a_3 = \left(\frac{m_1^2}{m^2} + \ln m^2 \right) \frac{1}{f(0)}
\eeq
The perturbation $W(q)$, up to second order in $q$, can be expressed
via the introduced functions  $a_{1,2}$ and $a'_{1,2}$ and the constant
$a_{3}$. After some calculations we find a part of $W$ coming
from the kinetic term in Hamiltonian in the form
\bea
W_{1}(l)=
 &&\frac{1}{2}\int d^2 l' \frac{f(l)}{f(l')} \Bigl\{
 \bigl( \frac{1}{f(l-l')}+\frac{1}{f(-l-l')} \bigr)
 \Bigl[ -a'_1 \qlp -\frac{a'_1}{4} q^2 + \bigl( {a'}^2_1-\frac{a'_2}{2}\bigr)
 \qlp^2 \Bigr] + \nonumber \\
 &&\Bigl[ -\frac{a_1 a'_1 \ql \qlp}{f(l-l')} + \frac{a_1 a'_1 \ql \qlp}
 {f(-l -l')} \Bigr] \Bigr\}+ A_0(l) \bigl[\frac{a_1}{4} q^2+
\frac{a_2}{2}\ql^2\bigr]
\eea
 The part of $W$ coming from the quasi-local part of the interaction
 can be written as
\beq
W_{2}(l,l')=
  \frac{1}{2} \bigl[ a_1 ( q \cdot l) -a'_1 (q \cdot l') \bigr]^2
  L^{(ql)}_{0}(l,l')
\eeq
and the one coming from the separable part as
\beq
  W_{3}(l,l')=\bigl[
  \bigl( a_3 - \frac{a_1+a'_1}{4} \bigr) q^2 -\frac{1}{2} (a_2-a_1^2)
  (q \cdot l)^2
  - \frac{1}{2} (a'_2-{a'}_1^{2}) (q \cdot l')^2 \bigr]
  L^{(sep)}_{0}(l,l')
\eeq
As mentioned, only isotropic solutions have the
intercept larger than one and are of interest. Then
the expression for $W(q)=\sum_{i=1,2,3} W_{i}$
has to be integrated over the azimuthal angles.
Thus integrated  values will be denoted
 $\hat{W}_i$, $i=1,2,3$.
Using (\ref{aveeq1})-(\ref{aveeq3}), they can be conveniently expressed via the
kernel $B_{n}$ (eq. (\ref{aveeq2})):
\beq
\frac {\hat{W}_1}{2 \pi q^2} =
  \frac{1}{4} \int d {l'}^2 \frac{f(l)}{f(l')} \Bigl\{
  \bigl[ - \frac{a'_1}{2} + \bigl({a'}_1^2-\frac{a'_2}{2} \bigr) {l'}^2\bigr]
  B_0(l,l') -a_1 a'_1 l l' B_1(l,l') \Bigr\} +
 \frac{A_0(l)}{4} \bigl( a_1+a_2 l^2 \bigr) \eeq\beq
\frac {\hat{W}_{2}}{2 \pi q^2} = - \frac{1}{2}
 \bigl( a_1^2 l^2 + {a'}_1^2 {l'}^2 \bigr)
  B_0(l,l') + a_1 a'_1 l l' B_1(l,l') \eeq\beq
\frac {\hat{W}_3}{2 \pi q^2} =
  \frac{f(0)}{f(l) f(l')} \Bigl[ 2 \pi \bigl(a_3 - \frac{a_1+a'_1}{4} \bigr)
     - \frac{1}{4} l^2 \bigl(a_2 - a_1^2 \bigr)
     - \frac{1}{4} {l'}^2 \bigl(a'_2 - {a'}_1^2 \bigr) \Bigr]
\eeq
The slope is given by the momentum average of the sum of these
 expressions, taken with a given isotropic wave function:
\beq
\alpha' = - \frac {(1/2)\int d l^2 d {l'}^2
   \psi(l) \psi(l') \bigl( \hat{W}_2 +\hat{W}_3 \bigr) +
	\int d l^2 \psi(l)^2 \hat{W}_1 }
	{   2\pi q^2 \int d l^2 \psi^2 (l)}
\eeq

%%%%%%%%%%%%%%%%%%%%%%%%%%%%%%%%%%%%%%%%%%%%%%%%%%%%%%%%%%%%%%%%%%%%
\subsection{Numerical procedure and results}
Eq. (\ref{eigeneq3}) was first changed to the variable $t=\ln q^{2}$
whereupon the wave function and the kernel transform according to
\beq
\psi(q^{2})\rightarrow{\tilde\psi}(t)=q\psi(q^{2})
\label{subst1}
\eeq
and
\beq
L(q^{2},q_{1}^{2})\rightarrow{\tilde L}(t,t_{1})=qq_{1}L(q^{2},q_{1}^{2})
\eeq
Performing a discretization on a grid the equation was reduced to a finite 
system of linear equations by approximating the integral by a sum
\beq
\int_{-\infty}^{\infty}dt\,F(t)\simeq\sum_{i=1}^{n}w_{i}F(t_{i})
\eeq
with the grid points $t_{i}$ and weights $w_{i}$ depending on the chosen
approximation scheme. The final equation is thus
\beq
\sum_{j=1}^{n}B_{ij}x_{j}=\epsilon x_{i},\ \ i=1,...n
\eeq
where
\beq
x_{i}=\sqrt{w_{i}}{\tilde \psi}(t_{i})
\label{subst2}
\eeq
and
\beq
B_{ij}=A(t_{i})\delta_{ij}+\sqrt{w_{i}w_{j}}{\tilde L}(t_{i},t_{j})
\eeq
The vector space for ${\bf x}$ has been chosen in order to obtain
a symmetric matrix $B$.
After determining the lowest eigenvalues $\epsilon$ and the corresponding
eigenvectors $x_{i}$ the wave function in the momentum space is directly
given by (\ref{subst1}) and (\ref{subst2}) at points $q^{2}=\exp t_{i}$. 
It should be normalized according to
\beq
\int\frac{d^{2}q}{(2\pi)^{2}}|\psi(q)|^{2}=1
\eeq
Note that this wave function is a partially amputated one (see eq. 
(\ref{semiamppsi})).
The full (non-amputated) wave function is given by $\Phi(q)=\psi(q)/\eta(q)$
(in the forward case).
It is this function that appears in the physical amplitudes.

\begin{figure}
\centering
\includegraphics[angle=-90, width=3.0in]{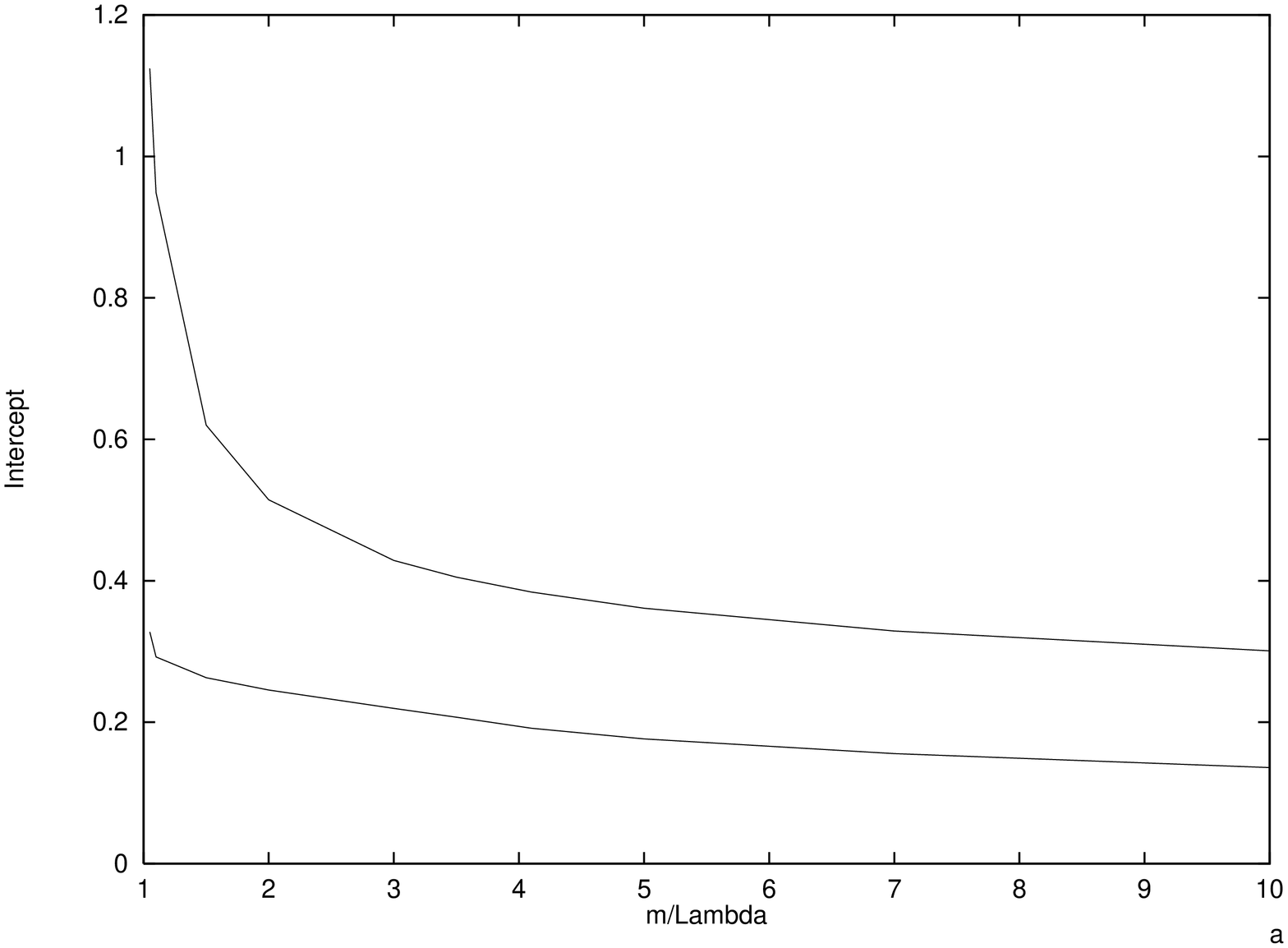}
\hspace{-0.2in}
\includegraphics[angle=-90,width=3.0in]{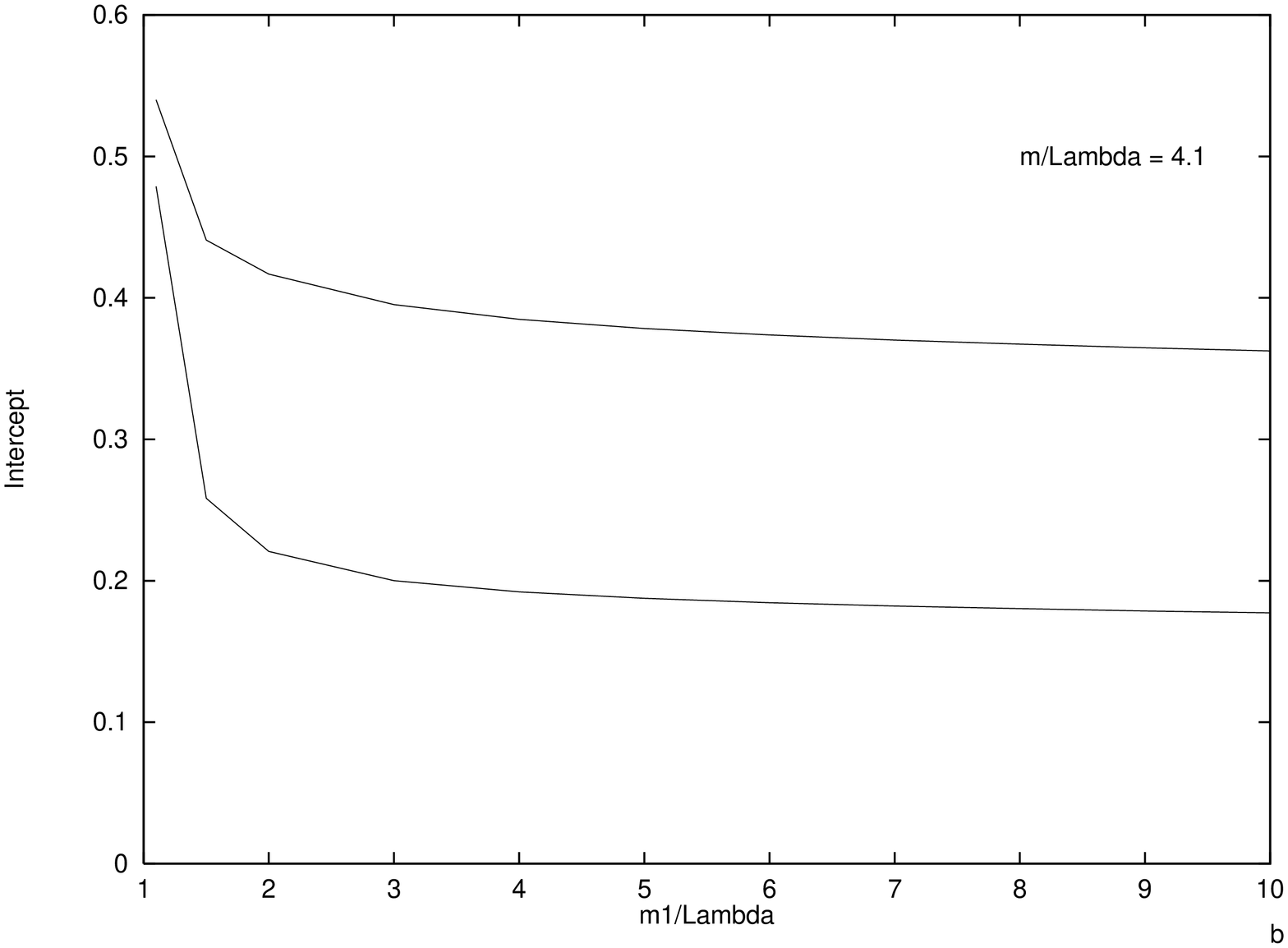}
\captionm{.Pomeron intercepts as a function of the infrared regulator mass
$m=m_{1}$; $\Lambda=0.2\ GeV$ (a) and for different values of the confinement 
parameter $m_{1}$ and the coupling freezing parameter $m$ (b).}
\label{figint}
\end{figure}     

The results for the lowest (and negative) eigenvalues of energy
for the case $n=0$ (isotropic pomeron) are
presented in Fig. \ref{figint}. Actually energies with an opposite sign are
shown, which according to (\ref{slope}) give precisely the intercepts 
(minus one).
As mentioned, the QCD scale here and in the following is taken to be
$\Lambda=0.2\ GeV$. In Fig. \ref{figint}.a the intercepts are shown for the 
case when the two scales $m$ and $m_{1}$ in (\ref{eta}) are equal. Fig. 
\ref{figint}.b illustrates the dependence of the intercepts on the ratio 
$m/m_{1}$.
The most interesting observation which follows from these figures at
once is that in all cases one observes two positive intercepts, which
correspond to two  different supercritical pomerons, the leading and
subleading ones. The single cut, characterizing the BFKL spectrum, is deformed
with the introduction of the running coupling and a discrete part of the
spectrum, giving some poles, appears. 
The intercept of the leading pomeron is found to be in accordance with the 
earlier calculations \cite{braun1,braun2}, performed by the variational
method (it is slightly larger, which was to be expected). For physically
realistic values of $m$ and $m_{1}$ in the interval $0.5\div 1.0\  GeV$
it takes on values in the region of $0.5\div 0.3$ falling with the masses
$m$ and $m_{1}$. The same trend is seen for the intercepts of the subleading
pomeron, which lie in the interval $0.25\div 0.15$.

\begin{figure}
\centering
\includegraphics[angle=-90, width=2.0in]{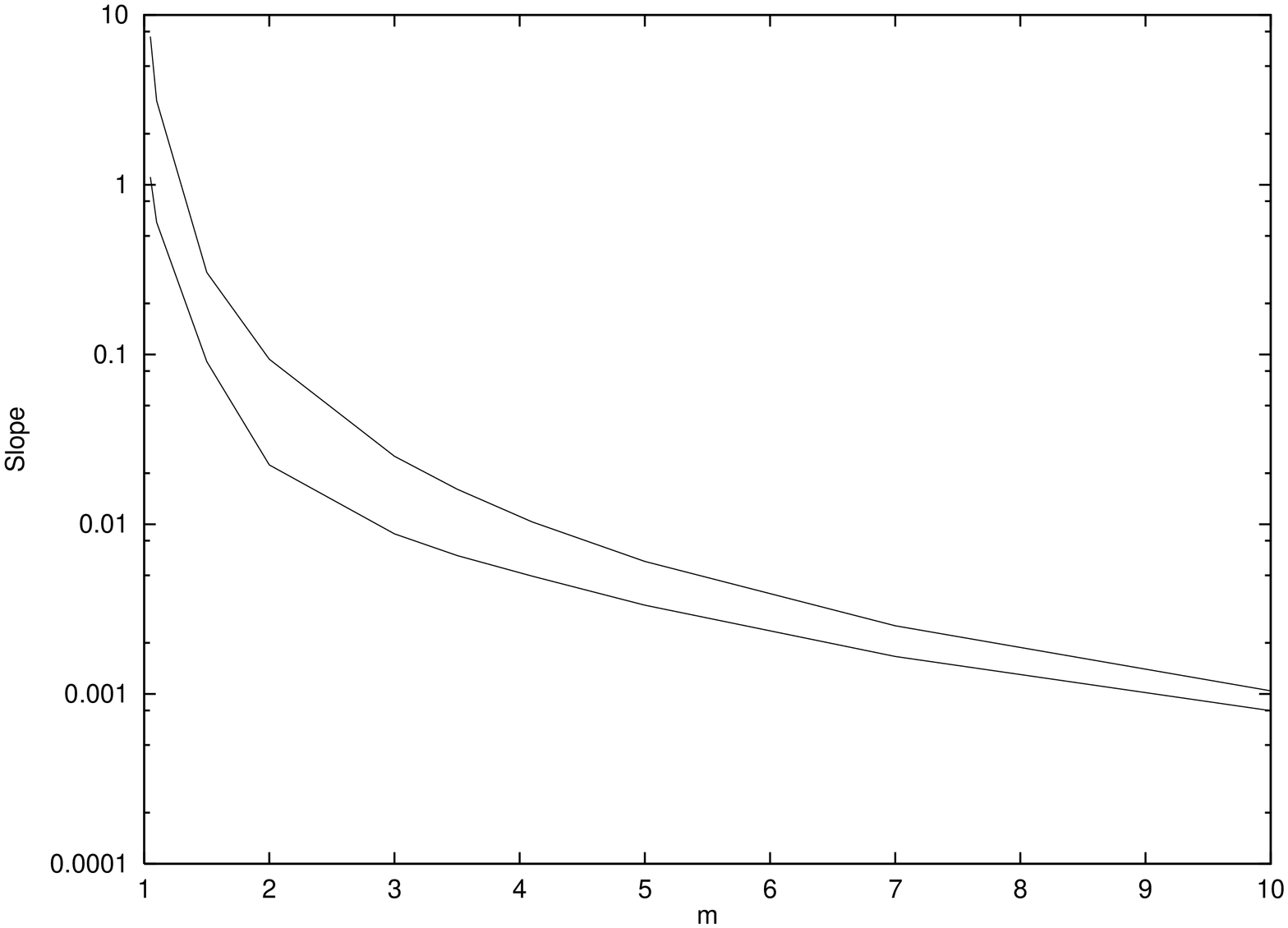}
\captionm{Pomeron slopes as a function of the infrared regulator mass
$m=m_{1}$; $\Lambda=0.2\ GeV$.}
\label{figslo}
\end{figure}  

\begin{figure}
\centering
\includegraphics[angle=-90, width=2.0in]{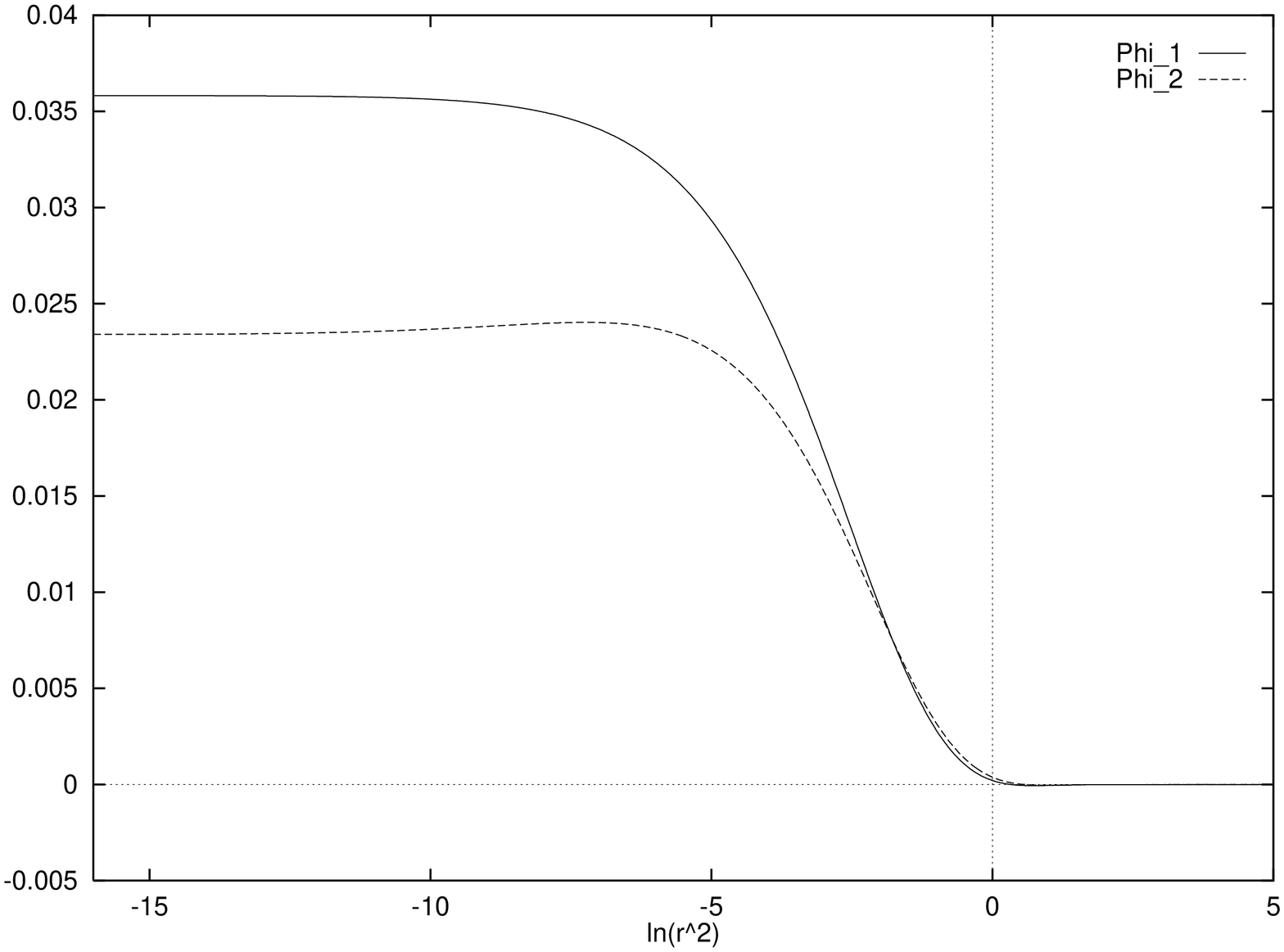}
\captionm{Coordinate space wave functions for the leading  ($\Phi_{0}(r)$) and
subleading ($\Phi_{1}(r)$) pomerons. Both $r$ and $\Phi$ are in units
$1/\Lambda\simeq 1\ fm$.}
\label{figwave}
\end{figure}

The slopes of the two found pomerons are shown in Fig. \ref{figslo} as a 
function of $m$ for the case $m=m_{1}$. The slopes depend very strongly on 
the value of the regulator mass. 
The physically reasonable slopes for the dominant pomeron
of the order of $\alpha'\sim 0.25\ (GeV/c)^{-2}$ restrict the values of $m$ to
the region $0.7-0.9\ GeV$. So finally we choose
\beq
m=0.82\ GeV\eeq
which leads to the following parameters of the leading (0) and subleading (1)
pomerons
\beq
\Delta_{0}=0.384,\ \ \alpha'_{0}=0.250\ (GeV/c)^{-2};\ \
\Delta_{1}=0.191,\ \ \alpha'_{1}=0.124\ (GeV/c)^{-2}
\eeq
In Fig. \ref{figwave} we show the coordinate space wave functions $\Phi(r)$  of
these two pomerons.

%%%%%%%%%%%%%%%%%%%%%%%%%%%%%%%%%%%%%%%%%%%%%%%%%%%%%%%%%%%%%%%%%%%
\subsection{Pomerons and the high-energy  scattering.}
To apply the found results to the actual physical processes one has to couple
the pomerons to the external sources corresponding to the colliding
particles. The only way to do it in a more or less rigorous manner is to
assume that both the projectile and target are highly virtual photons, or 
heavy onia, with
momenta $q$ and $p$ respectively, $-q^{2}=Q^{2}>>\Lambda^{2}$ and
$-p^{2}=P^{2}>>\Lambda^{2}$. Then the non-perturbative effects inside the
target and projectile can safely be neglected. 
We shall address to the very complicated
problem of coupling the gluons to the virtual photons in the running coupling
framework in section 2.3 and consider here the simple lowest order approach,
sufficient for the BFKL pomeron, where the coupling is small and fixed. This
corresponds to taking the contribution of a pure quark loop into which the
incoming photon goes. We shall then need of the colour densities of the virtual
photons whose explicitly form, in the forward case, was found in 
\cite{nikzak2} for both transverse and longitudinal photons. 
In the appendix A.2 we
present a generalization for the non-forward case, both in momentum and
coordinate representation, since we shall need of it, at least formally, in 
the third chapter, where we return to the fixed coupling formalism.

We shall try in the following to understand the role of the two found pomeron
states and analyze the situation in which more than one pomeron is exchanged.
The last situation can be seen as an attempt to understand some aspects 
which are related to higher order corrections needed to restore the unitarity.
Incidentally the true situation is more complicated, as is shown in 
chapter 3, since, for a two pomeron amplitude, one has to take into account 
the triple pomeron interaction, which can take a form which absorbs 
completely the double pomeron exchange.

Nevertheless the contribution of multipomeron scattering to some processes
is of interest and we shall try to analyze it as done in 
\cite{braun3} where the scattering
amplitude in the large colour number limit takes an eikonal form for fixed 
transverse dimensions of the projectile and target and leads to a cross-section
\beq
\sigma=2\int d^{2}Rd^{2}rd^{2}r'\rho_{q}(r)\rho_{p}(r')
\left(1-e^{-z(\nu,R,r,r')}\right)
\label{crosect}
\eeq
where
\beq
z(\nu,R,r,r')=(1/8)\int\frac{d^{2}qd^{2}q_{1}d^{2}q'_{1}}{(2\pi)^{6}}
G(\nu,q,q_{1},q'_{1})e^{iqR}\prod_{i=1,2}(1-e^{iq_{i}r})(1- e^{iq'_{i}r'})
\eeq
is essentially the Fourier transform of the (non-amputated) Green function of
eq. (\ref{eigeneq1}), $G(\nu,q,q_{1},q'_{1})$, considered as a function of 
the energetic variable $\nu=pq$ and with $q=q_{1}+q_{2}=q'_{1}+q'_{2}$. 
The functions $\rho_{q}$ and $\rho_{p}$ correspond to the colour densities of 
the projectile and target photons, respectively.

\begin{figure}
\centering
\includegraphics[angle=-90, width=3.0in]{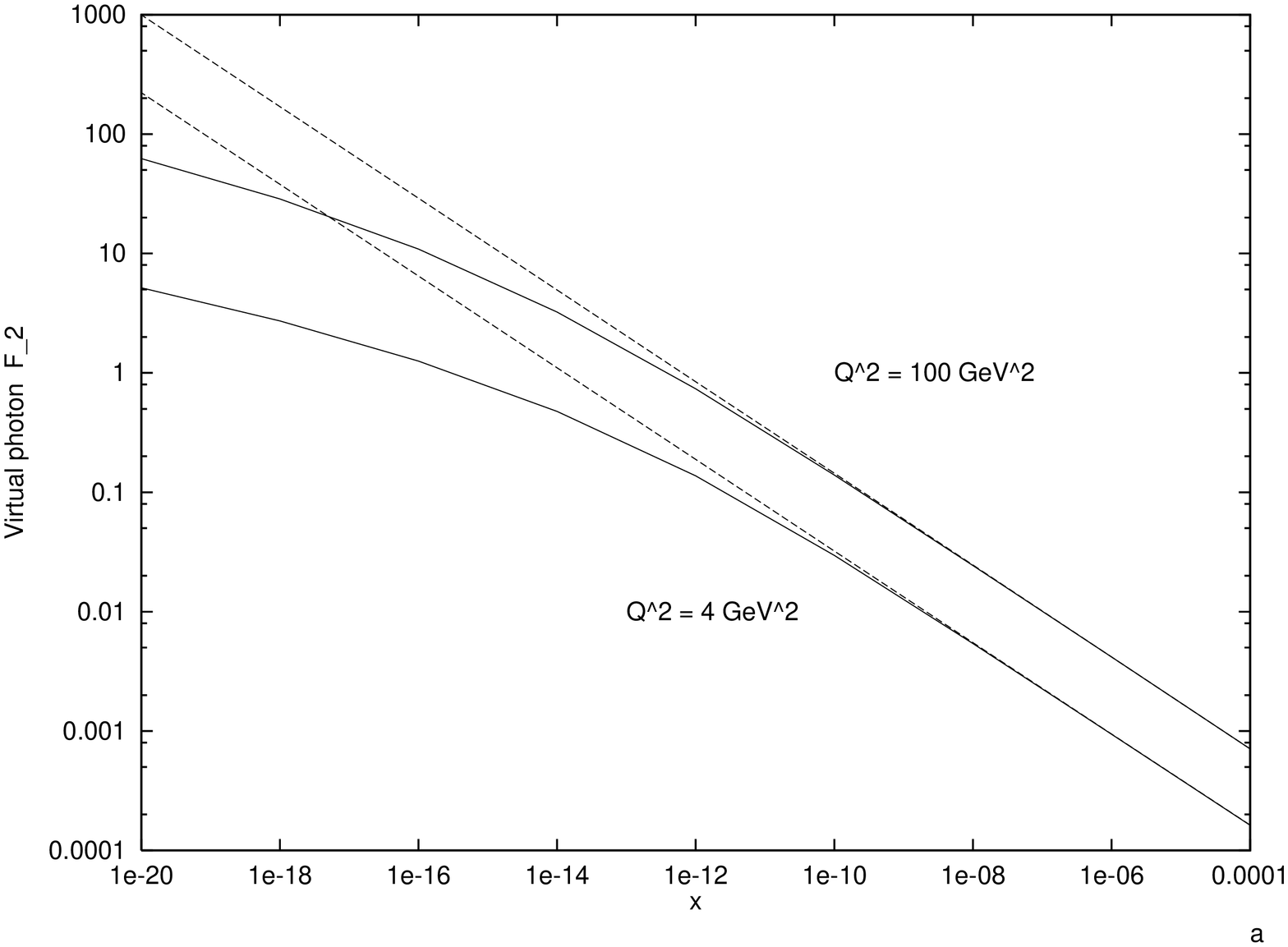}
\hspace{-0.2in}
\includegraphics[angle=-90,width=3.0in]{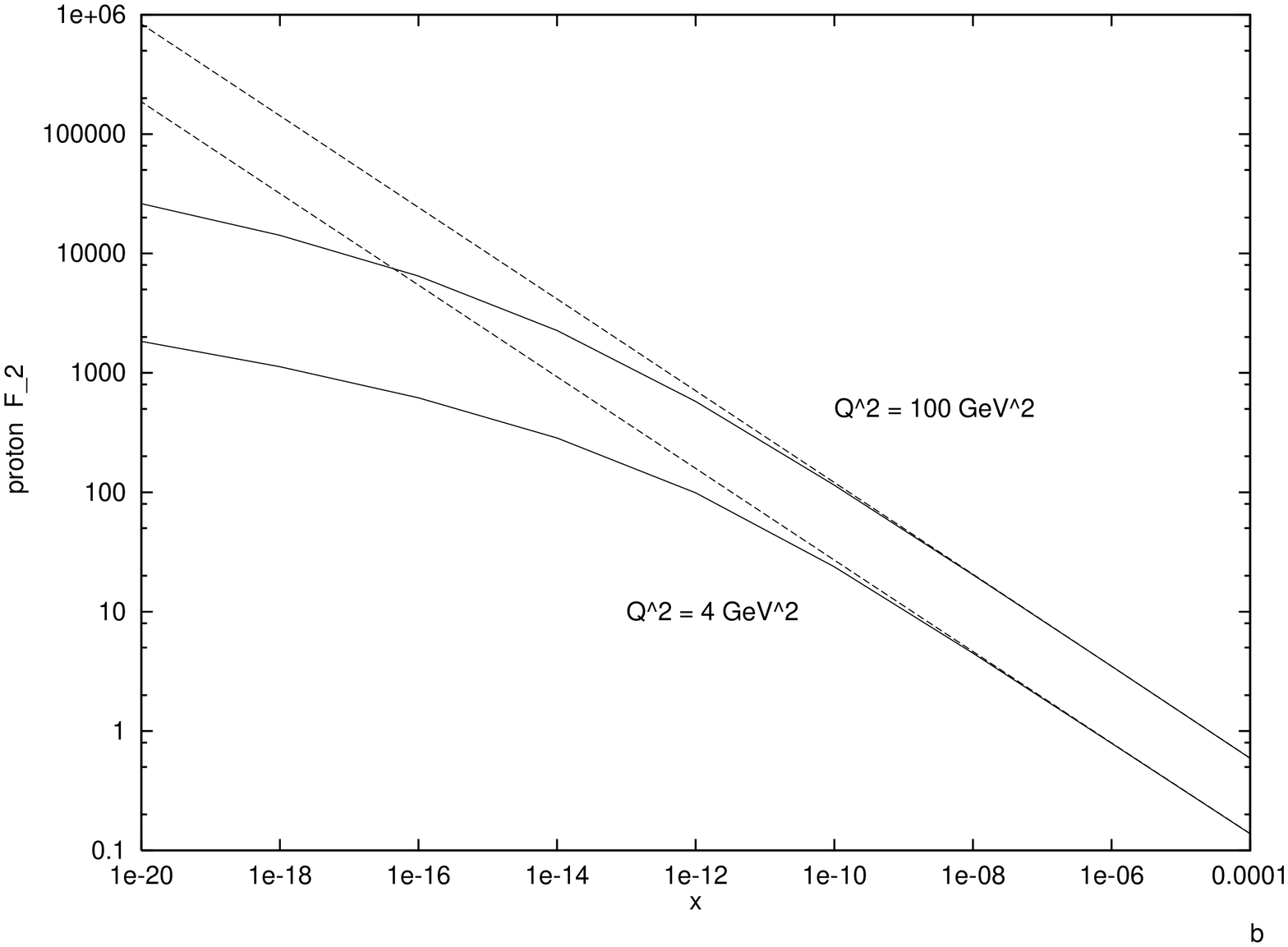}
\captionm{ Structure functions of a virtual photon ($P=1\ GeV/c$) at $Q^{2}=
4$ and $100\ (GeV/c)^{2}$ as a function of $x$ (a) and of the proton  at 
$Q^{2}=4$ and $100\ (GeV/c)^{2}$ as a function of $x$ (b) (solid curves). 
Dashed curves show the contribution of a single pomeron exchange.}
\label{figstru}
\end{figure}

The found supercritical pomerons represent a part of the total pomeron
spectrum which contributes to the Green function, in the high energy limit, 
with a term given by
\beq
G_{P}(\nu,q,q_{1},q'_{1})=\sum_{i=0,1}\nu^{\alpha_{i}(q)-1}
\Phi_{i}(q_{1},q_{2})
\Phi_{i}^{\ast}(q'_{1},q'_{2})
\label{green1}
\eeq
where $\alpha_{i}$ and $\Phi_{i}$ are the trajectories and wave functions of
the leading (0) and subleading (1) pomerons. At high $\nu$ we can neglect the
dependence on the total momentum $q$ of the wave functions, taking them at
$q=0$, and approximate the trajectories according to (\ref{slope}). 
Then all the quantities in (\ref{crosect}) become determined, so that we can 
calculate the
cross-sections for both the transversal and longitudinal projectile photon
and thus find the structure function of the virtual photon target. We have
taken for the latter a transversal photon with the lowest momentum
admissible of $P=1 \ GeV/c$. The resulting structure functions are presented
in Fig. \ref{figstru}.a for the interval of small $x$ which we extended to 
extraordinary small values to clearly see the unitarization effects.

To move closer to reality one has to consider hadronic target and
projectiles. The confinement effects make any rigorous treatment of
such a case impossible. Rather than to introduce arbitrary parameters (in
fact, functions) we extend the formula (\ref{crosect}) to hadronic target and 
projectile substituting the photonic colour densities by hadronic ones. 
For the latter
we take a Gaussian form and a normalization which follows from the comparison
to the electromagnetic densities with only the simplest quark diagrams taken
into account. In particular for the proton we take the Gaussian $\rho$, with
the observed electromagnetic proton radius and normalized to three. Such a
treatment, in all probability, somewhat underestimates the density, since it
does not include coupling to gluons.

\begin{figure}
\centering
\includegraphics[angle=-90, width=3.0in]{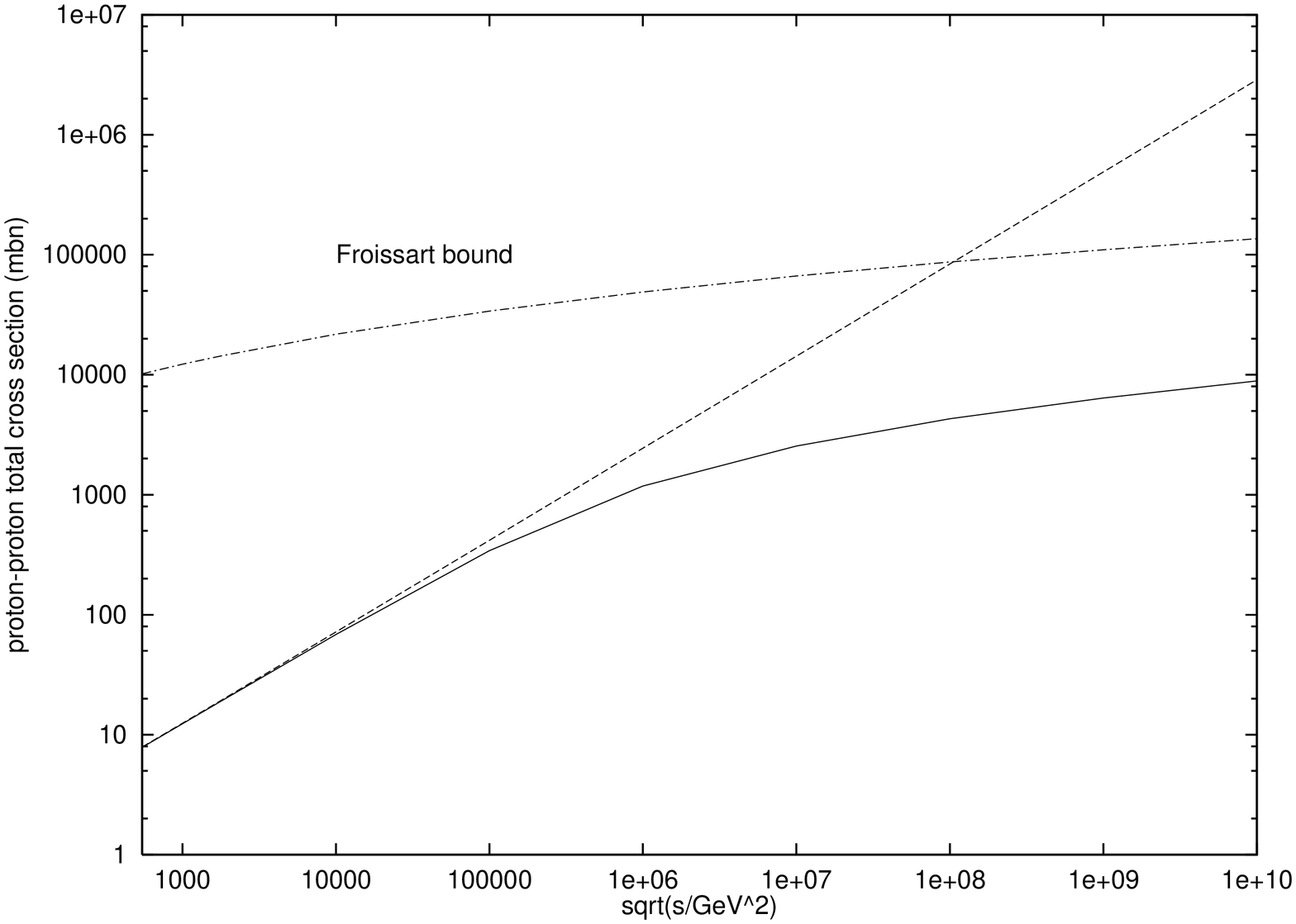}
\captionm{Proton-proton total cross-sections
as a function of c.m. energy $\sqrt{s}$ (the solid curve). The dashed curve
shows the contribution of a single pomeron exchange. The dash-dotted curve
marks the Froissart bound.}
\label{figfrois}
\end{figure}

The proton structure functions and the proton-proton total cross-sections
which follow from this approximation for the densities are shown in Figs.
\ref{figstru}.b and \ref{figfrois} respectively.
To see the unitarization effects we had again  to
consider extraordinary high values of $1/x$ and energies, well beyond our
present experimental possibilities.

In discussing these results, we have first to note that their overall
normalization is somewhat undetermined, since the exact scale at which $\nu$
enters into $\ln\nu$ factors is unknown. A second point to note is that
the subleading pomeron contribution is always very small: it amounts to a few
percent at smallest values of $1/x$ and $s$ considered and naturally gets
still smaller at higher $1/x$ or $s$.

As one observes from Figs. \ref{figstru}-\ref{figfrois}, the structure 
functions and cross-sections
monotonously rise with $1/x$, $s$ and $Q^{2}$. Studying the asymptotics of
the solutions of eq. (\ref{eigeneq1}) at high $q$ and of eq. (\ref{crosect}) 
one can show that this rise is logarithmic. 
In particular, the structure function of the virtual
photon rises as $\ln^{4}(1/x)$ and as $\ln^{\beta}(q^{2})$ with $\beta\sim
2.5$. The proton-proton cross-section eventually rise as $\ln^{2}s$, as
expected. Comparison to the Froissart bound (dash-dotted line in Fig. 7)
shows however that it remains far from being saturated.

The most interesting result that follows from Figs. 
\ref{figstru}-\ref{figfrois} is that the
unitarization effects become visible only at exceedingly very small values of
$x$ or very large values of $s$, well outside the range of the present
experiment. They appear earlier at lower $Q^{2}$. Still at the smallest value
$Q=2\ GeV/c$ considered, the exchange of more than one pomeron achieves
only 15\% of
the total for the  proton structure function at $x=10^{-10}$. Likewise the
relative contribution of many pomerons to the proton-proton cross-section
rises to 23\% only at $\sqrt{s}\sim 10^{5} GeV$.

Comparing the calculated proton structure functions and the cross-sections
with the experimental results at highest $1/x$ and $s$ available we observe
that our results are essentially smaller than the observed ones.
Experimental value of $F_{2p}(Q^{2},x)$ at $Q^{2}=8.5\ (GeV/c)^{2}$ and
$x=0.000178$ is $1.19\pm 0.05\pm 0.16$ \cite{H11995}. 
Our calculations only give a
value 0.17. The $\bar p p$ cross-section at $\sqrt{s}=1800\ GeV$ is around
$80\ mbn$ \cite{multi}, whereas our result is $18.5\ mbn$. Of course, 
having in mind
the uncertainties in the overall normalization and a very crude picture for
the pomeron coupling to the proton assumed, one cannot ascribe too much
importance to this fact. However one is tempted to explain this
underestimation of the experimental values by the simple reason that we are
too far from the pure asymptotic regime yet and that other solutions of
eq. (\ref{eigeneq1}) different from the found supercritical pomerons and 
having their intercepts around unity give the bulk of the contribution at 
present energies.
This may also explain the notorious discrepancy between a
high value of the hard pomeron intercept, of the order 0.35--0.5, and the
observed slow growth of the experimental cross-section, well described by the
"soft pomeron" with an intercept around 0.08.

Forgetting for the moment the triple pomeron interactions, if this picture is 
correct then we may expect that with the growth of energy
the cross-sections will grow faster and faster, until at $\sqrt{s}\sim
10\ TeV$ they will become well described by a pure hard pomeron with
the intercept  0.35--0.5. This power growth will
continue until energies of an order $1000\ TeV$ when finally the
unitarity corrections set in to moderate the growth in accordance with the
Froissart bound.

\section{Inclusive jet production.}

%%%%%%%%%%%%%%%%%%%%%%%%%%%%%%%%%%%%%%%%%%%%%%%%%%%%%%%%%%%%%%%%%%%%%%%%%%%
\subsection{General formalism.}
%%%%%%%%%%%%%%%%%%%%%%%%%%%%%%%%%%%%%%%%%%%%%%%%%%%%%%%%%%%%%%%%%%%%%%%%%%%
We shall consider in the following the scattering of two highly 
virtual photons (or heavy "onia") to make the derivation  
 more rigorous. To obtain the formula for the inclusive 
jet production let us recall  the total cross section 
for one pomeron exchange \cite{braun3}:  
\beq \sigma=\frac{1}{4}\int 
d^{2}rd^{2}r'\rho_{q}(r)\rho_{p}(r') 
\int\frac{d^{2}q_{1}d^{2}q'_{1}}{(2\pi)^{4}}
G(\nu,0,q_{1},q'_{1}) \prod_{i=1,2}(1-e^{iq_{i}r})(1- 
e^{iq'_{i}r'}) \lb{sigmatot} \eeq 
Here
$ q(p) $ is the momentum of the projectile (target);
$ \nu=qp=(1/2)s $; 
$G(\nu,q,q_{1},q'_{1})$ is the non-amputated pomeron Green 
function in the momentum space, $q$ being the total momentum of the
two gluons and $q_1\,(q'_1)$ being the initial (final) momentum of
 the first gluon;
 $\rho_{q(p)}$  
is the dipole colour density of the projectile 
(target).  For the virtual photons we take 
the colour dipole densities from \cite{nikzak1}.

In the high energy limit the dominant term of the Green
function comes from the two mentioned supercritical pomeron 
states (\ref{green1})
where $\alpha_{i}$ and $\Phi_{i}$ are the trajectories and wave functions of
the leading (0) and subleading (1) pomerons.

\begin{figure}
\centering
\includegraphics[angle=-90,width=4.0in]{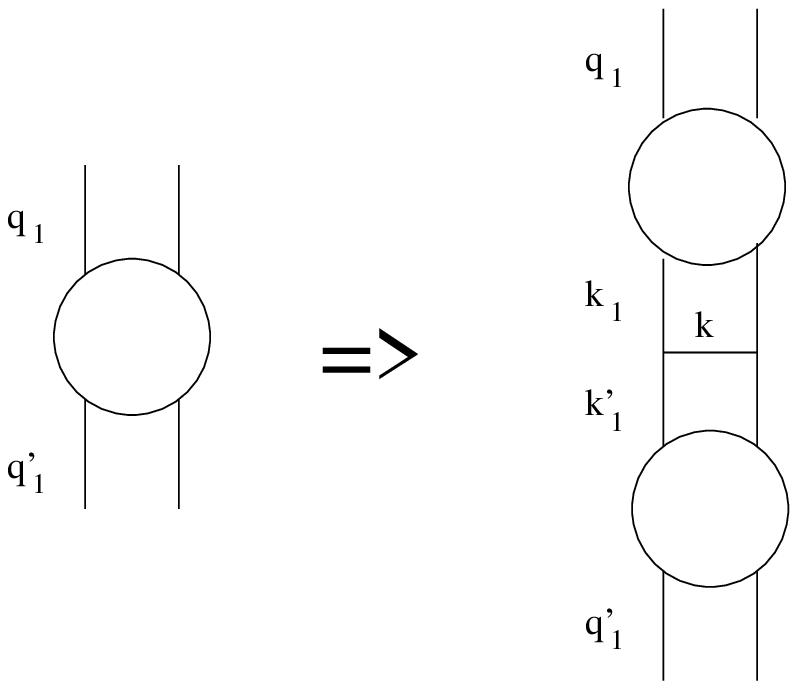}\\
\captionm{The substitution for the Green function used to calculate
the inclusive jet production cross section.}
\label{figgreen}
\end{figure}

Mini-jets appear as intermediate gluon states in the Green 
function (\ref{green1}). They possess arbitrary 
$ \kt $ subject to condition 
$ \kt^{2}<<s $. The inclusive cross-section for 
their production can be calculated by splitting the Green function in 
(\ref{sigmatot}) as indicated in Fig. \ref{figgreen} whihc means that we 
perform the substitution

\beq 
G(\nu,0,\qu,\qup) \Rightarrow \int \frac{d^2\ku d^2 \kup}{(2 
\pi)^4} G(\nu_1,0,\qu,\ku) V(\ku,\kup) G(\nu_2,0,\kup,\qup) 
\delta^{(2)} (\ku+\kt-\kup) \eeq where, for the running 
coupling case, \beq V(\ku,\kup)= 6 \frac{\eta(\ku) 
\eta(\kup)}{\eta(\ku-\kup)} - 3 \eta(0) 
\eeq
Remembering that $q_1+q_2=q'_1+q'_2=q=0$ we obtain
\bea
I(y,\kt)\equiv &&\frac{d^3\sigma}{dy d^2\kt} =
\int d^{2}rd^{2}r'\rho_{q}(r)\rho_{p}(r')
\int \frac{d^{2}q_{1}d^{2}q'_{1}}{(2\pi)^{4}} 
(1-e^{i \qu r})(1- e^{i \qup r'}) 
\int \frac{d^2\ku d^2 \kup}{(2 \pi)^4} \nonumber \\
&&\delta^{(2)} (\ku+\kt-\kup)
G(\nu_1,0,\qu,\ku) G(\nu_2,0,\kup,\qup) 
\bigl[6 \frac{\eta(\ku) \eta(\kup)}{\eta(\ku-\kup)} - 3 \eta(0) \bigr]
\lb{cross1}
\eea

 The appropriate kinematical variables  are the rapidity and 
 transverse momentum  of the observed gluon (jet), 
$y=\frac{1}{2} \ln \frac{k_+}{k_-}$ and $\kt$, respectively  
; we also have $s_i=2 \nu_i$ and $s_{1(2)}=\kt \sqrt{s} 
e^{-(+)y}$. It is convenient to use a mixed 
(momentum-coordinate) representation for the Green functions:  
\beq
G(\nu,0,r,k)= \int \frac{d^2\qu}{(2 \pi)^2} e^{i \qu r} G(\nu,0,\qu,k)
\eeq
Defining $\Delta\Phi_i(r)=\Phi_i(r)-\Phi_i(0)$ we can write
\beq
G(\nu,0,r,k)-G(\nu,0,0,k)=\sum_{i=0,1} \nu^{\Delta_i} \Delta\Phi_i(r)
 \Phi^{*}_i(k)
\eeq
We also introduce
\beq
R_i^{q(p)} \equiv {\langle \Delta\Phi_i \rangle}_{q(p)} =
\int d^2r \rho_{q(p)}(r) \Delta\Phi_i(r)
\eeq
In this notation and also using both the semi-amputated and full wave 
functions we find \beq I(y,\kt)= \sum_{i,j=0,1} R_i^q 
R_j^p \nu_1^{\Delta_i} \nu_2^{\Delta_j} \int 
\frac{d^2\ku}{(2\pi)^4} \Bigl[6 \frac{\psi_i(\ku) 
\psi_j(\ku+\kt)} {\eta(\kt)} - 3 \eta(0) \Phi_i(\ku) 
\Phi_j(\ku+\kt) \Bigr] \lb{cross2} 
\label{I1}
\eeq

Since the found supercritical pomerons are isotropic in the 
transverse space the inclusive cross-section 
$ I(y,\kt) $ also turns out to be isotropic and therefore we can 
integrate over azimuthal angles in (\ref{I1}).
Defining for the full and semi-amputated  wave 
functions the integrated quantities \bea \hat{\Phi} 
(k_1,\kt)&=&\int_0^{2 \pi} d\alpha \Phi(k_1+\kt) \nonumber \\ 
\hat{\psi} (k_1,\kt)&=&\int_0^{2 \pi} d\alpha \eta(k_1+\kt)
 \Phi(k_1+\kt)
\label{angaver}
\eea
where $\alpha$ is the angle between $k_1$ and $\kt$, we 
finally obtain the inclusive cross section 
\bea 
\frac{d^2 \sigma}{dyd\kt^2}= \frac{3}{4} \sum_{i,j=0}^1&&
\! \! e^{-y (\Delta_i-\Delta_j)}
(\frac{\kt^2 s}{4})^{\frac{1}{2}(\Delta_i+\Delta_j)} 
 R_i^q R_j^p \nonumber \\
&& \!\int \frac{d k_1^2} { (2 
 \pi)^3} \Bigl[ 2 \frac{\psi_i(k_1) 
\hat{\psi}_j(k_1,\kt)}{\eta(\kt)} - \eta(0) \Phi_i(k_1)  
   \hat{\Phi}_j(k_1,\kt) \Bigr] 
\lb{cross3}
\eea
The results of numerical calculations of the cross-section 
(\ref{cross3}) and also its generalization to more interesting cases of 
hadronic targets or/and projectiles will be discussed in the 
next section. In the rest of this section we shall study the 
asymptotic behaviour of the found inclusive jet production 
cross-section at very small and very large transverse momenta 
and also its $y$ -dependence.

As to the latter, all $ y $-dependence in (\ref{cross3}) comes from the factor 
$ \exp (-y(\Delta_{i}-\Delta_{j})) $ which has its origin in 
the existence of two different supercritical pomerons. 
Evidently in the limit $ s\to\infty $ this dependence dies out, since the 
relative contribution of the sub-dominant pomeron becomes negligible.
The model thus predicts an asymptotically flat 
$ y$ plateau at very high energies.

At small $ \kt $ the cross-section (\ref{cross3}) evidently goes down as
$ \kt^{\Delta_{1}} $, since all other factors  are finite in 
this limit. However one should remember that (\ref{green1}) gives the 
dominant contribution only while $ (\kt^{2}s) $ continues to be large.
At too small $ \kt $, when the above quantity becomes finite, all other 
states from the spectrum of the two-gluon equation (\ref{eigeneq1}), 
hitherto neglected, begin to give comparable or even 
dominant contribution, so that the found $ \kt^{\Delta_{1}} $
behaviour ceases to be valid. 

To find  the asymptotic behaviour of the 
inclusive cross section for $ \kt \to \infty$  
 we need to know  the behaviour of the pomeron wave functions
in the momentum space at 
$ q\to\infty $ and in the ordinary space at 
$ r\to 0 $. 
In the appendix A.1 we show that \bea \psi(q) && 
\mathop{\sim}_{q \to \infty} \quad \frac{1}{q^2} \bigl( \ln 
q^2 \bigr)^{\beta} \nonumber \\ \psi(r) && \mathop{\sim}_{r 
\to 0} \quad \bigl( \ln \frac{1}{r} \bigr)^{\beta+1} \eea
where $\beta = -1 -\frac{3}{b E}$ so that in the forward scattering case 
$\beta$ depends just on the intercept of the corresponding pomeron state.
Let us study now the behaviour of (\ref{cross2}) for 
$\kt \to \infty$.
Using  
$\Phi(q) = \psi(q)/\eta(q) \sim \frac{1}{q^4}\bigl( \ln q^2 \bigr)^{\beta-1}$
we find $\int d^2q \Phi(q) < \infty$, so that for the second term in the 
integrand of
(\ref{cross2}) we get
\beq
\int d^2 \ku \Phi_i(\ku) \Phi_j(\ku+\kt) 
\mathop{\longrightarrow}_{\kt^2 \to \infty} \Phi_j(\kt) \int d^2\ku \Phi_i(\ku)
\sim \frac{1}{\kt^4} \bigl( \ln \kt^2 \bigr)^{\beta_j-1}
\eeq

To analyze the first term we use the identity
\beq
\int \frac{d^2\ku}{(2 \pi)^2} \psi_i(\ku) \psi_j(\ku+\kt)=2 \pi
\int_0^{\infty} r dr J_0(\kt r) \psi_i(r) \psi_j(r)
\eeq
and the relation
\beq
\int_0^{\infty} r dr J_0(\kt r)f(r)=-\frac{1}{\kt}
\int_0^{\infty} r dr J_1(\kt r)f'(r)
\eeq
which is valid for naturally behaved f, such that 
$\Bigl[ r f(r) J_1(\kt r) \Bigr]_{r=0}^{r=\infty}=0$.

Putting $f(r) \sim \bigl( \ln \frac{1}{r} \bigr)^{\beta_i+\beta_j+2}$
we obtain the leading behaviour
\beq
\int d^2\ku\psi_i(\ku) \psi_j(\ku+\kt) \sim \frac{1}{\kt^2} 
\Bigl( \ln \kt^2 \Bigr)^{\beta_i+\beta_j+1}
\eeq
So at 
$ \kt\to\infty $ we find for the inclusive cross-section (\ref{I1}) 
\beq
I(y,\kt) \sim \sum_{i,j=0,1} R_i^q R_j^p 
 (\frac{\kt^2  s}{4})^{\frac{\Delta_i+\Delta_j}{2}}
   e^{-y (\Delta_i-\Delta_j)} \frac{1}{\kt^4} 
   \Bigl( \ln \kt^2 \Bigr)^{\beta_i+\beta_j}
\label{asymcross}
\eeq   
This asymptotics corresponds to the standard quark-counting 
rules behaviour ($\sim 1/\kt^{4}$), modified by a power factor 
due the pomeron energetic dependence and a logarithmic factor
coming from the pomeron wave function, i.e. from the running 
of the coupling. Note that in (\ref{asymcross}) the 
$ \kt $ and 
$ s $ dependencies are separated. As a result we find that the 
average 
$ \langle \kt \rangle $  is finite and independent of $ s $ and 
$ \langle \kt^{2} \rangle \sim s^\Delta $, since it formally
diverges for (\ref{asymcross}) and one has to restrict $k^2_{\bot}<s$.

For the the multiplicity
\beq
\langle n \rangle = \frac{1}{\sigma} \int \frac{d^3 \sigma}{dy d^2\kt}
		dy d^2\kt
\eeq
the standard asymptotic behaviour $ \langle n \rangle = a \ln s + b$
is obtained. Indeed integrating (\ref{cross2}) we get 
$\int dy d^2\kt \frac{d^3 \sigma}{dyd^2\kt}=B s^{\Delta_0}\ln s
+ C s^{\Delta_0}$.  Since the total cross-section has the form
$\sigma= A s^{\Delta_0}$ at large 
$ s $, we get the mentioned asymptotical expression for the 
multiplicity. Of course this result is valid only in the 
extreme limit 
$ s\to\infty $. At large but finite 
$ s $ the existence of two different pomerons leads to some 
additional non-trivial 
$ s $-dependence.

It is instructive to compare our asymptotic results with those 
obtained in the BFKL fixed coupling model. In the latter case 
the inclusive cross section is of course badly behaved in the 
$\kt \to 0$ limit due to scale invariance. It is also 
very different in the high $\kt$ limit.  At very large 
$ \kt $ such that 
$ \ln\kt\sim\sqrt{\ln s} $ one finds 
\beq \Bigl(\frac{d^2\sigma}{dyd\kt^2}\Bigr)_{BFKL} \quad
\mathop{\sim}_{\kt \to \infty} \quad a(y)\frac{ 
(\kt^2 s)^{\Delta}}{\kt^2} \frac{e^{- \ln^2 \kt^2/a^{2}(y)}} 
{\ln \kt^2\sqrt{\ln s}} \eeq
where $a^2 \sim (\ln s-4y^{2}/\ln s)$. Thus for  large 
$\kt$  the BFKL cross section goes down faster than any power.
Also one obtains that 
$ \langle \ln\kt\rangle\sim\sqrt{\ln s} $ so that both $ \kt $ and 
$ \kt^2 $ grow with $ s $. 

To bring these predictions in better correspondence with the
physical reality, as mentioned, various modifications of this
orthodox BFKL approach have been introduced. In particular
in \cite{glr83,lr89} the fusion of gluons via  
 "fan" diagrams was assumed to take place at high gluonic 
densities, which was considered as a way to partially restore 
the s-channel unitarity.  Then, under some additional 
assumptions, a $ 1/\kt^4 $ asymptotic behaviour similar to 
(\ref{asymcross}) was found.
In our model such a behaviour naturally follows  without
introducing additional assumptions or imposing the unitarity
restrictions.

%%%%%%%%%%%%%%%%%%%%%%%%%%%%%%%%%%%%%%%%%%%%%%%%%%%%%%%%%%%%%%%%%%%%%%%%%
\subsection{Numerical results.}
%%%%%%%%%%%%%%%%%%%%%%%%%%%%%%%%%%%%%%%%%%%%%%%%%%%%%%%%%%%%%%%%%%%%%%%%%
Taking the wave function evaluated numerically in \cite{bvv} we have
computed  the cross section (\ref{cross3}).

\begin{figure}
\centering
%\raisebox{4cm}{$\ln{\frac{a(t)}{a(t_0)}}$} \hspace {-0.2cm}
\includegraphics[angle=-90, width=3.0in]{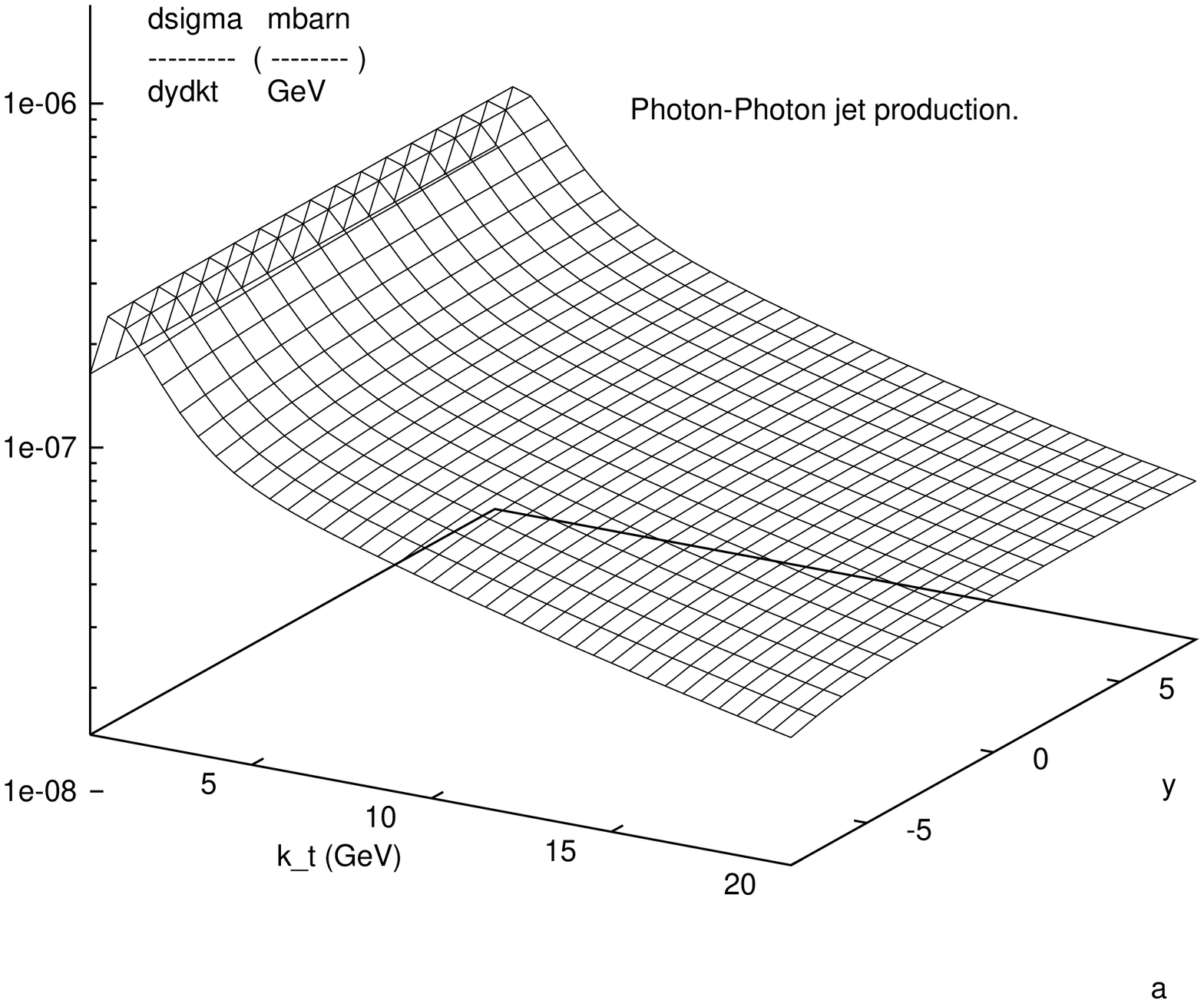}
\hspace{-0.2in}
%\raisebox{4cm}{$H$} \hspace {-0.3cm}
\includegraphics[angle=-90,width=3.0in]{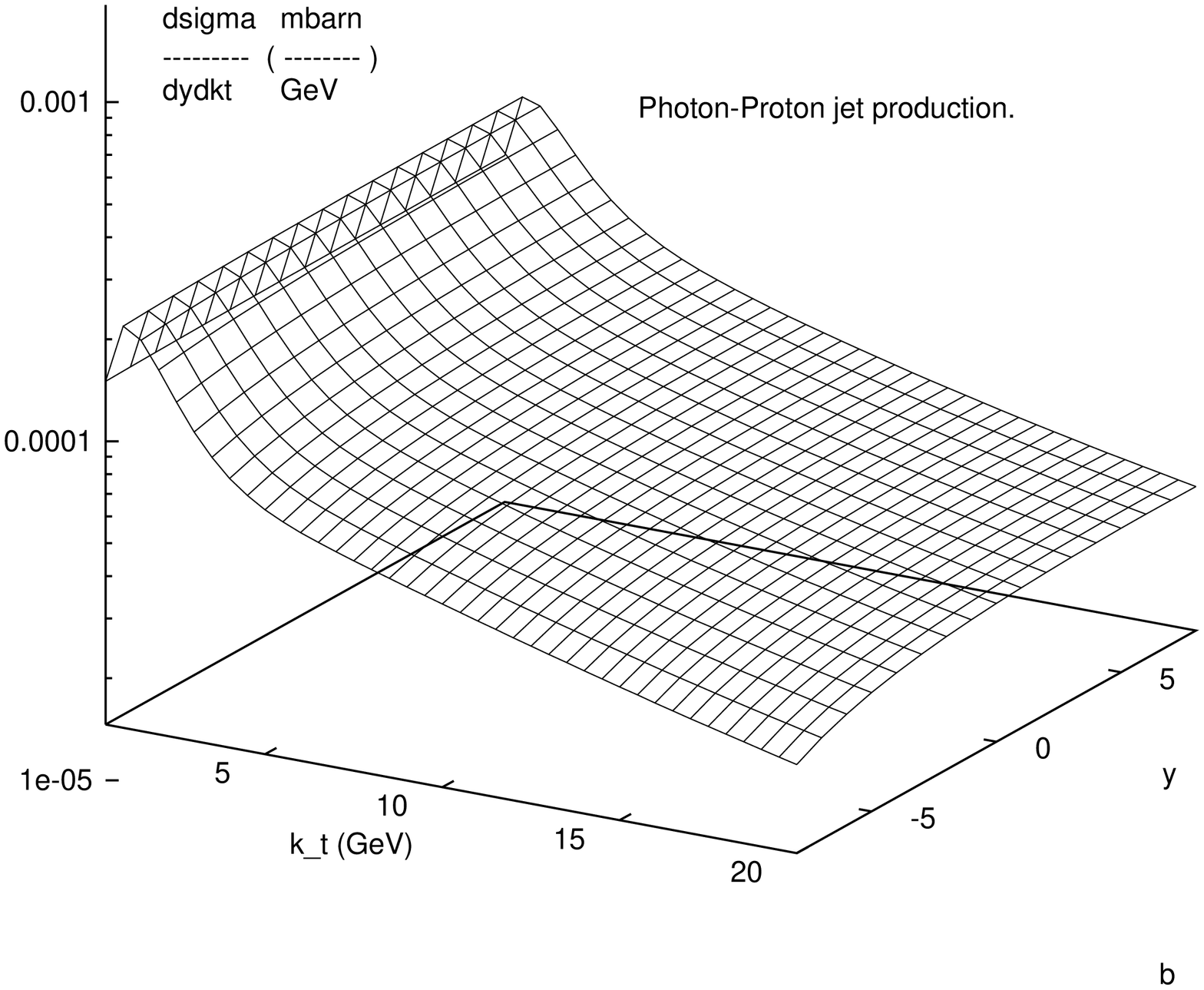}\\
%\raisebox{4cm}{$\rho$} \hspace {-0.2cm}
\includegraphics[angle=-90,width=3.0in]{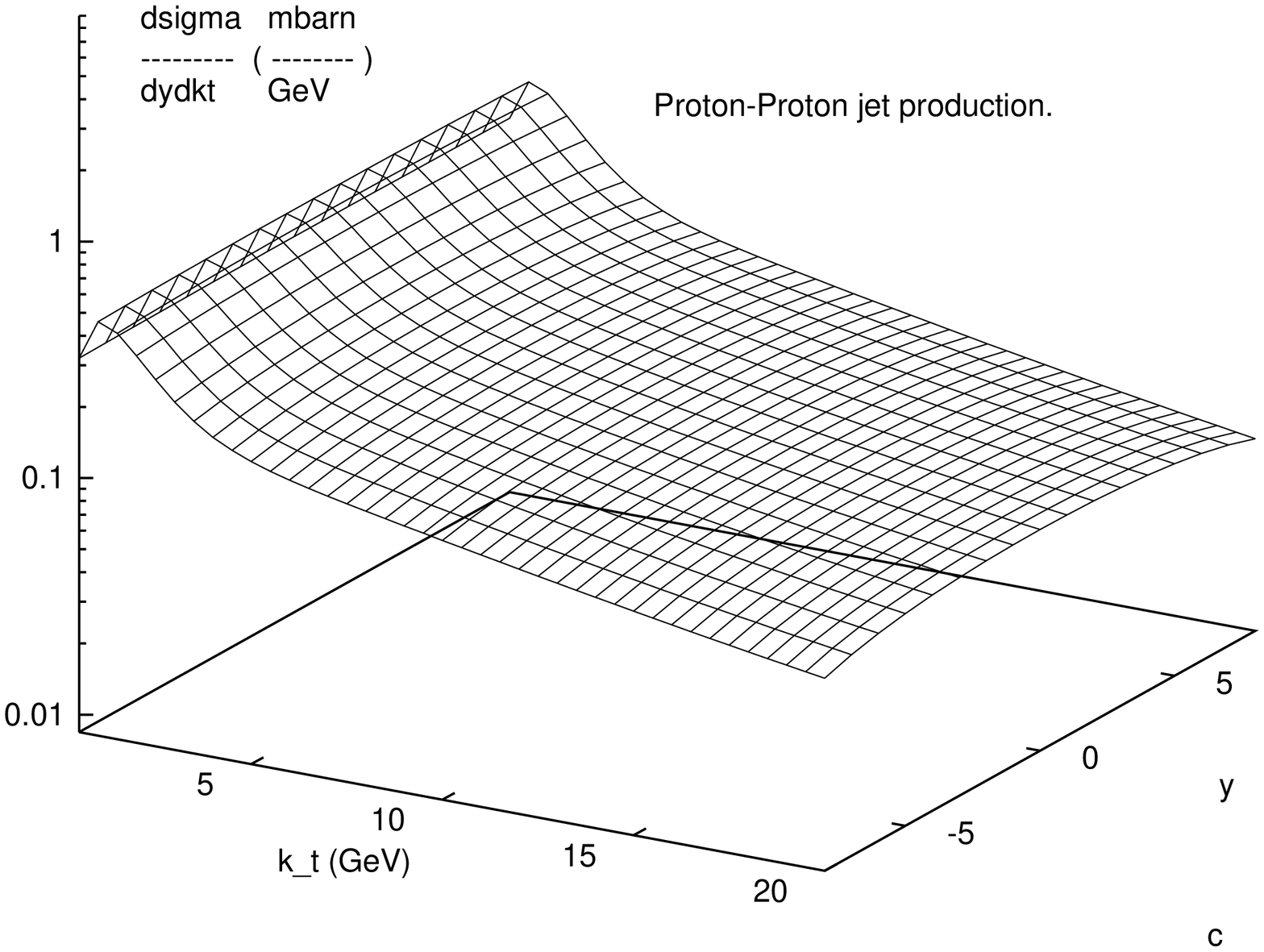}

\captionm{Inclusive jet production cross sections for the processes
 $\gamma^*\gamma^*$ at  $\sqrt{s}=540GeV$ with $Q^2=25(GeV/c)^2$
and $P^2=1(GeV/c)^2$ (a), for $\gamma^* p$ at  $\sqrt{s}=540GeV$ with 
$Q^2=25(GeV/c)^2$ (b) and for $pp$ at the same c.m.s. energy (c).}
\label{fig2d}
\end{figure}

In Fig. \ref{fig2d}.a we present $d^2\sigma/dyd\kt$ for the process
$\gamma^*\gamma^*$ (in 
units $c=1$).
We have chosen  the projectile photon to have virtuality  
$Q=5GeV/c$ and  the target one to have virtuality $P=1GeV/c$. 
The center of mass energy is $\sqrt{s}=540 GeV$.

Of course, processes involving hadronic targets or/and 
projectiles are much more interesting from the practical point 
of view. However these require some non-perturbative 
input for the colour densities of the colliding hadrons. 
A possible way to introduce it is evidently to convert 
eq. (\ref{eigeneq1}) into an evolution equation in 
$ 1/x $  and take initial conditions for it from the 
existing experimental data. Postponing this complicated 
procedure for future studies, we use here a simpler 
approach, taking for the hadron (proton) a Gaussian 
colour density with a radius corresponding to the 
observed electromagnetic one. Such an approximation, in 
all probability, somewhat underestimates the coupling of 
the hadron to the pomeron, since the coupling of the 
latter to constituent gluons is neglected. Nevertheless, we 
hope that it gives a reasonable estimate for the 
inclusive cross-section. Using this approach we get the 
inclusive jet production cross-sections 
for the $\gamma^*p$ and $pp$ scattering  shown in Fig. \ref{fig2d}.b and 
\ref{fig2d}.c respectively.

A common feature of jet production in all processes is that 
the cross-section reaches a maximum at 
$ \kt\approx 1\ GeV/c $ from which it monotonously goes down 
both for smaller and larger $ \kt $.

\begin{figure}
\centering
\includegraphics[angle=-90,width=3.0in]{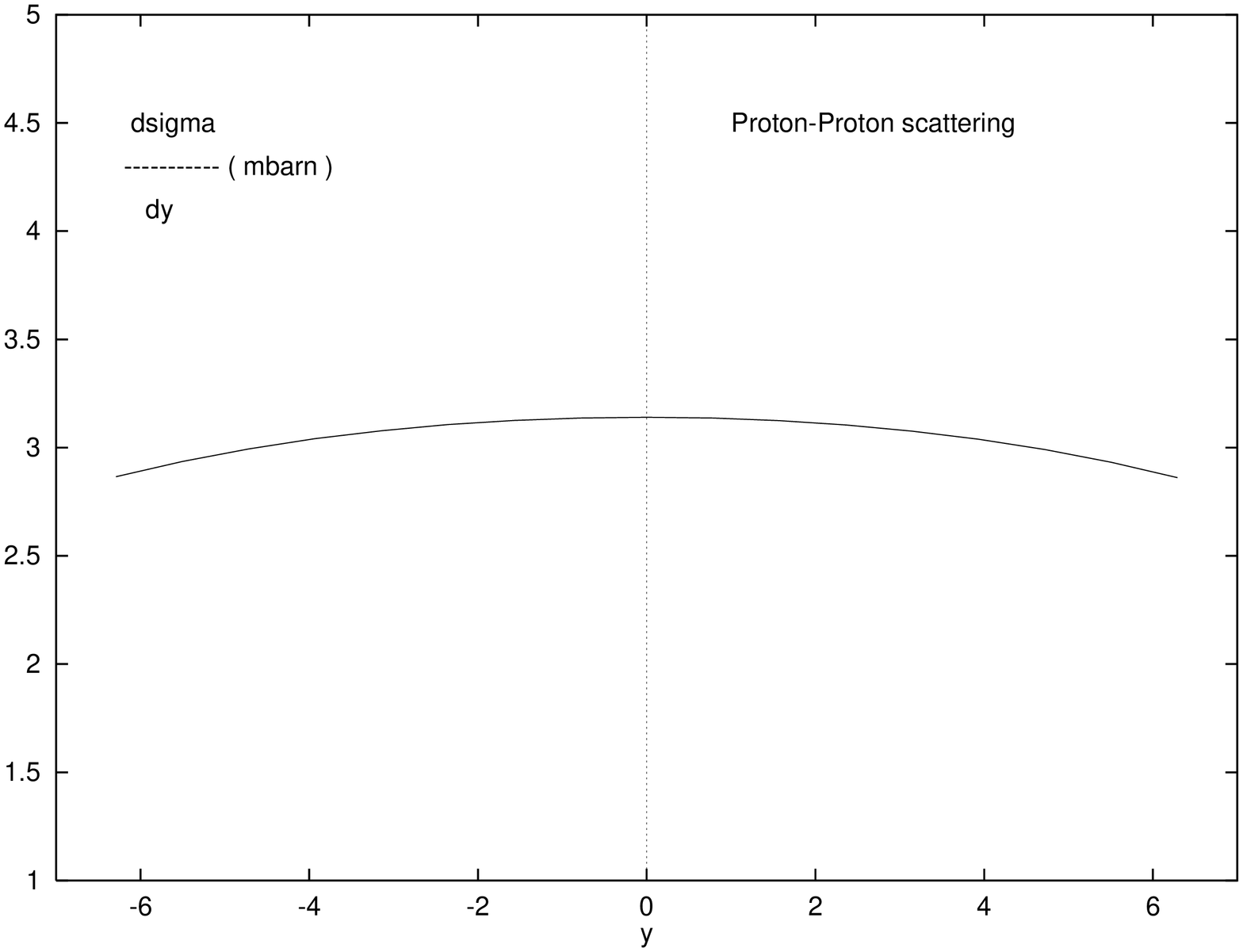}\\
\captionm{Cross-sections $d\sigma/dy$ for the $pp$ process
at $\sqrt{s}=540GeV$.}
\label{fig1d}
\end{figure}

In Fig. \ref{fig1d} the cross-section $d \sigma/dy$ integrated over 
$ \kt $ in the interval $0.5\div 20\ GeV/c$ is presented for the process 
$ pp $. The limitations in the numerical calculations of the 
 wave functions do not allow us to study higher values of 
$ \kt$, so that we cannot numerically reach the region where 
the asymptotical behaviour (\ref{asymcross}) is strictly valid.

\begin{figure}
\centering
\includegraphics[angle=-90,width=3.0in]{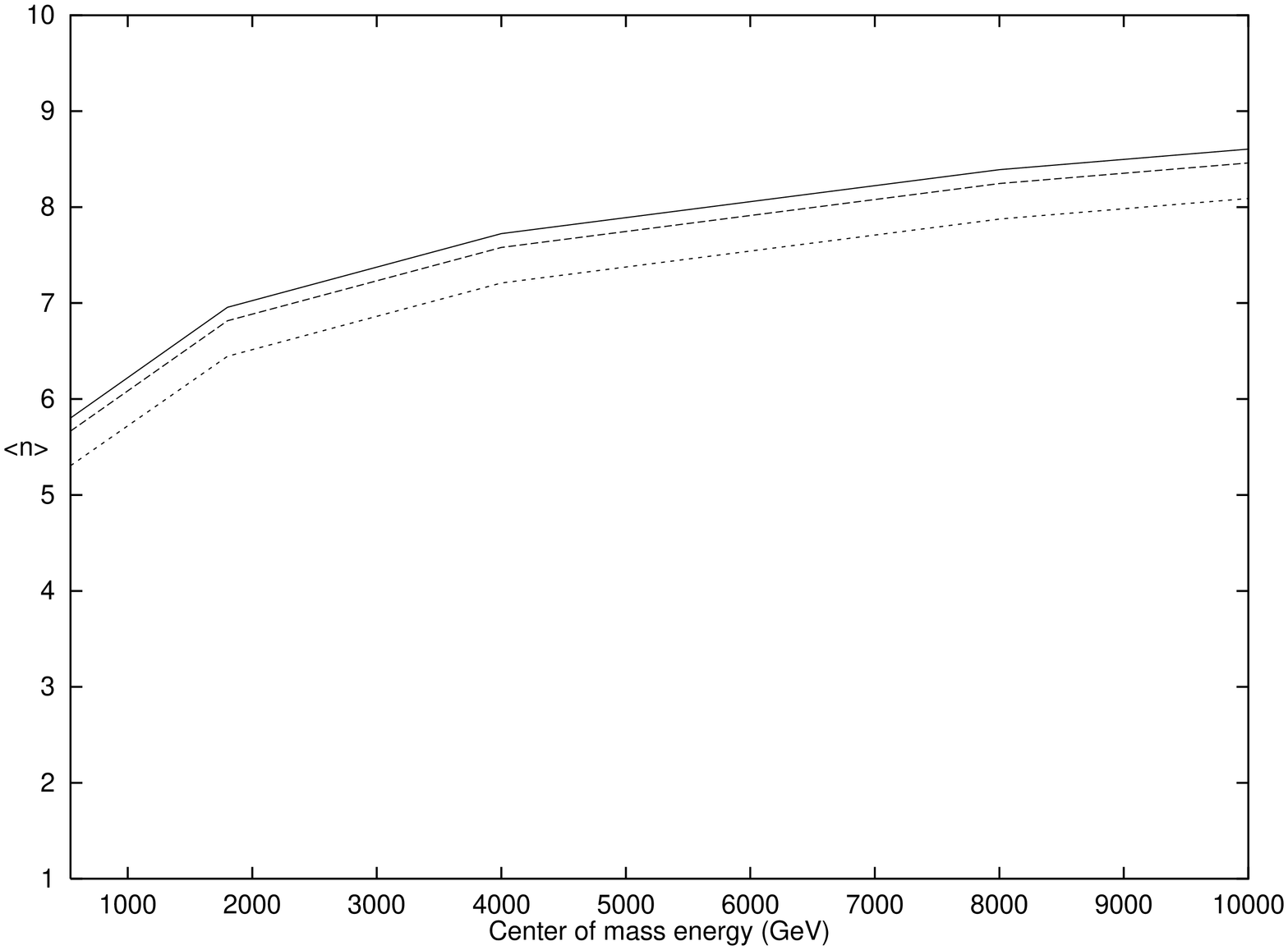}\\
\captionm{Multiplicities $\langle n \rangle$ as a function of the center
of mass  energy $\sqrt{s}$ for the processes $\gamma^*\gamma^*$
(the solid curve),$\gamma^*p$ (the dashed curve) and $pp$  (the dotted curve).}
\label{figmulti}
\end{figure}

In Fig. \ref{figmulti} we show jet multiplicities as a function of 
$ s $ for the three studied processes $\gamma^*\gamma^*$,
$\gamma^*p$ and $pp$. As one observes, their magnitude and 
behaviour prove to be quite similar. 

We also tried to estimate the average $ \langle \kt \rangle$.
Unfortunately, although it exists according to (\ref{asymcross}), its value
results very sensitive to the high momentum tail of the
pomeron wave function, poorly determined from our numerical
calculations. To avoid this difficulty we chose to calculate the
average $\langle \ln \kt/\Lambda \rangle$ instead. This average at 
fixed $y$ depends on $s$ and $y$ only due to the existence
of two different pomerons, so that this dependence should die
out at large enough $s$. In fact the found average
$\langle \ln \kt/\Lambda \rangle$ turns our to be practically
independent of $y$ and very weakly dependent on $s$ in the whole
studied range of $s$ and $y$, rising from  3.97 at $\sqrt{s}=20\ GeV$ to
4.19 at $\sqrt{s}=20\ TeV$. These values imply a rather high average
$\kt$ rising from 10.6 $GeV/c$ at $\sqrt{s}=20\ GeV$ to
13.2 $GeV/c$ at $\sqrt{s}=20\ TeV$.

Our calculations show that the introduction of a running 
coupling constant on the basis of the bootstrap condition 
cures all the diseases of the orthodox BFKL approach for jet 
production. At high $\kt$  the cross-section becomes well-behaved and more or 
less in accordance with the expectations based on the quark 
counting rules. At small $\kt$ no singularity occurs, although contributions 
from other states is expected  to dominate.

As to $s$-  and $y $ -dependence, our predictions are even simpler 
than in the BFKL approach, since the  running of the coupling
converts the branch point in the complex angular momentum 
plane, corresponding to the BFKL pomeron, into poles, of which 
only two are located to the right of unity and contribute at
high energies. As a result at super-high energies, when only 
the dominant pomeron survives, the 
$ y $-dependence completely disappears and the 
$ s $-dependence reduces to the standard 
$ s^{\Delta_{0}} $ factor. At smaller 
$ s$  some 
$ y $- and extra 
$ s $-dependence appears due to the existence of two 
supercritical pomerons.

With all these refinements, some basic predictions of the BFKL 
theory are reproduced. The cross-section for mini-jet 
production rises fast and saturates the total cross-section as 
$s\to\infty$. Jet multiplicities rise logarithmically.
 
However $ \langle\kt\rangle $ turns out to depend on $s$ weakly
and its calculated value results pretty high, of the order
of 10-12 $GeV/c$. This should be contrasted to the experimentally
observed much lower values of $\langle \kt\rangle$ rising with
energy. A natural explanation of this discrepancy follows from
the results \cite{bvv} shown in the first section of this chapter 
in particular from the study of the structure
functions and total cross-sections in our model: at present 
energies the contribution from the two supercritical pomerons only
covers a part of the observed phenomena because we are still rather far
from the real asymptotics. The bulk of the contribution comes
from other states, which produce a much softer spectrum of
particles. The observed rise of the $\langle\kt\rangle$ is then related
to the  dying out of these sub-asymptotical states. Our prediction
is then that the rise of $\langle\kt\rangle$ should saturate at the
level of 10-12 $GeV/c$.

This circumstance has also to be taken into account when discussing
the numerical results presented in Figs. 6-9. Probably they also
illustrate predictions for considerably higher energies than the 
present ones, at which, according to the estimate made in \cite{bvv},
they should account for $\sim 20$\% of the observed spectra.

To obtain predictions better suited for present energies
one should evidently take into account all states present
in the spectrum of the pomeron equation (\ref{eigeneq1}) and not only the
two supercritical pomerons. To realize this program an evolution
equation in $\nu$ following from (\ref{eigeneq1}) seems to be an appropriate 
tool. As mentioned, it could also effectively take into account the 
non-perturbative effects related to the pomeron coupling to physical hadrons.

%%%%%%%%%%%%%%%%%%%%%%%%%%%%%%%%%%%%%%%%%%%%%%%%%%%%%%%%%%%%%%%%%%%%%%%%%
\section{Evolution of the gluon density.}
For the forward scattering amplitude the pomeron equation (\ref{eigeneq1})
for the isotropic wave function with an inhomogeneous term reads
\beq
(H-E)\psi=\psi_{0}
\label{shorteq}
\eeq

Taking the Mellin transformation of (\ref{shorteq}) one converts it into an 
evolution equation in $1/x$:
\beq
\frac{\partial}{\partial\ln 1/x}\psi(x,k)=-H\psi(x,k)
\label{evoleq}
\eeq
which should be supplemented with an initial condition at some $x=x_{0}$
\beq
\psi(x_{0},k)=\psi_{0}(k)
\eeq
containing the non-perturbative input about the coupling to the hadronic 
target.

The physical interpretation of the pomeron wave function is provided by 
the fact that in the DLLA scheme eq. (\ref{evoleq}) reduces to an 
equation for the fully amputated function $\phi(x,k)=\eta(k)\psi(x,k)$:
\beq
\frac{\partial}{\partial\ln k^2}
\frac{\partial}{\partial\ln 1/x}\phi(x,k)=\frac{3\alpha_{s}(k^{2})}
{\pi}\phi(x,k)
\label{dlevoleq}
\eeq
which coincides with the standard equation for the unintegrated 
gluon density $xg(x,k^{2})$ in the DLLA limit. In fact, this circumstance 
lies at the root of our method to introduce a running coupling into the 
scheme. Thus we may identify
\beq
\phi(x,k)=cxg(x,k)
\label{phiglu}
\eeq
The normalizing factor $c$ cannot be determined from the asymptotic 
equation (\ref{dlevoleq}). We shall be able to fix it by studying the 
coupling of the pomeron to the incoming virtual photon in the next section.

%%%%%%%%%%%%%%%%%%%%%%%%%%%%%%%%%%%%%%%%%%%%%%%%%%%%%%%%%%%%%%%%%%%%%%%%%
\subsection{Coupling to the virtual photon}
Once the function $\phi$ proportional to the gluon density is 
determined, one has to couple it to the projectile particle to calculate 
observable quantities. In particular, to find the structure function of 
the target one has to couple the gluons to the incoming virtual photon, 
that is, to find the colour density $\rho(q,k)$ which connects the photon 
of momentum $q$ to the gluon of momentum $k$. This problem is trivial 
within the BFKL approach with a fixed small coupling. Then it is 
sufficient to take the colour density in the lowest order $\rho_{0}(q,k)$, 
which corresponds to taking for it the contribution of a pure quark loop
into which the incoming photon goes.

The problem complicates enormously when one tries to introduce a running 
coupling into $\rho$. Then one has to take into account all additional 
gluon and $q\bar q$ pair emissions which supply powers of the logarithms
of transverse momenta. Apart from making the coupling run, they will 
evidently change the form of $\rho(q,k)$. Unfortunately the bootstrap 
relation can tell us nothing about the ultimate form of the colour 
density with a running coupling, which essentially belongs to the 
$t$-channel with a vacuum colour quantum number. So we have to find a 
different way to introduce a running coupling into $\rho$.

A possible systematic way to do this consists in
applying to the photon-gluon coupling the DGLAP evolution equation.
One may separate the colour 
density from the rest of the amplitude by restricting its rapidity range 
to some maximal rapidity $y_{0}\sim\log Q^2$ (which, of course should be
much smaller than the overall rapidity $Y\sim\log Q^{2}/x$). Then the 
kinematical region of $\rho(q,k)$ will admit the standard DGLAP 
evolution in $Q^2$. Solving this equation one will find the quark 
density at scale $Q^2$ of the gluon with momentum $k$ (i.e. essentially 
the structure function of the gluon with the virtuality $k^2$). This is 
exactly the quantity needed to transform the calculated gluon density 
created by the target into the observable structure function of the target.
As a starting point for the evolution one may take the perturbative 
colour density $\rho_{0}$ at some low $Q^2$ when the logs of the 
transverse momenta might be thought to be unimportant.

This ambitious program, combining both evolution in $1/x$ and 
$Q^2$, does not, however, look very simple to realize. As a first step, 
to clearly see the effects of the introduction of a running coupling 
preserving the gluon reggeization, we adopt a more phenomenological approach 
here,  trying to guess a possible correct form for $\rho(q,k)$ on the basis of 
simple physical reasoning and also using the DLLA to fix its final form.

With a pure perturbative photon colour density one would obtain for the 
$\gamma^{*}p$ cross-section
\beq
\sigma(x,Q^{2})=\int\frac{d^{2}k\rho_{0}(q,k)\phi(x,k)}{(2\pi)^{2}\eta^{2}(k)}
\label{crosstot}
\eeq
In fact, the projectile particle should be coupled to the full pomeron 
wave function $\phi/\eta^2$. From the physical point of view this 
expression is fully satisfactory for physical particles. However it is 
not for a highly virtual projectile.

\begin{figure}
\centering
\includegraphics[width=4.0in]{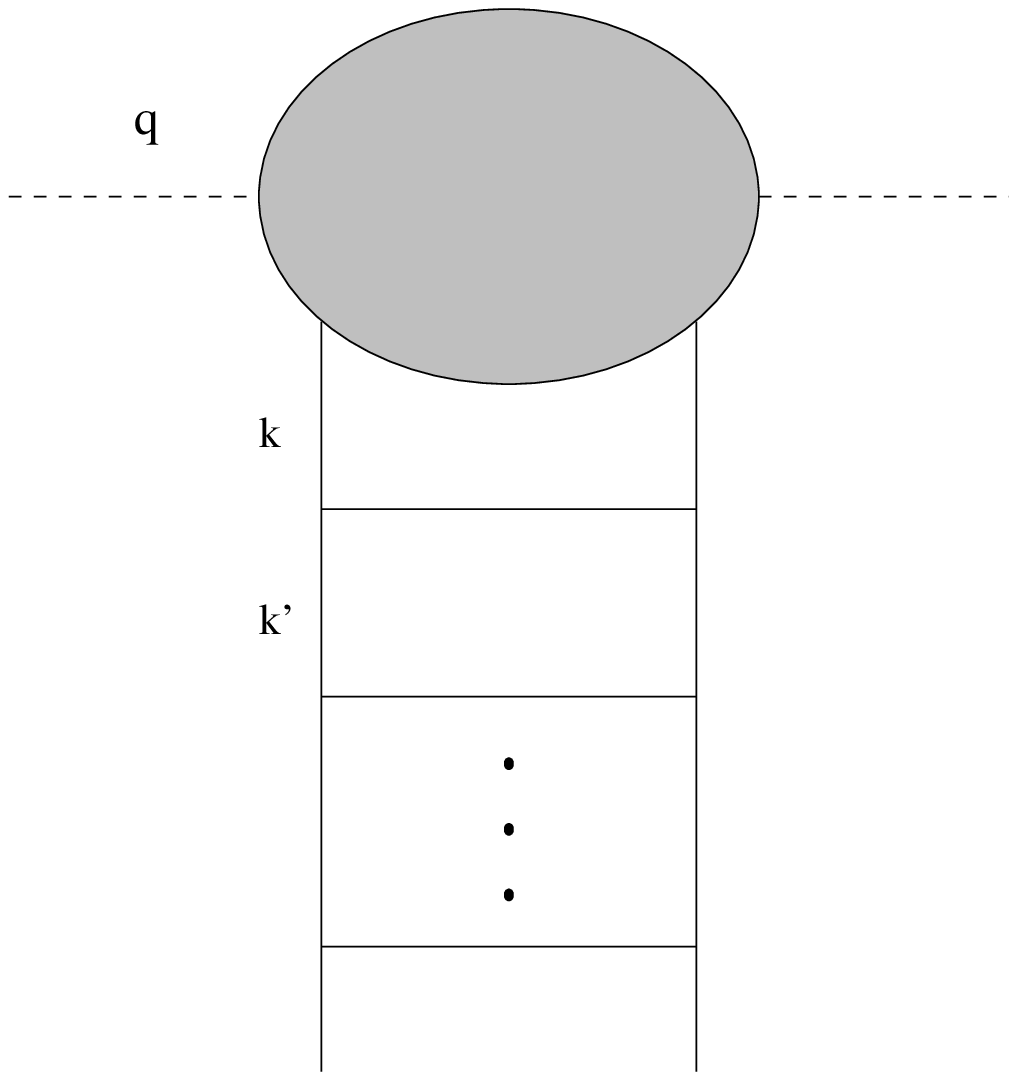}\\
\captionm{Forward amplitude for the scattering of a photon with a reggeized 
gluon..}
\label{figblob}
\end{figure}

To see this, we first note that for the forward amplitude our method of 
introducing a running coupling reduces to a very simple rule: the scale 
at which the coupling should be taken is given by the momentum of the 
emitted real gluon ($(k-k')^2$ in the upper rung in Fig. \ref{figblob}). 
Now take $Q^{2}$ very large and apply the DLLA approximation.
Then the momenta in the ladder  become ordered from top to bottom
\[Q^{2}>>k^{2}>>{k'}^{2}>>.....\]
In this configuration, as can be traced from (\ref{shorteq}) and 
(\ref{crosstot}), all $\alpha_{s}$'s
acquire the right scale (i.e. corresponding to the DGLAP equation) except
for the upper rung: $\alpha_{s}(k^{2})$ appears twice. 
In order to clearly see why this happens let us explicitly write eq. 
(\ref{shorteq}) which describes the interactions in the ladder of Fig. 
\ref{figblob} (the kinetic part given by the trajectories of the two 
reggeized gluons is denoted by $T$)
\beq
(E -T(k))\phi(k)=\phi_0(k)+2\int \frac{d^2k'}{(2\pi)^2} 
\frac{\eta(k)}{\eta(k') \eta(k-k')} \phi(k')
\label{itereq}
\eeq
Constructing by iteration the solution, for example at the second 
order, one has
\beq
\phi^{(2)}(k)=\frac{2}{E-T(k)}
\int \frac{d^2k'}{(2\pi)^2} \frac{\eta(k)}{\eta(k-k')} \frac{2}{E-T(k')}
\int \frac{d^2k''}{(2\pi)^2} \frac{1}{\eta(k'') \eta(k'-k'')} 
\frac{\phi_0(k'')}{E-T(k'')}
\label{dlitersol}
\eeq
and by taking the limit of the DLLA approximations one obtains
\beq
\phi^{(2)}(k)=\frac{2}{E-T(k)}
\int \frac{d^2k'}{(2\pi)^2} \frac{1}{\eta(k')} \frac{2}{E-T(k')}
\int \frac{d^2k''}{(2\pi)^2} \frac{1}{\eta(k'')} 
\frac{\phi_0(k'')}{E-T(k'')}
\label{itersol}
\eeq
It is evident that after taking this limit there is no cancellation of
one $\eta(k)$ term in (\ref{crosstot}).
Therefore this defect can be understood if one notices that the upper gluon 
is, in fact, coupled to a virtual particle. 
If this particle were a gluon, then the interaction $K_q$
in (\ref{tra_int}) would cancel one of the two $\alpha(k^{2})$'s and 
substitute it by an $\alpha$ taken at the scale corresponding to its own 
virtuality.
We assume that something similar should take place also for virtual 
quarks to which the gluon chain may couple. The scale of the particle
momenta squared which enter the upper blob in Fig. 10 should have the order 
$Q^{2}$ (this is the only scale that remains after these momenta are 
integrated out).
As a result  the lowest order density should be rescaled according to
\beq
\rho_{0}(q,k)\rightarrow\frac{\alpha(Q_{1}^{2})}{\alpha(k^{2})}\rho_{0}(q,k)
\label{ansatz}
\eeq
where $Q_{1}^{2}$ has the same order as $Q^2$.

The approximation we assume is that the substitution 
(\ref{ansatz}) is sufficient to correctly represent the photon colour density 
with a running coupling. We shall check its validity by studying the quark 
density which results from (\ref{ansatz}) in the DLLA approximation
and comparing it with the known result based on the DGLAP equation.

Explicitly the zeroth order density $\rho_{0}$ has the following forms 
for the transverse (T) and longitudinal (L) photons (see e.g. \cite{nikzak1} 
and do the integration in the quark loop momenta)
\beq
\rho^{(T)}_{0}(q,k)=\frac{3e^{2}}{8\pi^2}\sum_{f}Z_{f}^{2}
\int_{0}^{1}d\alpha\left((\alpha^{2}+(1-\alpha)^{2})
((1+2z^{2})g(z)-1)+\frac{\zeta}{\alpha(1-\alpha)+\zeta}(1-g(z))\right)
\label{rhotra}
\eeq
\beq
\rho^{(L)}_{0}(q,k)=\frac{3e^{2}}{2\pi^2}\sum_{f}Z_{f}^{2}
\int_{0}^{1}d\alpha\frac{(\alpha(1-\alpha))^{2}}{\alpha(1-\alpha)+\zeta}
(1-g(z))
\eeq
Here the summation goes over the quark flavours. The dimensionless 
variables $\zeta$ and $z$  are defined as
\beq
\zeta=\frac{m_{f}^{2}}{Q^2},\ \ 
z=\frac{k^2}{4Q^2}\frac{1}{\alpha(1-\alpha)+\zeta}
\eeq
and $m_{f}$ and $Z_{f}$ are the mass and charge of the quark of 
flavour $f$. The function $g(z)$ is given by
\beq
g(z)=\frac{1}{2z\sqrt{z^{2}+1}}\ln\frac{\sqrt{z^{2}+1}+z}{\sqrt{z^{2}+1}-z}
\label{gzeta}
\eeq
 The structure function is obtained from the cross-section 
by the standard relation
\beq
F_{2}(x,Q^{2})=\frac{Q^2}{\pi e^2}(\sigma^{(T)}+\sigma^{(L)})
\label{strufot}
\eeq

In the DLLA limit only the transverse cross-section contributes. We can 
also neglect the quark masses in this approximation. Then, with a 
substitution (\ref{ansatz}), from (\ref{crosstot}), (\ref{rhotra}) and 
(\ref{strufot}) we obtain an expression for the quark (sea) density of the 
target
\beq
xq(x)=\frac{3}{\pi^{2}b_{0}^{2}}\frac{Q^2}{\ln Q^2}
\int^{Q^2}\frac{dk^2\phi(x,k)}{k^{4}\ln k^2}
\int_{0}^{1}d\alpha(\alpha^{2}+(1-\alpha)^{2})
((1+2z^{2})g(z)-1)
\label{quarkint}
\eeq
where $g(z)$ is given by Eq. (\ref{gzeta}) and we assumed that large values of 
$k^{2}<Q^2$ contribute in accordance with the DLLA approximation.
In this approximation the asymptotics of the gluon density $xg(x,k^{2})$ 
and consequently of $\phi(x,k^{2})$ is known:
\beq
\phi(x,k^{2})=cxg(x,k^{2})\simeq c
\exp \sqrt{a\ln\frac{1}{x}\ln\ln k^{2}}
\label{dlgluo}
\eeq
where $a=48/b_{0}$. Putting (\ref{dlgluo}) into (\ref{quarkint}), after 
simple calculations we find the asymptotical expression for the quark density
\beq
xq(x,k^{2})\simeq\frac{4c}{\pi^{2}b_{0}^{2}}\sqrt{\frac{\ln\ln k^2}
{a\ln 1/x}}\exp \sqrt{a\ln\frac{1}{x}\ln\ln k^{2}}
\label{ourdist}
\eeq
On the other hand, let us consider the DGLAP equation in the small $x$ limit
(in such a case it is possible to neglect the quark contributions to the sea 
quarks)
\beq
\frac{\partial}{\partial \ln Q^2} q(x,Q^2)=
\frac{\alpha_s(Q^2)}{8\pi} \int_x^1 \frac{dy}{y} P_{qg}(\frac{x}{y})g(y,Q^2)
\quad ; \quad P_qg(z)=2[z^2+(1-z)^2]
\eeq
After a double integration we find
\beq
xq(x,k^{2})\simeq\frac{4}{3b_{0}^{2}}\sqrt{\frac{\ln\ln k^2}
{a\ln 1/x}}\exp \sqrt{a\ln\frac{1}{x}\ln\ln k^{2}}
\label{dglapdist}
\eeq
As we observe the approximation (\ref{ansatz}) for the colour density of the 
photon projectile leads to the correct relation between the quark and gluon 
densities in the DLLA limit. This justifies the use of (\ref{ansatz}), at 
least for high enough $1/x$ and $Q^{2}$. Comparing (\ref{ourdist}) and 
(\ref{dglapdist}) we also obtain the normalization factor $c$ which relates 
the pomeron wave function to the gluon density
\beq
c=\pi^{2}b_{0}/3
\eeq

%%%%%%%%%%%%%%%%%%%%%%%%%%%%%%%%%%%%%%%%%%%%%%%%%%%%%%%%%%%%%%%%%%%%%%%%%
\subsection{The initial distribution}
To start the evolution in  $1/x$ we have to fix the initial gluon 
density at some small value $x=x_{0}$. Evidently, the smaller
is $x_{0}$, the smaller is the region
where we can compare our predictions with the experimental data.
On the other hand, if $x_{0}$ is not small enough, application of the
asymptotic hard pomeron theory becomes questionable. Guided by these
considerations we choose $x_{0}=0.01$ as our basic initial $x$
although we also tried $x=0.001$ to see the influence of possible 
sub-asymptotic effects.

The initial wave function $\phi(x_{0},k^{2})$ has to be chosen in 
accordance with the existing data at $x=x_{0}$ and all $k^2$ available.
The experimental $F_{2}$ is a sum of the singlet and non-singlet parts, 
the latter giving a relatively small contribution at $x=0.01$. Our 
theory can give predictions only for the singlet part (and one of the 
criteria for its applicability is precisely the relative smallness of 
the non-singlet contribution). The existing experimental data at $x=0.01$
give values for $F_{2}$ averaged over rather large intervals of $x$ 
and $Q^2$. For all these reasons, rather than to try to adjust our 
initial $\phi(x_{0},k^{2})$ to the pure experimental data, we have 
preferred to match it with the theoretical predictions for the 
gluon density  and the singlet 
part of $F_{2}$ given by some standard parametrization fitted to the 
observed $F_{2}$ in a wide interval of $Q^2$ and small $x$. As such we 
have taken the GRV LO parametrization \cite{grv}. The choice of LO has been 
dictated by its comparative simplicity and the fact that at $x=0.01$ the 
difference between LO and NLO is insignificant.

Thus, for the initial distribution we have taken  the GRV
LO gluon density with an appropriate scaling factor. Putting this
density into eqs. (\ref{phiglu}),(\ref{crosstot}) and (\ref{strufot}) one 
should be able to reproduce the sea quark density and thus the singlet part 
of the structure function.
In the GRV scheme the relation between the gluon density and the quark 
density is much more complicated and realized through the DGLAP 
evolution. Since the DGLAP evolution and the pomeron theory are not 
identical, one should not expect that our initial gluon density 
should exactly coincide with the GRV one to give the same singlet
structure function. One has also to have 
in mind the approximate character of our colour density $\rho$ at small
$Q^2$. In fact, with the initial $\phi$ given by (\ref{phiglu}) and the
gluon density exactly taken from the GRV parametrization at $x=0.01$
we obtain a 30\% smaller values for the singlet part of the structure 
function as given by the same GRV parametrization, the difference 
growing at low $Q^2$.
To make the description better we used a certain arbitrariness in the scale 
$Q_{1}^{2}$ which enters (\ref{ansatz}) and also the scale at which the 
coupling freezes in the density $\rho$.
A good choice to fit the low $Q^2$ data is to take
\beq
\alpha (Q_{1}^{2})=\frac{4\pi}{b_0} 
\frac{1}{\ln ((0.17*Q^{2}+0.055\ (GeV/c)^{2})/ \Lambda^{2})}
\eeq
With this $\alpha(Q_{1}^{2})$ the obtained singlet structure function at 
$x=0.01$ has practically the same $Q^{2}$ dependence as the GRV one,
although it results 30\% smaller in magnitude. 
This mismatch can be interpreted in two different ways. Either we may 
believe that the gluon density given by the GRV is the correct one and
the deficiency in the singlet part of the structure function is caused
by our approximate form of the colour density $\rho$ (which is most 
probable). Or we may think that the colour density to be used in the
DGLAP should coincide with ours only for large enough $Q^2$ and $1/x$
and at finite values they may somewhat differ (our relation 
(\ref{phiglu}) was established strictly speaking only in the DLLA limit).
Correspondingly we may either take the relation (\ref{phiglu}) as it stands 
and use the GRV LO gluon density at $x=0.01$ in it, or introduce a correcting 
scaling factor 1.3 which brings the structure function calculated with 
the help of (\ref{crosstot})-(\ref{strufot}) into  agreement with the GRV 
predictions.
In the following we adopt the second alternative, that is we assume that 
our initial gluon distribution at $x=0.01$ is 30\% higher that the
one given by the GRV parametrization.
Of course to strictly compare the rate of growth of the distribution under
evolution one should take into account the differences at the starting point. 

\begin{figure}
\centering
\includegraphics[angle=-90,width=3.0in]{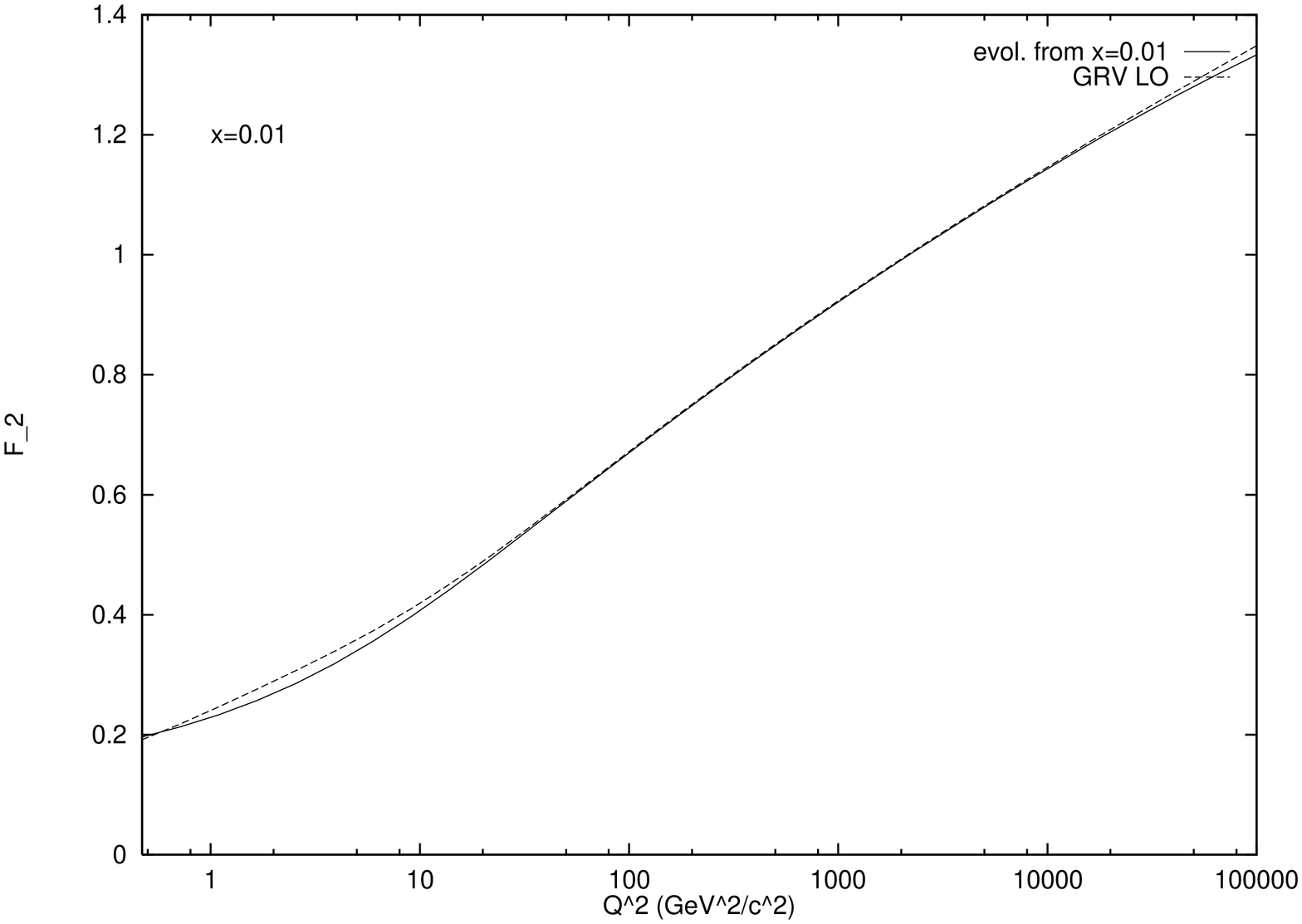}\\
\captionm{The singlet part of the structure function of the proton at $x=0.01$.
The continuous line is the result of our calculation while the dashed line
correspond to the GRV prediction.}
\label{figinit}
\end{figure}

The singlet part of the structure function at $x=0.01$
calculated from (\ref{crosstot})-(\ref{strufot}) with this choice is shown in 
Fig. \ref{figinit} together with the GRV predictions. 
However one can easily pass to the first alternative by simply reducing our 
results by factor 1.3.

%%%%%%%%%%%%%%%%%%%%%%%%%%%%%%%%%%%%%%%%%%%%%%%%%%%%%%%%%%%%%%%%%%
\subsection{Numerical results}
With the initial wave function $\phi(x=0.01,k^{2})$ chosen as indicated 
in the preceding section we solved the evolution equation for 
$10^{-8}<x<10^{-2}$. 

\begin{figure}
\centering
\includegraphics[width=6.0in]{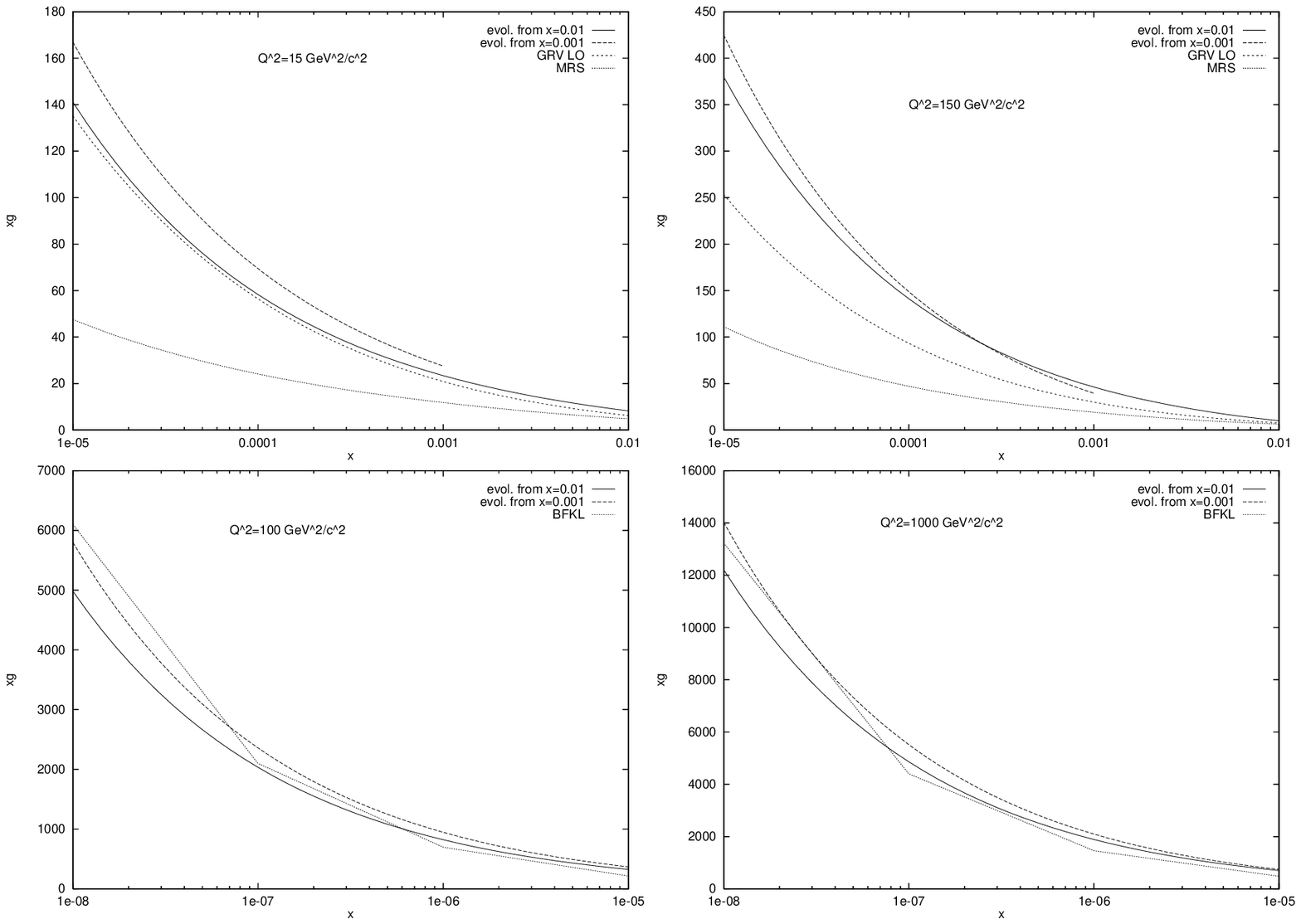}
\captionm{The gluon distributions as a function of $x$ evolved from
$x=0.01$ and $x=0.001$ for the experimentally accessible kinematical
region and for asymptotically high values of $Q^2$ and $1/x$.
Standard DGLAP evolved parametrizations (GRV-LO and
MRS) and the BFKL evolved distributions from \cite{kms} (we report only few 
points connected by lines) are shown for comparison. For the latter
distribution one should take into account a further factor which actually has 
been chosen in \cite{kms} to reduce the initial distribution (see text).}
\label{figevglu}
\end{figure}

The adopted computational scheme consists in diagonalizing 
the Hamiltonian in (\ref{shorteq}), reduced to one dimension in the 
transverse momentum space after angular averaging, and represent the initial 
wave function as a superposition of its eigenvectors.
To discretize $k^2$ a grid was introduced, after which the problem is reduced 
to a standard matrix one.
To check the validity of the obtained results we have also repeated the
evolution using a Runge-Kutta method, resulting in a very good agreement.
The final results obtained for the gluon distribution $xg(x,Q^{2})$ as 
a function of $x$ for various $Q^2$ are shown in Fig. \ref{figevglu} where the
first two plots correspond to $x$ and $Q^2$ presently available, 
whereas the last two plots show the behaviour of the calculated gluon 
density in the region up to very small $x$ and very high $Q^2$,
well beyond the present possibilities.
For comparison we have also shown the gluon densities for the 
GRV LO parametrization \cite{grv}, for the MRS parametrization \cite{mrs} 
and also for the BFKL evolution as calculated in \cite{kms}.

\begin{figure}
\centering
\includegraphics[width=6.0in]{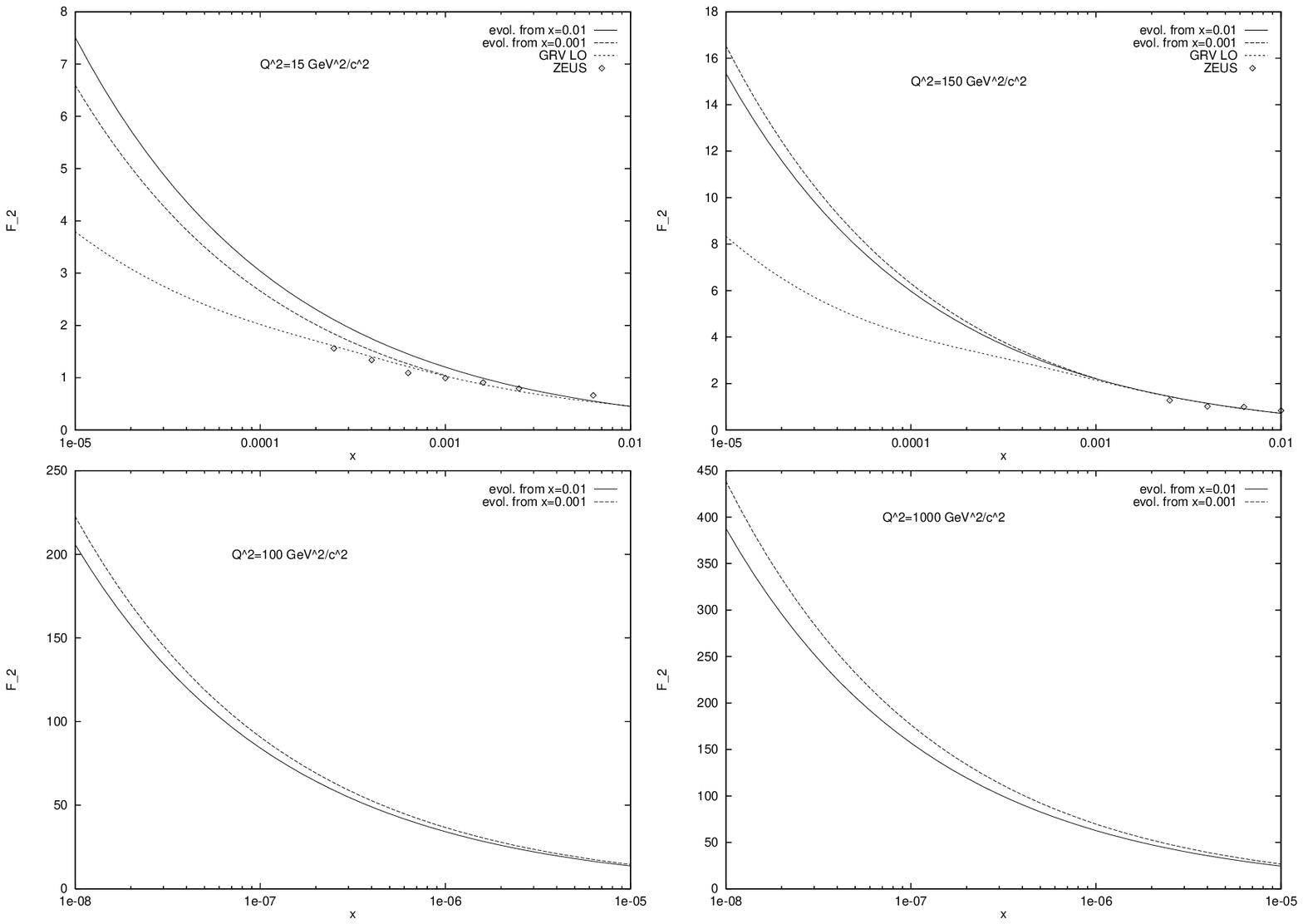}\\
\captionm{$x$ dependence of the singlet part of the proton structure function
obtained by evolution from
$x=0.01$ and $x=0.001$, compared to the GRV prediction and the ZEUS 94 data.
Values for asymptotically high values of $Q^2$ and $1/x$ are also shown.}
\label{figevstr}
\end{figure}

Putting the found gluon densities into eqs. (\ref{crosstot})-(\ref{strufot})
we obtain the (singlet part of) proton structure function $F_{2}(x,Q^{2})$.
The results are illustrated in Fig. \ref{figevstr} for the $x$ dependence. 
As for the gluon densities, the experimentally
investigated region is shown in the first two plots, in which the
existing experimental data from \cite{ZEUS2} are also presented.

Finally, to see a possible influence of sub-asymptotic effects, we
have repeated the procedure taking as a starting point for the evolution 
a lower value $x=0.001$. The resulting gluon distributions and structure 
functions are also presented in the above figures.

To discuss the results obtained we have to remember that they involve 
two quantities of a different theoretical status. One is the pomeron 
wave function $\phi$ which can be identified with the gluon distribution 
(up to a factor). The other is the quark density (which is equivalent to the 
structure function), for which we actually have no rule for the introduction 
of a running constant and which in the present calculation involves a 
semi-phenomenological ansatz (\ref{ansatz}). 
Evidently the results for the latter are much less informative as to the 
effect of the running coupling introduced in our way.
Therefore we have to separately discuss our prediction for the gluon 
distribution, on the one hand, and for the structure function, on the other.

Let us begin with the gluon distribution. Comparing our results with
those of GRV, which correspond to the standard DGLAP evolution,
we observe that at high enough $Q^2$ and low enough $x$ our distributions
rise with $Q$ and $1/x$ faster than those of GRV. This difference is,
of course, to be expected. The hard pomeron theory in any version
predicts a power rise of the distribution with $1/x$ to be compared with
(\ref{dlgluo}) for the DGLAP evolution. As to the $Q$-dependence, the 
fixed coupling (BFKL) hard pomeron model predicts a linear rise, again much 
stronger than (\ref{dlgluo}). Our running coupling model supposedly leads
to a somewhat weaker rise. From our results it follows that it is still
much stronger than for the DGLAP evolution. However one can observe that
these features of our evolution become clearly visible only at quite
high $Q$ and $1/x$. For moderate $Q<10\, GeV/c$ and/or $x>10^{-4}$
the difference between our distributions and those of GRV is 
insignificant. As to the DGLAP evolved MRS parametrization, it
gives the gluon distribution which lies systematically below the
GRV one and, correspondingly, below our values, the difference
growing with $Q$ and $1/x$.

To compare our gluon evolution dynamics with the BFKL one, as presented in 
\cite{kms}, we have to remark that we should take a huge correcting factor
that depends in $Q^2$ for the following reasons: we must remember of the 
factor 1.3 discussed in section 4 and we should note that the starting
values used in \cite{kms} for the gluon distribution are lower than the
GRV values, by a factor $1.92$ at $Q^2=4\ GeV^2/c^2$ and by a factor $1.1$
at $Q^2=1000\ GeV^2/c^2$.
According to this to compare just the evolution in this two models we should
multiply our distribution by a factor that is $0.4$ at $Q^2=4\ GeV^2/c^2$ and
that is $0.7$ at $Q^2=1000\ GeV^2/c^2$.
Taking into account this factors looking at Fig. \ref{figevglu} we see that 
there is a good agreement for not too small $x$ values between the two 
evolutions, while looking at the values at very small $x$ in all the $Q^2$ 
interval our model predicts a quite slower growth of the gluon distribution.
We note also that the spectral method we have used shows how the dominant 
components of the spectrum contribute to the evolution of the gluon density;
some small changes of the initial gluon distribution may results in different
projections onto the spectral basis, but cannot lead to a very different 
behaviour at very small $x$, even if, as it is, the initial gluon distribution
is well known in a restricted $Q^2$ interval.
This is because the first two dominant states are precisely giving an 
important contribution in this interval.

Passing to the structure functions we observe in Fig. \ref{figevstr} that our 
results give a somewhat too rapid growth with $1/x$ in the region 
$10^{-3}<x<10^{-2}$ as compared to the experimental 
data (and also to the parametrizations GRV fitted to these data). 
With the scaling factor 1.3 introduced to fit the data at 
$x=0.01$ we overshoot the data at $x<10^{-3}$ by $\sim$25\%. Without this 
factor we get a very good agreement for $x<10^{-3}$ but are below 
experiment at $x=0.01$ by the same order. This discrepancy may be 
attributed either to sub-asymptotic effects or to a poor quality of our 
ansatz (\ref{ansatz}). Comparison with the result obtained with a lower 
starting point for the evolution $x=0.001$ shows that sub-asymptotic effects
together with a correct form of coupling to quarks
may be the final answer.

% cap3

\chapter{Unitarization in the large $N_c$ limit: the three pomeron vertex}
In the first chapter we have reviewed and analyzed some aspect and properties
of the BFKL approximation to the pomeron which is based on the application
of the perturbation theory to QCD in the Regge kinematical region 
and corresponds to small $x_B$ in processes such as DIS.
Specifically the BFKL pomeron is based on the resummation of the leading 
large logarithms in $1/x$ and presents the important feature of a steep 
increase of the cross sections as $x$ decreases, with a power like behavior
in $1/x$. This is, as we already noted in the first chapter, in contrast
with the bound set by unitarity. In fact this is the well known 
Froissart bound \cite{froiss}, which sets the limit to the increase behavior 
with energy of the total hadronic cross section to $\ln^2 s$ and which, 
for DIS, would lead to the growth bound of $\ln^2 1/x$ \cite{butchm,ayala}.
Taking the double logarithmic limit,
where one resums powers of $(\ln 1/x  \ln Q^2)$ and which
thus come out also from the small $x$ limit of the DGLAP approach, one has
a somewhat less dramatic increase, in accord to an $\exp{(\sqrt{1/x})}$
behavior, but still unitarity is strongly violated. 

Therefore, to extend the kinematical region of validity of the approach 
one has to go beyond the leading logarithmic approximation. This
is in principle a formidable task and could also not be very
satisfactory in presence of hadrons which cannot be studied only with 
perturbative methods.
Here we shall be interested only in a perturbative analysis, which 
nevertheless is better established for the study of processes like
onium-onium scattering,  and one hopes that any result could give some new 
insight to the complete theory which surely would need of a non-perturbative
treatment. 

A systematic approach to take into account a minimal set of corrections
necessary to restore the unitarity has been studied by Bartels \cite{bart2}.
The higher order amplitudes, which contain any number of (reggeized) gluons 
in the $t$-channel, are constructed using unitarity and dispersion relations
starting from the lower order ones and one has a coupled system of equations
for the $n$ reggeized gluon state amplitudes.
The main feature is the presence of
transitions vertices which change the number of reggeon in the $t$-channel
and thus leads to a reggeon field theory.

The study of this problem is far from being complete. The first step has been
to consider the coupled equations for the amplitudes relative to $2$, $3$ and
$4$ gluons in the $t$-channel \cite{bartwue}.
The most interesting object which appears is an effective vertex for the 
transition between $2$ and $4$ gluon in the $t$-channel. This vertex has been 
found to be conformal invariant \cite{baliwu} and in the appropriate color 
subspace is related to the triple pomeron vertex. 
Another element of some interest is the $4$-gluon interacting state (and its
generalizations to $n$ gluons) satisfying the so called BKP equation 
\cite{bart1,bart3}.
This amplitude, characterized by a constant number of gluons in the 
$t$-channel, has been studied in the large $N_c$ limit and the system has 
been proved to be complete integrable and equivalent to an XXX Heisenberg 
zero spin model \cite{lip2,fadkorch,tatafa}.   

The BFKL amplitudes and the $2\to4$ effective reggeon vertex have been
interpreted in terms of correlation functions of a $2$D conformal field theory
\cite{pertqcd,korch97} hoping to obtain in the future such a 
formulation for the effective QCD in the Regge limit.
In this context some work \cite{lotter} has been done to understand the 
spectrum 
of the $4$ gluon state in terms of the scaling dimension of some operators 
appearing in the twist expansion of the amplitude in the short distance limit.
The idea is based on the fact that, due to the strong symmetry of a 
bidimensional conformal field theory, all the informations are contained in 
the scaling dimensions of the operators present in the OPE 
(operator product expansion) and in the expansion coefficients.

We shall be interested here in the study of the coupled system of equations
up to $4$ reggeized gluons in the $t$-channel in the large $N_c$ limit
\cite{braun4, bv4}.
The analysis will be performed at the leading and next to leading order (NLO) 
in $1/N_c$ since it is at the latter level that the double pomeron amplitude 
will appear.
A link to some results coming from the dipole colour approach let arise
the possibility of a general constructions of the $n$ pomeron amplitudes
in the large $N_c$ limit. In order to better understand this one should
go beyond the $4$ gluon system. Already the $6$ gluon system in the large 
$N_c$ limit is capable to show if some equivalence with the A.H. Mueller 
approach to dipoles is maintained, since it seems that the latter leads to a 
system of fan diagrams with only a triple pomeron interactions.
 
We shall start introducing the formal general system of equations for the
$4$ gluons system and its main properties.
Written the system for the large $N_c$ limit case the solutions in the
leading order are given and at the NLO in $1/N_c$ system studied.
The double pomeron amplitude at NLO is found to separate in a direct
double pomeron exchange (DPE) contribution and in a triple pomeron 
interaction (TPI) term. The TPI term can be written in terms of 3 point 
function building blocks which are shown to be conformally invariant so 
that the same property, as already known, has to be shared by the triple 
pomeron vertex.

It is later shown in detail the interesting fact that the DPE contribution 
can be substituted by a completely equivalent TPI term (the inverse is not 
true). The absence of DPE terms was already noted for $N_c=3$ in 
\cite{bartwue} and, in fact, in this form
the final vertex interactions left essentially coincides in the two 
approaches, provided one takes the limit $N_c\rightarrow\infty$.
Coupled to pomerons, this vertex effectively reduces to the one found 
by R. Peschanski in \cite{pesh}, in a colour dipole framework.
In particular this fact has been contemporaneously also shown by
Korchemsky \cite{korch97}. However this
does not mean that the double dipole density in the dipole approach
coincides with the diffractive amplitude in the $s$-channel unitarity
approach: there are certain terms in the latter which are missing in the
double dipole density and an asymmetry factor which distinguish, in the
rapidity evolution, between the original pomeron and the two which 
originate from the split.

Finally a section is devoted to study some properties of higher order dipole
densities in the A.H. Mueller approach which are found to be represented by
a set of pomeron fan diagrams with only a triple pomeron coupling.

%%%%%%%%%%%%%%%%%%%%%%%%%%%%%%%%%%%%%%%%%%%%%%%%%%%%%%%%%%%%%%%%%%%%%%
\section{General unitarization approach}
A program devoted to considering a minimal subset of next to leading 
logarithmic contributions necessary to construct a unitary amplitude
has been carried out in J. Bartels work \cite{bart2}.
The basic object under consideration is the amplitude $D_n$ for the coupling
of $n$ reggeons (gluons) to two external elementary particles in a colour
singlet state. These amplitudes
result from a multiple discontinuity analysis from which emerges the feature
that the number of reggeized gluons in the $t$-channel is not conserved.
This fact can be also analyzed in the physical $t$-channel region ($t>0$) 
looking for the discontinuities along the $t$-cut associated with the threshold
for the reggeon production.
One nice feature due to the gluon reggeization property is the absence
of contributions, in the adjoint colour representation, to the 
partial wave amplitude cut related to $n$-particle production with $n >2$.
 
Let us write the formal system of equations in cascade for the $D_n$ 
amplitudes (seen as vectors in the appropriate colour spaces) up to $n=4$: 
\bea
\Bigl[ j-1- \sum_{i=1}^2 \omega(k_i) \Bigr] D_2(\{k_i\}) &=&
            D_{20}(\{k_i\})+[K_{2\to2}\otimes D_2](\{k_i\}) \nonumber \\
\Bigl[ j-1- \sum_{i=1}^3 \omega(k_i) \Bigr] D_3(\{k_i\}) &=& D_{30}(\{k_i\}) +
   [K_{2\to3}\otimes D_2](\{k_i\}) + \sum [K_{2\to2}\otimes D_3](\{k_i\}) 
\nonumber \\
\Bigl[ j-1- \sum_{i=1}^4 \omega(k_i) \Bigr] D_4(\{k_i\}) &=& D_{40}(\{k_i\}) +
   [K_{2\to4}\otimes D_2](\{k_i\}) + \sum [K_{2\to3}\otimes D_3](\{k_i\}) +
\nonumber \\
     && \sum [K_{2\to2}\otimes D_4](\{k_i\})
\label{unibart}
\eea
In the above equations $D_{n0}$ is the lowest order term with the gluons
coupled directly to the $q \bar q$ loop. $K_{2\to2}$ is the BFKL interaction
kernel with the gluon propagators included. Remembering the expression
for its momentum dependence
\beq
V(k_1,k_2;k'_1,k'_2)=\frac{k_{1}^{2}{k'_{2}}^{2}+k_{2}^{2}{k'_{1}}^{2}}
{{k'_{1}}^{2}{k'_{2}}^{2}(k_{1}-k'_{1})^{2}}-\frac{(k_{1}+k_{2})^{2}}
{{k'_{1}}^{2}{k'_{2}}^{2}} 
\label{vert22}
\eeq 
it is possible to give the expression for the kernels $K_{2\to3}$ and 
$K_{2\to4}$ in a simple way. In fact, defining
\beq
W(k_1,k_2,k_3;k'_1,k'_3)= V(k_2,k_3;k'_1-k_1,k'_3)-V(k_1+k_2,k_3;k'_1,k'_3) 
\label{gluo23}
\eeq
one has, on including the colour structure,
\bea
K_{2\to3}(k_1,k_2,k_3;k'_1,k'_3)&=&g^3 
f^{a'_{1}a_{1}c}f^{ca_{2}d}f^{da_{3}a'_{3}} W(k_1,k_2,k_3;k'_1,k'_3) 
\nonumber \\
K_{2\to4}(k_1,k_2,k_3,k_4;k'_1,k'_4)&=&g^4 
f^{a'_{1}a_{1}c}f^{ca_{2}d}f^{da_{3}e} f^{ea_{4}a'_{4}}
W(k_1,k_2+k_3,k_4;k'_1,k'_4)
\label{vertices}
\eea
Moreover in the coupled system of equations (\ref{unibart}) the sums run 
over all possible pairwise interactions and the symbol $\otimes$ means an
integration over primed variables with a weight $(2\pi)^{-3}\delta^{(2)}
(\sum k-\sum k')$.

The analysis of the colour structure for the system of $4$ gluons was
performed by Bartels in \cite{bart2}, separating the gluons in two pairs, 
each in the same colour quantum state belonging to the set ( $1$, $8_A$, $8_S$,
$\bar{10}+10$, $27$). Therefore the amplitude has been characterized by
a 5-dimensional colour vector for the three different signature
assignments $(-,-)$, $(+,+)$, $(+,-)$
and it has been possible to reduce the general operator $K_{4 \to 4}$, 
which should appear in principle in the last term of the equation for $D_4$,
in a sum of $K_{2 \to 2}$ operators acting on different colour subspaces.

%%%%%%%%%%%%%%%%%%%%%%%%%%%%%%%%%%%%%%%%%%%%%%%%%%%%%%%%%%%%%%%%%%%%%%%%
\section{Large $N_c$ limit for the system of four reggeized gluons}
The set of equations (\ref{unibart}) for the system of four reggeized 
gluons, coupled to a fermion loop in a colour singlet state, found by 
J. Bartels and M. Wuesthoff \cite{bartwue} for the real case $N_c=3$ is 
quite complicated because of the colour structure.

The large $N_c$ limit \cite{venez,thooft} is interesting because in the 
leading order the amplitude for the system is based on a single BFKL pomeron 
exchange, i.e.  one has a fully reggeizing contribution. 
Performing a perturbative analysis at the NLO in $1/N_c$ one finds a 
subleading diffractive amplitude which consists of a sum of a DPE and a 
TPI terms, in full correspondence with the Regge-Gribov picture.  

The results presented in this section where studied in \cite{braun4} and 
\cite{bv4} respectively 
for the forward and for the more general non-forward direction case which 
is necessary to study also the conformal invariance properties of the 
$2\to4$ effective gluon vertex.

%%%%%%%%%%%%%%%%%%%%%%%%%%%%%%%%%%
\subsection{Leading order in $1/N_c$}
The results for a system of two and three gluons are not sensitive to the
$N_c\to\infty$ limit, which instead affects the four gluon case.

Let us first consider the two gluon system.
In the lowest (zero) order approximation the 2 gluon amplitude
discontinuity (Mellin transformed to the complex angular
momentum $ j$) is given by the $ q\bar q $ loop with the two gluons
attached to it in all possible ways (4 diagrams in all). Their colour
indices $ a_{1} $ and $ a_{2} $ enter into the colour trace 
$Tr\{t_{a_{1}}t_{a_{2}}\}=(1/2)\delta_{a_{1}a_{2}} $, where 
$t_{a} $ is the colour of the quark. Separating this trace and the
coupling $ g^{2} $ and projecting on the vacuum colour state
$|0\rangle=(1/N)\delta_{a_{1}a_{2}}$ 
we write the zero order contribution in the large $N_c$ limit as 
\beq
D_{20}(1,2)=D_{20}(2,1)=g^2N_c\left(f(1+2,0)-f(1,2)\right)
\eeq                                                                       
where we use the notation in which only the
number of the gluon is indicated whose momentum enters as a variable and
$f(1,2)=f(2,1)$ is a contribution of the $q\bar q$ loop with gluon 1 attached
to $q$ and gluon 2 attached to $\bar q$. Its explicit form can be easily found
(see Appendix A.2) but has no importance for the following.

The basic quantity is the full amplitude $D_2$. It corresponds to the exchange
of two reggeized gluons (the BFKL amplitude) and satisfies the BFKL equation
\beq
S_{20}D_{2}=D_{20}+g^{2}N_cV_{12}D_{2}
\label{eq2glu}
\eeq
where
$ S_{20} $ is the 2 gluon  "free"  Schr\"odinger operator for the energy
$ 1-j $
\beq
S_{20}=j-1-\omega(1)-\omega(2)
\eeq
$ \omega(k) $ is, as usual, the gluon Regge trajectory,
$ V_{12} $ is the BFKL interaction and the previous found $D_{20}$ acts
as an inhomogeneous term. For simplicity we leave here the $\otimes$ symbol
understood.  
 
For the number of exchanged gluons $n=3,4$ amplitudes $D_n(j)$ are defined
as integrals of $D_n(j_1,...j_{n-1})$, depending on $n-1$ partial $t$-channel
angular momenta, over all $j_i$ subject to condition $\sum_{i=1}^{n-1}(j_i-1)=
j-1$ corresponding to conservation of "reggeon energies".
Moreover we extend the notation for the $n$ reggeized gluon free Schr\"odinger
operator defining $S_{n0}=j-1-\sum_i \omega(i)$.

The three gluon system is described by an equation which can be solved
in terms of the $D_2$ amplitudes thanks to the gluon reggeization property.
It is worth to study its colour structure. The colour factor associated to the
lowest order amplitude of three gluons attached to the quark loop is given by
the trace $ Tr\{t_{a_{1}}t_{a_{2}}t_{a_{3}}\}=(1/4)h_{a_{1}a_{2}a_{3}}$,
where $h_{a_{1}a_{2}a_{3}}=d_{a_{1}a_{2}a_{3}}+if_{a_{1}a_{2}a_{3}}$ is 
cyclic symmetric and satisfies the properties
\beq   
h^{*}_{a_{1}a_{2}a_{3}}=h_{a_{2}a_{1}a_{3}} \quad , \quad
\sum_{cd}h^{*}_{acd}h_{bcd} =\delta_{ab} 2N_c(1-2/N_c^{2}) \quad , \quad
\sum_{cd}h_{acd}h_{bcd} =-\delta_{ab}(4/N_c) 
\eeq
One may introduce the two colour wave functions 
\beq
|123\rangle=\frac{1}{\sqrt{2N_c^{3}}}h_{a_{1}a_{2}a_{3}},\  \
  |213\rangle=\frac{1}{\sqrt{2N_c^{3}}}h_{a_{2}a_{1}a_{3}}
\eeq     
orthonormal in the large $N_c$ limit and project the above lowest order 
amplitude on to these states. The result can be expressed in terms of the 
function $f$ because the momenta of the gluons attached to the same fermion 
line sum together; there is also an overall sign depending on the parity
of the number of gluons attached to the quark or to the antiquark lines.
One has, for $N_c \to \infty$,
\beq
D_{30}^{(123)}=-D_{30}^{(213)}=g\sqrt{\frac{N_c}{8}}
\Bigl[ D_{20}(2,1+3)-D_{20}(1,2+3)-D_{20}(3,1+2) \Bigr]
\eeq     
Due to the global colour singlet state, each pair is in an adjoint 
representation and the interaction cannot change the colour configuration.
Remembering the bootstrap property each two reggeon state in an adjoint
representation gives one reggeon state so that the considered three gluon 
amplitude must reduce to a superposition of BFKL pomeron amplitudes.

In more detail one can write the equation for the colour state 
$|123\rangle$
\beq
S_{30}D_{3}^{(123)}=D_{30}^{(123)}+D_{2\rightarrow 3}^{(123)}+
\frac{g^{2}N_c}{2}(V_{12}+V_{23}+V_{31})D_{3}
\label{eq3glu}
\eeq       
The second inhomogeneous term $D_{2\rightarrow 3}^{(123)}$ is related to the
contribution which arises for the transition from 2 to 3 gluons.
In fact, using the vertex $ K_{2\rightarrow 3} $ recalled in the previous 
section and projecting onto the state $|123\rangle$ one finds
\beq
D_{2\rightarrow 3}^{(123)}=g^{3}\sqrt{\frac{N_c^{3}}{8}}W(1,2,3;1'3')\otimes
D_{2}(1',3')\equiv g^{3}\sqrt{\frac{N_c^{3}}{8}}W_{2}(1,2,3)
\label{w2def}
\eeq
In the equation for the orthogonal state $|213\rangle$ the inhomogeneous terms
have opposite sign.
As expected the solutions of the equations can simply be obtained promoting
the $D_{20}$ in the zero order term to a full BFKL pomeron amplitudes $D_2$,
i.e.
\beq 
D_{3}^{(123)}=-D_{3}^{(213)}=g\sqrt{\frac{N_c}{8}}
\Bigl[ D_{2}(2,1+3)-D_{2}(1,2+3)-D_{2}(3,1+2) \Bigr]
\label{sol3glu}
\eeq 
This can be easily proved (see appendix A.4) directly using (\ref{gluo23}) to 
transform $D_{2\rightarrow 3}^{(123)}$ and by applying the bootstrap condition
\beq
\frac{g^2N_c}{2} \int\frac{d^2k'_1}{(2\pi)^3}V(k_1,k_2;k'_1,k'_2)\equiv 
\frac{g^2N_c}{2} V_{12} \otimes 1=
\omega(1)+\omega(2)-\omega(1+2)
\eeq
to obtain contributions coming from  BFKL equations for amplitudes where 
two reggeized gluons have fused into one.

Let us now consider the four gluon system. 
In the leading order the colour configuration is
much more conveniently described by the order of gluons along
the cylinder surface 1234, 1324, etc.
If one takes the lowest order 
amplitude given by the contributions of 16 diagrams with the 4 gluons attached
in all possible ways to the $q \bar q$ loop, one has for each case a color
factor given by the trace 
$Tr\{t_{a_{1}}t_{a_{2}}t_{a_{3}}t_{a_{4}}\}=
(1/8)h_{a_{1}a_{2}b}h_{a_{3}a_{4}b}+(1/4N_c)\delta_{a_{1}a_{2}}
\delta_{a_{3}a_{4}}$. In the $N_c\to \infty$ limit only the first term,
cyclic symmetric, surveys. It is useful to define the colour state
\beq   
|1234\rangle =\frac{1}{2N_c^{2}}h_{a_{1}a_{2}b}h_{a_{3}a_{4}b}
\eeq                                                       
together with others, differing for a permutation, orthonormalized in the large
$N_c$ limit. The zero order amplitude projected on to these states gives
the components
\beq
D_{40}^{(1234)}=D_{40}^{(4321)}=\frac{g^{2}N_c}{4}\Bigl[D_{20}(1,2+3+4)+
D_{20}(4,1+2+3)-D_{20}(1+4,2+3)\Bigr]
\label{gluo401}
\eeq
\beq
D_{40}^{(2134)}=D_{40}^{(4312)}=\frac{g^{2}N_c}{4}\Bigl[D_{20}(2,1+3+4)+
D_{20}(3,1+2+4)-D_{20}(1+2,3+4)-D_{20}(1+3,2+4)\Bigr]
\label{gluo402}
\eeq  
The gluons lie on a cylinder attached to the $q\bar q$ loop and all the
neighbouring pairs are globally in an adjoint representation. 
Moreover in the leading order in $1/N_c$ all the interactions across the 
cylinder are suppressed and only the neighbouring gluons interact.

The equations for the above amplitudes decouple and are given by \cite{braun4}
\bea
S_{40}D_{4}^{(1234)}&=&D_{40}^{(1234)}+D_{2\rightarrow 4}^{(1234)}+
D_{3\rightarrow 4}^{(1234)}+\frac{g^{2}N_c}{2}(V_{12}+V_{23}+V_{34}+V_{41})
D_{4}^{(1234)}
\nonumber \\
S_{40}D_{4}^{(2134)}&=&D_{40}^{(2134)}+
D_{3\rightarrow 4}^{(2134)}+\frac{g^{2}N_c}{2}(V_{21}+V_{13}+V_{34}+V_{42})
D_{4}^{(2134)}
\eea
The inhomogeneous terms related to the transitions which change the number of 
gluons from 2 to 4 are
\beq
D_{2\rightarrow 4}^{(1234)}=-\frac{g^{4}N_c^{2}}{4}W_{2}(1,2+3,4)
\eeq   
while the transitions between 3 and 4 gluons are due to the terms
\bea
D_{3\rightarrow 4}^{(1234)}=g^{3}\sqrt{\frac{N_c^{3}}{8}}
\Bigl[W(2,3,4;2',4')\otimes
D_{3}^{(124)}(1,2',4')+W(1,2,3;1',3')\otimes D_{3}^{(134)}(1',3',4)\Bigr] 
\nonumber \\
D_{3\rightarrow 4}^{(2134)}=
-g^{3}\sqrt{\frac{N_c^{3}}{8}}\Bigl[W(1,2,4;1',4')\otimes
D_{3}^{(134)}(1',3,4')+W(1,3,4;1',4')\otimes D_{3}^{(124)}(1',2,4')\Bigr]
\nonumber \\
\eea            
It can be shown that also the solutions for the 4 gluon amplitudes can be
written in terms of the two reggeon amplitudes just by substituting the 
$D_{20}$ terms in (\ref{gluo401}) and (\ref{gluo402}) by $D_2$ ones.
Thus in the leading order in $1/N_c$ the 4 reggeized gluon system reduces
to a single pomeron.

%%%%%%%%%%%%%%%%%%%%%%%%%%%%%%%%%%%% 
\subsection{Next to leading order in $1/N_c$}
The corrections at the NLO in $1/N_c$ force to consider the two subsystems
of gluons (12) and (34). The state which relates to the diffractive amplitude
and contributes to the triple pomeron interaction is characterized by both
the subsystems in a colour singlet state so that we shall project the
zero order state of the 4 gluons attached to the quark loop onto
$|0\rangle=(1/N^{2})\delta_{a_{1}a_{2}}\delta_{a_{3}a_{4}}$, obtaining
\beq
D_{40}^{(0)}=\frac{1}{2}g^2(\sum_{i=1}^4 D_{20}(i,1+2+3+4-i)-
\sum_{i=2}^4D_{20}(1+i,2+3+4-i))
\label{zerodpe}
\eeq
which is explicitly down by a factor $1/N_c$  respect to the 4 gluons 
amplitudes at leading order (its order is $g^4N_c$ compared to $g^4N_c^2$ of
the amplitudes $D_{40}^{(1234)}$ and $D_{40}^{(2134)}$).       
The full diffractive amplitude $D_4^{(0)}$ (directly related to the 
diffractive cross-section integrated over the diffractive mass) 
results to be the solution of the following equation 
\beq
S_{40}D_{4}^{(0)}=D_{40}^{(0)}+D_{2\rightarrow 4}^{(0)}+D_{3\rightarrow
4}^{(0)}+D_{4\rightarrow 4}^{(0)}+g^{2}N_c (V_{12}+V_{34})D_{4}^{(0)}
\label{eqd40first}
\eeq
where there is also an inhomogeneous term $D_{4\rightarrow 4}^{(0)}$
which comes from a LO 4 gluon cylindric configuration due to the 
interactions between adjacent gluons as well as across the cylinder. 
In the Bartels approach \cite{bart2} the corresponding transitions are 
related to the matrix elements 12 and 13 (which turns out to be zero) of 
the matrix $K^T_{4 \to 4}$.
The contributions to the two colour singlet amplitude at the NLO 
have explicitly been found \cite{braun4}:
\bea 
D_{2\rightarrow4}^{(0)}&=&-g^{4}N_c W_{2}(1,2+3,4) \nonumber \\
D_{3\rightarrow4}^{(0)}&=&g^{3}\sqrt{2N_c} \Bigl[ W(1,2,3;1'3')\otimes
D_{3}^{(134)}(1',3',4)-W(1,2,4;1',4')\otimes D_{3}^{(134)}(1',3,4')+
\nonumber \\
&&W(2,3,4;2',4')\otimes D_{3}^{(124)}(1,2',4)-W(1,3,4;1',4')\otimes
D_{3}^{(124)}(1',2,4')\Bigr] \nonumber \\
D_{4\rightarrow4}^{(0)}&=&g^{2}(V_{23}+V_{14}-V_{13}-V_{24})
\Bigl[D_{4}^{(1234)}-D_{4}^{(2134)}\Bigr]  
\eea                                    
From the inspection of these inhomogeneous terms one can observe that each
one is given by the action of some operator on the BFKL pomeron amplitude
$D_2$. It is then possible to define an operator $Z$ associated to the triple
pomeron vertex and acting on the BFKL pomeron
\beq
Z\, D_2 \equiv D_{2\rightarrow 4}^{(0)}+D_{3\rightarrow
4}^{(0)}+D_{4\rightarrow 4}^{(0)}
\eeq
and rewrite eq. (\ref{eqd40first}) as
\beq
S_{40}D_{4}^{(0)}=D_{40}^{(0)}+Z \, D_2+g^{2}N_c(V_{12}+V_{34})D_{4}^{(0)}
\label{eqd40}
\eeq

\begin{figure}
\centering
\includegraphics[width=4.0in]{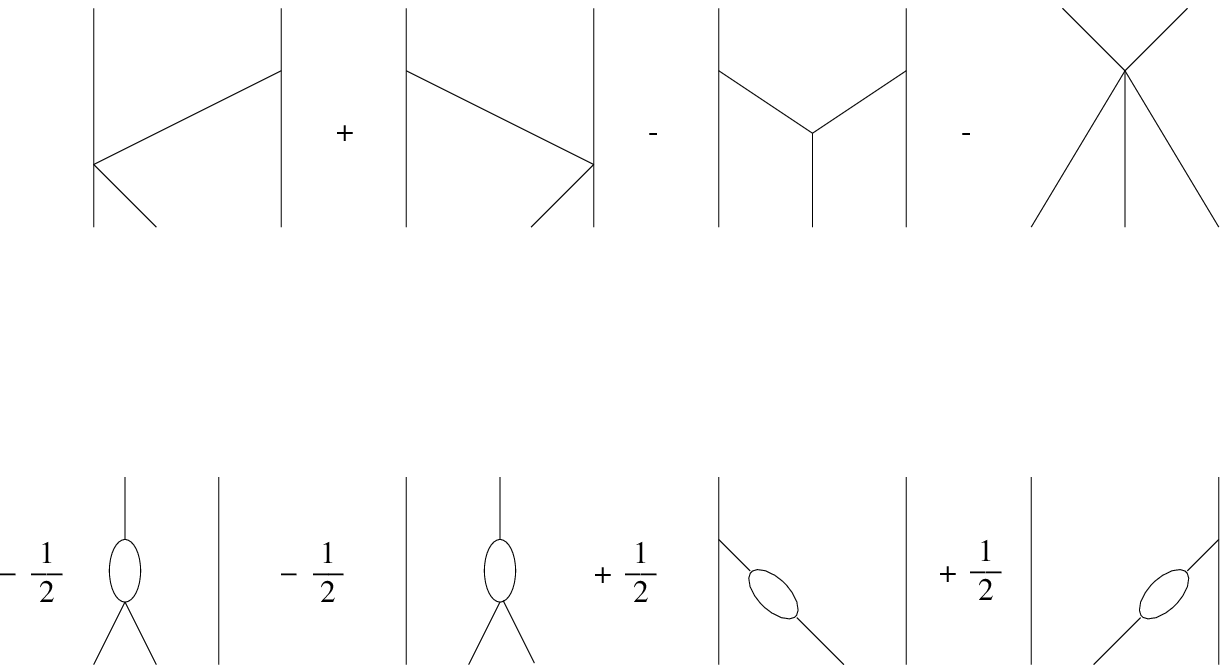}
\captionm{Graphical representation of the conformal invariant function $G$ 
in term of which the vertex $Z$ is constructed: 
for each line one has to write the corresponding propagator, for each vertex 
the square of the sum of the momenta above or below that vertex and the 
propagators of the lower lines are omitted.}
\label{figG}
\end{figure}

The three-pomeron vertex operator $Z$ can be conveniently reexpressed
in terms of a new function $G(k_1,k_2,k_3)$ (see also appendix A.3) which is 
defined as the vertex $K_{2\rightarrow 3}$ for the transition of 2 to 3
gluons, integrated with the BFKL pomeron and regularized in the infrared by 
terms proportional to the gluon trajectory in the same manner as in the total 
BFKL kernel
\bea
G(k_1,k_2,k_3)&=&G(k_3,k_2,k_1)=-g^2 N_c\, W_2(k_1,k_2,k_3)-
D(k_1,k_2+k_3)(\omega(k_2)-\omega(k_2+k_3))-\nonumber \\
&&D(k_1+k_2,k_3)(\omega(k_2)-\omega(k_1+k_2))
\label{Gfunc}  
\eea
It is a generalization of a similar function of two momenta introduced
in \cite{bartwue} for the forward case and in the graphical notation
introduced by Bartels it can be written as illustrated in Fig. \ref{figG}.

If we write $\dtf^{(0)}$ using $G$ we obtain (as usual we denote $k_i$ by 
writing just $i$ for brevity),
\bea
\dtf^{(0)} &=& \frac{g^2}{2} \Bigl[  
2 \ G(2, 3 + 4, 1) + 2 \ G(3, 1 + 2, 4) -
 G(1 + 3, 2, 4) - G(1 + 4, 2, 3) -\nonumber \\
&& G(1 + 4, 3, 2)- G(2 + 4, 3, 1) +\nonumber \\
&&D_2(1, 2 + 3 + 4)\ (2 \omega(3+4) -\omega(3)-\omega(2+3+4)) + \nonumber \\
&&D_2(2, 1 + 3 + 4)\ (2 \omega(3+4) -\omega(3)-\omega(1+3+4)) + \nonumber \\
&&D_2(3, 1 + 2 + 4)\ (2 \omega(1+2) -\omega(2)-\omega(1+2+4)) + \nonumber \\
&&D_2(4, 1+2 + 3)\ (2 \omega(1+2) -\omega(2)-\omega(1+2+3)) + \nonumber \\
&&D_2(1 + 3, 2 + 4)\ (\omega(2+4) -\omega(2)-\omega(3)) + \nonumber \\
&&D_2(1 + 4, 2 + 3)\ (\omega(2+3) -\omega(2)-\omega(3))
\Bigr]
\eea
This expression is not symmetric in gluons (12) nor in gluons (34) and,
since the function 
$ D_{4}^{(0)} $ is symmetric in both gluon pairs, one should 
symmetrize it. The resulting symmetric expression is
\bea
\dtf^{(0)} &=& \frac{g^2}{4} \Bigl[
2 \ G(1,3+4,2)+2 \ G(2,3+4,1)+2 \ G(3,1+2,4)+2 \ G(4,1+2,3) - \nonumber \\
&& G(1+3,2,4)-  G(1+3,4,2) - G(2+3,1,4)-G(2+3,4,1) - \nonumber \\
&& G(1+4,2,3)-G(1+4,3,2)-G(2+4,1,3)-G(2+4,3,1)+ \nonumber \\
&& D_2(1, 2 + 3 + 4) \ (2 \ \omega(2+3+4)-2 \ \omega(2+4)-2 \ \omega(2+3) + 
\omega(3) + \omega(4) )+ \nonumber \\
&& D_2(2, 1 + 3 + 4) \ (2 \ \omega(1+3+4)-2 \ \omega(1+4)-2 \ \omega(1+3) + 
\omega(3) + \omega(4) )+ \nonumber \\
&& D_2(3, 1 + 2 + 4) \ (2 \ \omega(1+2+4)-2 \ \omega(2+4)-2 \ \omega(1+4) + 
\omega(1) + \omega(2) )+ \nonumber \\
&& D_2(4, 1 + 2 + 3) \ (2 \ \omega(1+2+3)-2 \ \omega(1+3)-2 \ \omega(2+3) + 
\omega(1) + \omega(2) )  \nonumber \\
&& D_2(1+3,2+4) \ ( 2 \ \omega(1+3)+2 \ \omega(2+4) 
-\omega(1)-\omega(2)-\omega(3)-\omega(4)) + \nonumber \\
&& D_2(1+4,2+3) \ ( 2 \ \omega(1+4)+2 \ \omega(2+3) 
-\omega(1)-\omega(2)-\omega(3)-\omega(4))
\Bigr]
\eea
The term $ \ddf^{(0)} $ can be rewritten as 
\beq 
\ddf^{(0)}=g^{2}\Bigl[ G(1,2+3,4)+D_{2}(1,2+3+4)(\omega(2+3)-\omega(2+3+4))+
D_{2}(4,1+2+3)(\omega(2+3)-\omega(1+2+3))\Bigr]
\eeq
and, after an analogous symmetrization procedure, it becomes
\bea
\ddf^{(0)}&=&(1/4)g^{2} \Bigl[
G(1,2+3,4)+G(1,2+4,3)+G(2,1+3,4)+G(2,1+4,3)+\nonumber \\
&&D_2(1,2+3+4) (\omega(2+3) + \omega(2+4)-2 \ \omega(2+3+4))+\nonumber \\
&&D_2(2,1+3+4) (\omega(1+3) + \omega(1+4)-2 \ \omega(1+3+4))+\nonumber \\
&&D_2(3,1+2+4) (\omega(1+4) + \omega(2+4)-2 \ \omega(1+2+4))+\nonumber \\
&&D_2(4,1+2+3) (\omega(1+3) + \omega(2+3)-2 \ \omega(1+2+3)) \Bigr]
\eea
We finally come to the term $ \dff^{(0)} $.  The operator 
$ V_{23}+V_{14}-V_{13} -V_{24} $ is antisymmetric under the interchange 
of 1 and 2 and/or 3 and 4. Since the two-pomeron state is symmetric 
under these substitutions, only antisymmetric parts of 
$ D_{4}^{(1234)} $ and 
$ D_{4}^{(2134)} $ give a nonzero contribution. Taking this into 
account we find
\beq  
\dff^{(0)}=\frac{g^{4}N_c}{4} (V_{23}+V_{14}-V_{13} -V_{24} )[D_{2}(1+3,2+4)-
D_{2}(1+4,2+3)]
\eeq
As we have previously noted these terms are due to the 
interaction between different gluons (e.g 
$ V_{13} D_{2}(1+4) $) or within the same gluon (e.g. 
$ V_{13}D_{2}(1+3) $).

The first type gives the contribution
\beq
\dff^{(01)}=\frac{g^{4}N_c}{4} [(V_{13}+V_{24})D_{2}(1+4)+(V_{23}+V_{14})
D_{2}(1+3)]
\eeq
To reduce it to the expression of the same type as before we use the 
identity (see appendix A.4):
\beq
V_{13}D_{2}(1+4)=W_{2}(2+3,1,4)+W_{2}(2,3,1+4)-W_{2}(2,1+3,4)
-W_{2}(2+3,0,1+4)
\label{relaz1}
\eeq
Introducing then functions $ G $ we find for this part
\bea
\dff^{(01)}&=& \frac{g^4 N_c}{4} \Bigl[
G(1,2+3,4)+G(1,2+4,3)+G(2,1+3,4)+G(2,1+4,3) -\nonumber \\
&& G(1,3,2+4)-G(1,4,2+3)-G(2,3,1+4)-G(2,4,1+3)-\nonumber \\
&& G(3,1,2+4)-G(3,2,1+4)-G(4,1,2+3)-G(4,2,1+3)+\nonumber \\
&& 2 \ G(1+3,0,2+4) + 2 \ G(1+4,0,2+3) +\nonumber \\
&&D_2(1,2+3+4) (\omega(2+3) + \omega(2+4)- \omega(3)-\omega(4))+\nonumber \\
&&D_2(2,1+3+4) (\omega(1+3) + \omega(1+4)- \omega(3)-\omega(4))+\nonumber \\
&&D_2(3,1+2+4) (\omega(2+4) + \omega(1+4)- \omega(1)-\omega(2))+\nonumber \\
&&D_2(4,1+2+3) (\omega(1+3) + \omega(2+3)- \omega(3)-\omega(4))-\nonumber \\
&&(D_2(1+3,2+4)+D_2(1+4,2+3))\sum_{i=1}^{4}\omega(i)
\Bigr]
\eea
The second part leads to the contribution
\[
\dff^{(02)}=-\frac{g^{4}N_c}{4} [(V_{13}+V_{24})D_{2}(1+3,2+4)+
\]\beq (V_{23}+V_{14})D_{2}(1+4,2+3)]
\eeq
Applying the bootstrap relation one immediately has
\bea
\dff^{(02)}&=&\frac{g^{2}N_c}{2} \Bigl[D_{2}(1+3,2+4)(\sum_{i=1}^{4}\omega(i)-
\omega(1+4))-\omega(2+3) +\nonumber \\
&&D_{2}(1+3,2+4)(\sum_{i=1}^{4}\omega(i)-\omega(1+3)-\omega(2+4))\Bigr]
\eea
The final three-pomeron vertex is obtained after summing all the contributions,
explicitly $ZD_2=\ddf^{(0)} + \dtf^{(0)}+ \dff^{(01)}+\dff^{(02)}$
and is given by:
\bea
ZD_2 &=& -\frac{g^2}{2} \Bigl[
G(1+3,2,4)+G(1+3,4,2)+G(1,3,2+4)+G(3,1,2+4)+ \nonumber \\
&& G(1+4,2,3)+G(1+4,3,2)+G(1,4,2+3)+G(4,1,2+3)- \nonumber \\
&&2G(1,3+4,2)-2G(3,1+2,4)-G(1,2+3,4)-G(2,1+4,3)-\nonumber \\
&&G(1,2+4,3)-G(2,1+3,4)- G(1+3,0,2+4)-G(1+4,0,2+3) \Bigr]
\label{tripover}
\eea
One may note from the form of the $K_{2\to3}$ vertex that, after multiplying 
by the gluon propagators, the $G$ function vanishes if the first or the third
argument is zero, but the property in not true for the second argument.
However at the global vertex level, after having performed the sum
(\ref{tripover}) of all the contributions, the validity of the Ward 
identities is restored.

Let us end this section looking at the form of the solution of the equation
(\ref{eqd40}). In fact it can be easily solved and, evidently, the solution 
may be constructed as a sum of two terms corresponding to the two parts of 
the inhomogeneous term $D_{40}^{(0)}$ and $ZD_2$:
\beq
D_4^{(0)}=D_4^{DPE}+D_4^{TPI}
\label{gensol}
\eeq
The DPE term $D_4^{DPE}$ comes from the inhomogeneous term $D_{40}^{(0)}$ and
its explicit form can be conveniently written using the
quark loop density in the transverse coordinate space defined by the Fourier
transform (see Appendix A.2 for explicit expressions)
\beq
f(1,2)=\int d^{2}r \rho_l(r)e^{ik_1r}
\eeq
where $l=k_1+k_2$ and $r$ is the dipole transverse dimension.
Thus one has a contribution of the form
\beq
D_{4}^{DPE}=(1/4)g^{4}N_c\int d^{2}r \rho_l(r)D_{4}^{(r)}
\eeq
Here $D_{4}^{(r)}$ is a convolution in the "energy" 
$ 1-j $ of two independent BFKL pomerons
\beq
D_{4}^{(r)}=\int dj_{12} 
dj_{34}\delta(j+1-j_{12}-j_{34})D_{2,j_{12}}^{(r)}(1,2) 
D_{2,j_{34}}^{(r)}(3,4)
\eeq 
and to restore the structure of the inhomogeneous term $D_4^{(0)}$ the
amplitudes must belong to two equations with a proper inhomogeneous part.
$ D_{2,j}^{(r)}(1,2) $ satisfies the equation
\beq
S_{20}D_{2,j}^{(r)}=\prod_{j=1}^{2}(e^{ik_{j}r}-1) 
+g^{2}N_cV_{12}D_{2,j}^{(r)}
\eeq
and a similar relation regards the second pomeron.
In the rapidity space one can directly write this contribution as a product
of two BFKL amplitudes relative to the two colour singlet pairs
\beq
D_4^{(r)}(y)=D_2^{(r,12)}(y) \, D_2^{(r,34)}(y)
\eeq
The part $D_4^{TPI}$ is the TPI contribution.  It can be written
as a convolution in the rapidity space:
\[
D_4^{TPI}(1,2,3,4;Y)=
\]\beq\int_{0}^YG_2(1,2;1'2';Y-y)G_2(3,4,;3'4';Y-y)
\otimes Z(1',2',3',4';1'',2'')\otimes D_2(1'',2'';y)
\eeq
where $G_2$ is the BFKL Green function and the symbols $\otimes$ mean
integrations over intermediate momenta. This equation clearly shows that
$Z$ is just the three-pomeron vertex.

%%%%%%%%%%%%%%%%%%%%%%%%%%%%%%%%%%%%%%%%%%%%%%%%%%%%%%%%%%%%%%%%%%%%%%%%%
\section{Conformal invariance}

In \cite{bartwue} it was introduced the vertex $V(1234)$. 
For a non-zero momentum flow in the $t$-channel it can be written in a way 
similar to (\ref{tripover}),
\[
V(1234)D_2=\frac{1}{2}g^2 \Bigl[ G(1,2+3,4)+G(2,1+3,4)+G(1,2+4,3)+G(2,1+4,3)
\]\beq-
G(1+2,3,4)-G(1+2,4,3)-G(1,2,3+4)-G(2,1,3+4)+G(1+2,0,3+4) \Bigr]
\label{verbart}
\eeq
This vertex is conformal invariant \cite{baliwu} in the following sense.
If one transforms
$ VD_2 $ to the transverse coordinate space and integrates it over the 4 
gluon coordinates with a conformal invariant function, the resulting integral 
is invariant under conformal transformation of gluon coordinates. 
Comparing (\ref{verbart}) and (\ref{tripover}) one immediately see that our
vertex $Z$ can be written as
\beq
Z=V(1324)+V(1423)
\eeq
and thus is conformal invariant. 

 However a stronger conformal property is satisfied:
 not only the combination (\ref{verbart}) of functions $G$ is 
 conformal invariant, but each function $G(1,2,3)$ is conformal invariant 
 by itself. This is related to the fact that the function $G$ defined in
 (\ref{Gfunc}) represents a natural generalization of the 
 BFKL kernel not only in respect to its infrared stability but also in its 
 conformal properties.

 To prove the previous statement one needs a representation of $G$
 in the coordinate space. We give all the technical details in
 appendix A.3 and use here the computed final expressions.
 
 Denote the integral part of $G$, given by the $W_2$ term in (\ref{Gfunc}) 
 as $G_1$ and the remaining terms related to the gluon trajectories as $G_2$.
 In A.3 the following expression is derived: 
 \beq
 G_1(r_1,r_2,r_3)=A_1D_2(r_1,r_3)
 \eeq
 where $A_1$ is an operator in the coordinate space
 \[
 A_1=\frac{g^2N_c}{8\pi^3}\Bigl[
  2\pi\delta^2(r_{23})\partial_3^2(c-\ln r_{13})\partial_3^{-2}+
  2\pi\delta^2(r_{12})\partial_1^2(c-\ln r_{13})\partial_1^{-2}\]\beq
  -2\frac{{\bf r}_{12}{\bf r}_{23}}{r_{12}^2r_{23}^2}-
  2\pi (c-\ln r_{13})(\delta^2(r_{12})+\delta^2(r_{23})) -
  4\pi^2\delta^2(r_{12})\delta^2(r_{23})(\partial_1+\partial_3)^2
\partial_1^{-2} \partial_3^{-2}\Bigr]
\label{a1}
\eeq
 Here $r_{ij}=r_i-r_j$ and $c=\ln (2/m)+\psi(1)$ with $m$ the gluon mass
 acting as an infrared cutoff.

 The transformation of the part $G_2$ requires introduction of an
 ultraviolet cutoff $\epsilon$ due to the presence of the gluon trajectory
 terms. Of course the final results do not depend on $\epsilon$. One obtains
 \beq
 G_2(r_1,r_2,r_3)=A_2D_2(r_1,r_3)
 \eeq
 where $A_2$ is another operator in the coordinate space
\beq
 A_2=-\frac{g^2N_c}{8\pi^3}\Bigl[\frac{1}{r_{23}^2}-2\pi c\delta^2(r_{23})
\Bigr] +\delta^2(r_{23})\omega(-i\partial_3)
     -\frac{g^2N_c}{8\pi^3}\Bigl[\frac{1}{r_{12}^2}-2\pi c\delta^2(r_{12})
\Bigr] +\delta^2(r_{12})\omega(-i\partial_1)
\label{a2}
\eeq
The four terms in $A_2$ correspond to the four respective terms in $G_2$.
We remember that the singular operators $1/r_{12}^2$ and $1/r_{23}^2$ are
in fact defined with the help of $\epsilon$ as
\beq
\frac{1}{r^2}\equiv\frac{1}{r^2+\epsilon^2}+2\pi\delta^2(r)\ln\epsilon,
\ \ \epsilon\rightarrow 0
\label{regul}
\eeq
which is the expression in (\ref{regular}) and does not depend on $\epsilon$.
Summing $A_1$ and $A_2$ the terms containing $\ln m$ cancel, thus the
dependence on the gluon mass desappears and $G(r_1,r_2,r_3)$ is infrared
stable.

We want to check the conformal invariance of the integral
\beq
I=\int d^2r_1d^2r_2d^2r_3 \Phi(r_1,r_2,r_3) G(r_1,r_2,r_3)
\eeq
where function $\Phi$ is conformal invariant.  We shall demonstrate that
the integral $I$ does not change under conformal transformations. In doing so
we shall use the fact that the BFKL solution $\Psi(r_1,r_2)=
\partial_1^{-2}\partial_2^{-2}D_2(r_1,r_2)$ is conformal invariant,
as it has been shown in the first chapter.

It is sufficient to study the behaviour of the function $G$ only under the
inversion since the invariance under translations and rescaling is obvious.
The complex notation (see section 3 of the first chapter) is more convenient
for this aim: under inversion one has 
\beq
r\rightarrow 1/r,\ \ k\equiv -i\partial\rightarrow r^2k
\eeq
Therefore (for real $r$) 
\beq
d^2r\rightarrow d^2r/r^4
\eeq
and
\beq
D_2(r_1,r_2)\rightarrow
r_1^4r_2^4D_2(r_1,r_2)
\eeq
There are some contributions which are evidently invariant
under inversion. For example taking the last term from $A_1$ it is easy to
check that it leads to the integral
\beq
I_1=\frac{g^2N_c}{2\pi}\int d^2r\Phi(r,r,r)\partial^2\Psi(r,r)
\eeq
which is conformal invariant, since both functions $\Phi$ and
$\Psi$ are invariant, and the factor $r^{-4}$ from $d^2r$ is cancelled
by the factor $r^4$ from $\partial^2$.

Terms with the denominators $r_{12}^2$ and/or $r_{23}^2$ from $A_1+A_2$
combine into an integral
\beq
I_2=-\frac{g^2N_c}{8\pi^3}
\int d^2r_1d^2r_2d^2r_3\Phi(r_1,r_2,r_3)\frac{r_{13}^2}{r_{12}^2
r_{23}^2}
D_2(r_1,r_3)
\eeq
in which the regularization (\ref{regul}) is implied. Under inversion the
ultraviolet cutoff $\epsilon$ is transformed into
$\epsilon_1=r_1r_2\epsilon$ and $\epsilon_2=r_2r_3\epsilon$ in the
denominators $r_{12}^2$ and $r_{23}^2$ respectively. Consequently one finds
a change of $I_2$:
\beq
\Delta I_2=\frac{g^2N_c}{4\pi^2}\int d^2r_1d^2r_3
\Bigr[\Phi(r_1,r_1,r_3)\ln r_1^2+\Phi(r_1,r_3,r_3)\ln r_3^2\Bigr]D_2(r_1,r_3)
\eeq

All the other terms in $A_1+A_2$, proportional either to $\delta^2(r_{12})$
or to $\delta^2(r_{23})$ can be divided into two parts. On one side one has
terms in which the $\delta$-function is multiplied either by a constant or by
$\ln r_{13}$. It will give rise to a part of the integral $I_3$. Terms with a
constant are evidently invariant under inversion. However those containing
$\ln r_{13}$ are not and lead to the corresponding change of $I_3$
\beq
\Delta I_3=
-\frac{g^2N_c}{8\pi^2}\int d^2r_1d^2r_3\Bigr[\Phi(r_1,r_3,r_3)+
\Phi(r_1,r_1,r_3)\Bigr]\ln (r_1^2r_3^2)
D_2(r_1,r_3)
\eeq

The second part contains differential operators acting on $D_2$ of an
Hamiltonian type. It has a form
\beq
-\frac{g^2N_c}{8\pi^2}\Bigl[a_1\delta^2(r_{12})+a_3\delta^2(r_{23})\Bigr]
\label{pseudoexpr}
\eeq
where
\beq
a_1=\partial_1^2\ln r_{13}^2\,\partial_1^{-2}+\ln(-\partial_1^2)\equiv
\partial_1^2\,\tilde{a}_1\,\partial_1^{-2}
\eeq
and $a_3$ is obtained by interchange $1\leftrightarrow 3$. As shown in the
section 1.3 the operator $a_1$ can be transformed into a different form 
more convenient to analyze its properties under inversion. 
Indeed, in the complex notation, one writes the expression:
\beq
\tilde{a}_1=\ln r_{13} +\ln k_1 +c.c
\eeq
A useful relation, between pseudo-differential operators,
\beq
 \ln r_{13} +\ln k_1 =\ln (r_{13}^2 k_1)-k_1^{-1}\ln r_{13}\, k_1
 \eeq
can be deducted from the eq. (\ref{pseudoide2}). On applying it one finds
\beq
a_1=k_1\ln (r_{13}^2k_1)\,k_1^{-1}-\ln r_{13}+c.c
\eeq
Since under the inversion we have
\beq
\ln r_{13}^2k_1\rightarrow \ln r_{13}^2k_1-2\ln r_3
\eeq
the change in $a_1$ is
\beq
\Delta a_1=-2\ln r_{13}+\ln(r_1r_3)+c.c.=\ln\frac{r_1^2}{r_3^2}
\eeq
and the variation of the last part of the integral $I_4$,
which comes from (\ref{pseudoexpr}), is
\beq
\Delta I_4=-\frac{g^2N}{8\pi^2}\int d^2r_1d^2r_3
\Bigl[\Phi(r_1,r_1,r_3)-\Phi(r_1,r_3,r_3)\Bigr]
D_2(r_1,r_3)\ln\frac{r_1^2}{r_3^2}
\eeq
Summing all the changes one has $\Delta I_2 +\Delta I_3 +\Delta I_4 =0$
and this proves that the total integral $I$ is indeed invariant under 
inversion and, hence, under a global conformal transformation.

%%%%%%%%%%%%%%%%%%%%%%%%%%%%%%%%%%%%%%%%%%%%%%%%%%%%%%%%%%%%%%%%%%%%%%%%
\section{Double pomeron exchange and triple pomeron interaction}
We already know the general form of the solution of the 4-gluon equation for
the diffractive amplitude
\beq
S_{40}D_4^{(0)}=D_{40}^{(0)}+ZD_2+g^2N_c(V_{12}+V_{34})D_4^{(0)}
\label{eqd4nlo}
\eeq
It can be constructed as a sum
of the DPE and a TPI parts as in eq. (\ref{gensol}) from the sum of various
inhomogeneous terms.

However a more detailed analysis is instructive.
Let us separate from this known exact solution some arbitrary 
function $f$, which may depend on the angular momentum $j$:
\beq
D_4^{(0)}(j)=f(j)+\tilde{D}_4^{(0)}(j)
\eeq
This choice leads to an equation for the new 4-gluon function
$\tilde{D}_4^{(0)}$
\beq
S_{40}\tilde{D}_4^{(0)}=D_{40}^{(0)}+ZD_2-[S_{40}-g^2N_c(V_{12}+V_{34})]f
+g^2N_c(V_{12}+V_{34})\tilde{D}_4^{(0)}
\label{neweqd4}
\eeq
where the total inhomogeneous part has changed
\beq
D_{40}^{(0)}+ZD_2\rightarrow D_{40}^{(0)}+ZD_2-X
\quad , \quad X=[S_{40}-g^2N_c(V_{12}+V_{34})]f
\eeq

This procedure seems quite trivial but, as we shall see, gives some insight 
into the structure of the eq. (\ref{eqd4nlo}) and of its solution. 
To this end we choose $f$ to be the BFKL function which depends on some gluon 
momenta. The idea of separating a reggeization piece which correspond to BFKL 
pomerons has already been used in \cite{bartwue} and we shall try here to 
see in detail how it works; in this way we shall also be able to compare the 
results obtained in the two approaches.
The separation of this simple term $f$ in the amplitude leads to
the subtraction of a new inhomogeneous term $X$ which acquires the structure 
of a triple pomeron term.
This means also that one can calculate some specific triple pomeron 
contributions expressing them in terms of simple functions.

Let us see how this procedure works in some important cases, differing for
the momenta which enter in the separated BFKL function.
We shall find that, with a particular simple choice for $f$, one is able to
completely cancel the $D_{40}^{(0)}$ which generates the DPE contribution.
In the following, for simplicity, only the forward case is considered.

Let $f=(1/2)g^2D_2(1)=(1/2)g^2D_2(2+3+4)$. One has
\beq
X=\frac{g^2}{2}[S_{40}-g^2N_c(V_{12}+V_{34})]D_2(1)
\eeq
Using eq. (\ref{eq2glu}) we can express the $j-1$ term in $S_{40}$ 
in terms of the forward BFKL interaction $V_0$ and $\omega$ and have
\beq
X=\frac{g^2}{2}D_{20}+\frac{g^2}{2}[g^2N_c(V_0-V_{12}-V_{34})+2\omega(1)-
\sum\omega(i)] D_2(1)
\eeq
Making use of the relation
\beq
(V_0-V_{12})D_2(1)=-W_2(1,2,3+4)
\eeq
where $W_2$ is defined by (\ref{w2def}) and (\ref{gluo23}), 
applying the bootstrap for the (34) gluon system
\beq
g^2N_cV_{34}D_2(1)=2(\omega(3+4)-\omega(3)-\omega(4))D_2(1)
\eeq
and introducing the function $G(1,3)$ with two arguments \cite{bartwue},
which coincides with $G(1,2,3)$ for the forward direction case ($1+2+3=0$),
we finally obtain
\beq
X=\frac{g^2}{2}[D_{20}(1)+G(1,3+4)+D_2(1)(\omega(3)+\omega(4)-2\omega(3+4))+
D_2(1+2)(\omega(2)-\omega(1+2)]
\eeq
Thus we find that the changed function $\tilde{D}_4^{(0)}$ will satisfy an
equation with a new inhomogeneous term
\[
D_{40}+ZD_2+\frac{g^2}{2}[-D_{20}(1)-G(1,3+4)\]\beq
-D_2(1)(\omega(3)+\omega(4)-2\omega(3+4))-
D_2(1+2)(\omega(2)-\omega(1+2))]
\eeq
Note that the additional term $-(1/2)g^2D_{20}$ will cancel the identical
term in $D_{40}^{(0)}$. As a result, one has converted the double pomeron 
exchange contribution coming from $(1/2)g^2D_{20}(1)$ into a triple pomeron 
contribution corresponding essentially to $G(1,3+4)$ and in the same time has 
explicitly separated the term $(1/2)g^2D_2(1)$ in the amplitude. 
In other words, one can calculate the triple pomeron contribution 
corresponding to a vertex
\[
\frac{g^2}{2}[-G(1,3+4)-D_2(1)(\omega(3)+\omega(4)-2\omega(3+4))-
D_2(1+2)(\omega(2)-\omega(1+2))]
\]
as a sum of the double pomeron exchange coupled to $-(1/2)g^2D_{20}(1)$ and
a term $(1/2)g^2D_2(1)$.

Evidently this result is trivially generalized for $f=D_2(i)$, $i=2,3,4$
by simple permutation of indexes 1,2,3 and 4 and one may note that the
separated terms found for the amplitude have the same momentum dependence
of the corresponding zero order cancelled terms. Let us see if this is true
also for the last three terms in (\ref{zerodpe}).

Thus we consider the case $f=(1/2)g^2D_2(1+2)$ for which we have  
\beq
X=\frac{g^2}{2}D_{20}+\frac{g^2}{2}[g^2N_c(V_0-V_{12}-V_{34})+2\omega(1+2)-
\sum\omega(i)]D_2(1+2)
\eeq
In terms of $W$ we have
\beq
V_0D_2(1+2)=-W_2(1+2,0,3+4)
\eeq
The bootstrap gives
\beq
g^2N_c(V_{12}+V_{34})D_2(1+2)=2D_2(1+2)(2\omega(1+2)-\sum\omega(i))
\eeq
so that in terms of $G$ we obtain
\beq
X=\frac{g^2}{2}[-D_{20}(1+2)+G(1+2,3+4)-D(1+2)(4\omega(1+2)-\sum\omega(i))]
\eeq
Again we see that the term with the double pomeron exchange
coupled to $g^2(1/2)D_{20}(1+2)$ can be transformed into a triple pomeron
vertex, essentially, into $-G(1+2,3+4)$ term.

Finally we study a more complicated case with $f=(1/2)g^2D_2(1+3)$.
In this case we find
\beq
V_0D_2(1+3)=-W_2(1+3,0,2+4) 
\eeq
Calculation of $V_{12}$ or $V_{34}$ applied to $D_2(1+3)$ is done using the
formula (see appendix A.4)
\beq
V_{12}D_2(1+3)=W_2(2+4,1,3)+W_2(4,2,1+3)-W_2(3,1+2,4)-W_2(1+3,0,2+4)
\eeq
and
\beq
V_{34}D_2(1+3)=W_2(2+4,3,1)+W_2(2,4,1+3)-W_2(1,3+4,2)-W_2(1+3,0,2+4)
\eeq
Using these results one has
\[
X=\frac{g^2}{2}\Bigl[ D_{20}(1+3)+G(1,2+4)+G(2,1+3)+G(3,2+4)+G(4,1+3)\]\[
-G(1,2)-G(3,4)-G(1+3,2+4)-D_2(1)(\omega(3+4)-\omega(3))\]\beq
-
D_2(2)(\omega(3+4)-\omega(4))-D_2(3)(\omega(1+2)-\omega(1))-
D_2(4)(\omega(1+2)-\omega(2)) \Bigr]
\eeq
The result for $f=(1/2)g^2D_2(1+4)$ is obtained from this after the
permutation of 3 and 4.

Inspecting these results and comparing them with the form of our
triple pomeron vertex, we see that only four terms
\beq
\frac{g^2}{2}[G(1,3)+G(1,4)+G(2,3)+G(2,4)]
\eeq
are not changed under these transformations and thus correspond to a true
triple pomeron interaction. All the rest can be transformed into terms 
which are essentially double pomeron exchange contribution.
Conversely, one can eliminate terms from the double pomeron exchange
substituting them by equivalent triple pomeron contributions.

The most radical result follows if one takes
\beq
f=D_{40}^{(0)}(D_{20}\rightarrow D_2)
\eeq
In this case all the double exchange becomes cancelled and the whole amplitude
is given by a sum of two terms (in an evident symbolic notation)
\beq
D_4^{(0)}=D_{40}^{(0)} (D_{20}\rightarrow D_2)+
\int_0^YG_2(Y-y)G_2(Y-y)\tilde{Z}D_2(y)
\eeq
with a new vertex
\[
\tilde{Z}D_2=\frac{g^2}{2}\Bigl[G(1,3)+G(1,4)+G(2,3)+G(2,4)+G(1+2,3+4)\]\beq-
G(1,3+4)-G(2,3+4)-G(3,1+2)-G(4,1+2)\Bigr]
\eeq
Comparing to (\ref{verbart}) for the forward case, we observe that
it coincides with the part $V(1234)$ of the vertex introduced in 
\cite{bartwue}, being its leading part in the large $N_c$ limit.

%%%%%%%%%%%%%%%%%%%%%%%%%%%%%%%%%%%%%%%%%%%%%%%%%%%%%%%%%%%%%%%%%%%%%%

\section{Relation to the dipole approach}
According to the previous section we shall consider the DPE part 
completely transferred into the TPI one and study the triple pomeron vertex
in the simpler form (\ref{verbart}) for a non-forward direction.
It is important to note that the coupling to BFKL pomerons of this vertex
will lead to further simplifications.
In fact in the coordinate space of 4 gluons, the dependence on only the sum of
the momenta of two gluons, say, 1+2, is translated into a factor
$\delta^2(r_{12})$, so that the two gluons have to be taken at the
same point. Since the wave functions $\Psi(r_1,r_2)$
and $\Psi(r_3,r_4)$ of the two final pomerons, coupled to the vertex, 
vanish if $r_1=r_2$ and $r_3=r_4$ respectively, all terms in
eq.(\ref{verbart}) which depend either only on the sum 1+2 and/or only on the 
sum 3+4 give zero, coupled to the two pomerons.
Thus the effective triple pomeron vertex coupled to pomerons reads
\beq
\tilde{Z}D_2=\frac{g^2}{2}
\Bigl[ G(1,2+4,3)+G(1,2+3,4)+G(2,1+4,3)+G(2,1+3,4) \Bigr]
\label{truevert}
\eeq
Since both pomeron functions $\Psi(r_1,r_2)$ and $\Psi(r_3,r_4)$ are symmetric
in their respective arguments, due to the positive signature of the pomeron,
all terms in (\ref{truevert}) give identical contributions and one has 
\beq
\tilde{Z}D_2=2g^2 G(1,2+3,4)
\label{oneGvert}
\eeq
Moreover let as consider in the coordinate space the ``improper part''
of the function $G(r_1,r_2,r_3)$, which includes all the terms proportional
to $\delta^2(r_{12})$ or/and $\delta^2(r_{23})$; since (\ref{oneGvert}) is
also proportional to $\delta^2(r_{23})$ we find that in the
``improper part'' at least three gluons, either 123 or 234, are to be
taken at the same point in the transverse space. Then these terms will
vanish due to the mentioned property of the pomeron wave function.

Thus only the contribution of the ``proper part'' is left and it is
explicitely given by
\beq
G^{pr}(r_1,r_2,r_3)=-\frac{g^2N_c}{8\pi^3}\frac{r_{13}^2}{r_{12}^2
r_{23}^2}D_2(r_1,r_3)
\label{propGvert}
\eeq
(with the regularization (\ref{regul}) implied).

Coupling this triple pomeron vertex to the two final pomerons,
one obtains the following expression for the
triple pomeron contribution to the diffractive (non-forward) amplitude:
\beq
D_4^{TPI}(Y)=-\frac{g^4N_c}{4\pi^3}
\int_0^Y dy\int d^2r_1d^2r_2d^2r_3\frac{r_{13}^2}{r_{12}^2r_{23}^2}
D_2(r_1,r_3;y)\Psi_1(r_1,r_2;Y-y)\Psi_2(r_2,r_3;Y-y)
\eeq
If one writes the initial pomeron amplitude $D_2$ via the non-amputated 
function $\Psi$, which is the conformal invariant one, the amplitude reads
\beq
D_4^{TPI}(Y)=-\frac{g^4N_c}{4\pi^3}\int_0^Y dy
\int \frac{d^2r_1d^2r_2d^2r_3}{r_{13}^2r_{12}^2r_{23}^2}
\Psi_1(r_1,r_2;Y-y)\Psi_2(r_2,r_3;Y-y)r_{13}^4\partial_1^2\partial_3^2
\Psi(r_1,r_3;y)
\label{difframp}
\eeq
In this form it is evident that the triple pomeron vertex is not
symmetric with respect to the initial pomeron  and two final ones:
there appears an extra operator $r_{13}^4\partial_1^2\partial_3^2$ acting on
the initial pomeron. Note that this operator is essentially a product of
the Casimir operators of the conformal group for the holomorphic and
antiholomorphic parts.
Therefore it is convenient to expand the $\Psi$ functions in terms of the
conformal basis, introduced in the first chapter.
Using (\ref{bfklyexp}) we do this for all the three pomerons in 
(\ref{difframp}) and compute the action of the Casimir operators by
means of (\ref{eqconfbas}). 
We then obtain, after the integration over $y$
\[
D_4^{TPI}(Y)=-\frac{g^4N_c}{4\pi^3}\sum_{\mu,\mu_1,\mu_2}
\langle\mu|\Psi_0\rangle\langle\mu_1|\Psi_{10}\rangle
\langle\mu_2|\Psi_{20}\rangle
\frac{e^{Y(\omega_{\mu_1}+\omega_{\mu_2})}-e^{Y\omega_{\mu}}}
{\omega_{\mu_1}+\omega_{\mu_2}-\omega_{\mu}}
\]\beq
\frac{4\pi^8}{a_{n-1,\nu}a_{n+1,\nu}}
\int \frac{d^2r_1d^2r_2d^2r_3}{r_{13}^2r_{12}^2r_{23}^2}
E_{\mu_1}(r_1,r_2)E_{\mu_2}(r_2,r_3)E_{\mu}(r_1,r_3)
\label{tpiterm}
\eeq
In this form the triple pomeron contribution can be compared to the 
double dipole density found by Peschanski \cite{pesh} in  A.H.Mueller's
colour dipole approach. One  observes
that the two expressions differ only in the sign and by a factor 
\[
\frac{4\pi^8}{a_{n-1,\nu}a_{n+1,\nu}}
\]
which in our approach distinguishes the initial pomeron from the two
final ones. The integral over the coordinates of the three pomerons is
the same. So essentially the three-pomeron contribution to the diffractive
amplitude found in our approach coincides with the double dipole density
in the dipole approach.

However one should not forget that in our $s$-channel unitarity approach
the TPI term (\ref{tpiterm}) does not exhaust all the diffractive amplitude. 
In fact we have seen that
\beq 
D_4^{(0)}=D_{40}(D_{20}\rightarrow D_2)+D_4^{TPI}
\eeq
At high energies the TPI term behaves essentially as $s^{2\Delta}$ and the 
first one as $s^{\Delta}$ where $\Delta$ is the BFKL intercept. 
So one might think that the first term could be neglected. 
However the correct region of the validity of the leading log approximation, 
implicit in the hard pomeron theory, is $g^2\ln s\sim 1$ when the two terms 
in (80) have the same order of magnitude.
Moreover the BFKL separated term goes as $1/N_c$ and the TPI as $g^2 N_c$ and 
also these factors are of the same order.   
The dipole approach uses essentially the same leading log approximation but 
it leads to the double dipole density which coincides only with the TPI term 
in the $s$-channel unitarity approach and shows no trace of the first term.
Therefore this fact points to certain differences between the two 
approaches.

%%%%%%%%%%%%%%%%%%%%%%%%%%%%%%%%%%%%%%%%%%%%%%%%%%%%%%%%%%%%%%%%%
\section{Higher order densities in the dipole approach}
Keeping in mind this substantial agreement between the results obtained in the
$s$-channel unitarity approach and in the colour dipole picture for the
two pomerons (dipoles) distribution, let us investigate the general structure 
of the higher order dipole densities given by the latter. 
To see if the previous equivalence found is maintained,
one should study the multipomeron distribution in the former approach.
In particular the first step would regard the study of the structure of the
system of up to six reggeized gluons to look at the transition from 1 to 3
pomerons. In fact, as we shall show, this coupling does not directly appear 
in the standard dipole picture which instead leads to a set of fan diagrams
with a single ``elementary'' three-pomeron coupling.
If in such a context a future analysis will show the presence of pomeron
vertices of order higher than 3 then an extension of the dipole picture,
for example as the one proposed by Peshanski \cite{pesh}, should
be required.

Following the approach of A.H. Mueller, as briefly reviewed in section $2.4$,
in the  colour dipole formalism the k-fold inclusive dipole density is obtained
as the $k$-th functional derivative of the functional $D\{u(r_i,r_f)\}$,
taken at $u(r_i,r_f)=1$ \cite{mueller}. The arguments $r_i$ and
$r_f$ are the dipole endpoints in the transverse plane. 
We remember here the simple equation the functional $D$ satisfies
\[
D(r_1,r_0,y,u)=u(r_1,r_0)e^{2y\omega(r_{10})}\]\beq+
\frac{g^2N_c}{8\pi^3}
\int_0^y dy' e^{2(y-y')\omega(r_{10})}\int d^2r_2\frac{r_{10}^2}
{r_{12}^2r_{20}^2}D(r_1,r_2,y',u)D(r_2,r_0,y',u)
\label{genDeq}
\eeq
Here $r_1$ and $r_0$ are the end points of the $q\bar q$ pair which
determine the initial dipole; moreover we define $y=\ln z_{10}$ where $z_{10}$
is the ratio of the scaling factors of the antiquark and of the quark;
$\omega(r)$ {\it is not} a Fourier transform of the trajectory,
but just $\omega(k)$ with $k/m$ formally substituted by $r/\epsilon$,
where $\epsilon$ is an ultraviolet cutoff. This cutoff is also implied in
the singular kernel of the integral operator in $r_2$. Let as note that
eq. (\ref{genDeq}) is compatible with the normalization condition $D(u=1)=1$.

The $k$-th derivative of eq. (\ref{genDeq}) give an equation for the $k$-fold
dipole density if computed for $u=1$. For $k>1$ we obtain
\[
n_k(r_1,r_0;\rho_1,...\rho_k;y)=
\frac{g^2N_c}{8\pi^3}
\int_0^y dy' e^{2(y-y')\omega(r_{10})}\int d^2r_2\frac{r_{10}^2}
{r_{12}^2r_{20}^2}n_k(r_1,r_2;\rho_1,...\rho_k;y')+(r_1 \leftrightarrow r_0)\]\[
+\frac{g^2N_c}{8\pi^3}
\int_0^y dy' e^{2(y-y')\omega(r_{10})}\int d^2r_2\frac{r_{10}^2}
{r_{12}^2r_{20}^2}\sum_{l=1}^{k-1}(n_l(r_1,r_2;\rho_1,...\rho_l;y')
n_{k-l}(r_2,r_0;\rho_{l+1},...\rho_k;y')\]\beq+\ symmetrization\ terms)
\label{genkfold}
\eeq
where the symmetrization terms (ST) are obtained from the explicitly shown one
by taking all different divisions  of arguments $\rho_1,...\rho_k$ into
two groups with $l$ and $l-k$ arguments.
In the $k=1$ case an inhomogeneous term whose form is clear from 
(\ref{genDeq}) is present.
One should note that the operator on the right-hand side acts nontrivially 
only on the first argument of the density $n_k$. Its action on the rapidity
variable $y$, on the contrary, is rather simple and it is better treated
in a transformed associated equation. Thus it is convenient to multiply the
equation by $ e^{-2y\omega(r_{10})}$, differentiate then with respect
to $y$ and pass to the $j$-space by the standard Mellin transformation.
After having performed these operations one obtains
\[
(j-1) n_k(r_1,r_0;\rho_1,...\rho_k;j)=
\frac{g^2N_c}{4\pi^3}
\int d^2r_2L(r_1,r_2,r_{20})n_k(r_1,r_2;\rho_1,...\rho_k;y)\]\[+
\frac{g^2N_c}{8\pi^3}\int\frac{dj_1dj_2}{(2\pi i)^2(j+1-j_1-j_2)}
\int d^2r_2\frac{r_{10}^2}
{r_{12}^2r_{20}^2}\]\beq
\sum_{l=1}^{k-1}(n_l(r_1,r_2;\rho_1,...\rho_l;j_1)
n_{k-l}(r_2,r_0;\rho_{l+1},...\rho_k;j_2)+\ ST)
\label{neqj}
\eeq
where we introduced the BFKL kernel in the coordinate
space
\beq
L(r_{12},r_{20})=\frac{r_{10}^2}
{(r_{12}^2+\epsilon^2)(r_{20}^2+\epsilon^2)}-2\pi\ln\frac
{r_{10}}{\epsilon} \Bigl[\delta^2(r_{12})+\delta^2(r_{20})\Bigr]
\eeq
Comparing with (\ref{regul}) we see that it does not depend on $\epsilon$ and
is ultraviolet stable.
The solution of the equation (\ref{neqj}) is quite easy to obtain if one
describes the dependence of the densities on their first two arguments in 
terms of the conformal basis:
\beq
n_k(r_1,r_0)=\sum_{\mu}E_{\mu}(r_1,r_0)n_k^{\mu}
\eeq
Here we have suppressed all other arguments in $n_k$ irrelevant for the
time being. The densities $n_k^{\mu}$ in a given conformal state are
obtained from $n_k(r_1,r_0)$ by the inverse transformation which follows
from orthogonality property (\ref{orthorel}) and a relation between 
$E_{n,\nu}$ and $E_{-n,-\nu}$ (see \cite{pertqcd})
\beq
n_k^{\mu}=\int \frac{d^2r_1d^2r_0}{r_{10}^4}E^*_{\mu}(r_1,r_0)
n_k(r_1,r_0)
\label{nproj}
\eeq
Therefore, to pass to the conformal basis, we integrate eq. (\ref{neqj})
over $r_1$ and $r_0$ as indicated in (\ref{nproj}).
The first term on the right-hand 
side can be simplified due to the property of the BFKL kernel
\beq
\frac{g^2N_c}{4\pi^3}
\int d^2r_2L(r_{12},r_{20})E_{\mu}(r_1,r_2)=\omega_{\mu}E_{\mu}(r_1,r_0)
\eeq
Therefore after the integration we obtain 
\[
(j-1-\omega_{\mu})n_k^{\mu}(\rho_1,...\rho_k,j)=
\frac{g^2N_c}{8\pi^3}
\int\frac{dj_1dj_2}{(2\pi i)^2(j+1-j_1-j_2)}\]\beq
\int \frac{d^2r_1d^2r_2d^2r_0}
{r_{12}^2r_{20}^2r_{10}^2}E^*_{\mu}(r_{10})
\sum_{l=1}^{k-1}(n_l(r_1,r_2;\rho_1,...\rho_l;j_1)
n_{k-l}(r_2,r_0;\rho_{l+1},...\rho_k;j_2)+\ ST)
\eeq
To find the final form of the equation we have only to present also
the densities $n_l$ and $n_{k-l}$ as functions of their first arguments
in the form (\ref{nproj}). Then one has
\[
(j-1-\omega_{\mu})n_k^{\mu}(\rho_1,...\rho_k,j)=
\int\frac{dj_1dj_2}{(2\pi i)^2(j+1-j_1-j_2)}\]\beq
\sum_{\mu_1,\mu_2}V_{\mu,\mu_1\mu_2}
\sum_{l=1}^{k-1}(n_l^{\mu_1}(\rho_1,...\rho_l;j_1)
n_{k-l}^{\mu_2}(\rho_{l+1},...\rho_k;j_2)+\ ST)
\label{neqjproj}
\eeq
where
\beq
V_{\mu\mu_1\mu_2}=\frac{g^2N_c}{8\pi^3}\int \frac{d^2r_1d^2r_2d^2r_0}
{r_{12}^2r_{20}^2r_{10}^2}E^*_{\mu}(r_{10})E_{\mu_1}(r_{12})E_{\mu_2}(r_{20})
\label{vert1}
\eeq
is just one half of the three-pomeron vertex introduced by Peschanski.
The general structure of this equation allows one immediately to imagine
a tree-like expansion of the colour dipole densities as we shall show more
in detail. At each tree node a vertex (\ref{vert1}) will be present.
The integral of its explicit expression has been calculated
\cite{korch97, bialpesh} with the leading quantum numbers $n=\nu=0$ for the
related three pomerons and it turns out to have a quite huge
value (7766.679). Of course there will be present some correcting factors
but still this means that such a perturbative expansion in pomeron
vertices may need a complete resummation.

Eq. (\ref{neqjproj}) allows to obtain successively dipole densities for any 
number of dipoles starting from the lowest order one-dipole density, for which
\beq
n_1^{\mu}(\rho)=\frac{E^*_{\mu}(\rho)}{j-1-\omega_{\mu}}
\frac{1}{\rho^4}
\eeq
(we recall that in this notation $\rho$ includes two endpoints of the
dipole $\rho_i$ and $\rho_f$; $\rho^2\equiv \rho_{if}^2$).

Putting this into (\ref{neqjproj}) for $k=2$ and integrating over $j_1$ and 
$j_2$ we arrive at the expression obtained by Peschanski
\beq
n_2^{\mu}(\rho_1,\rho_2;j)=\frac{1}{\omega-\omega_{\mu}}
\sum_{\mu_1,\mu_2}\frac{1}{\omega-\omega_{\mu_1}-
\omega_{\mu_2}}V_{\mu,\mu_1,\mu_2}E^*_{\mu_1}(\rho_1)E^*_{\mu_2}(\rho_2)
\frac{1}{\rho_1^4\rho_2^4}
\eeq
where $\omega=j-1$. (To compare with \cite{pesh} one should take into 
account that factors $1/(2a_{\mu})$ are included in the definition of sums 
over $\mu$'s in our notation).

As a further step we study the density for three dipoles.
Eq. (\ref{neqjproj}) for $k=3$ reads
\[
n_3^{\mu}(\rho_1,\rho_2,\rho_3;j)=\frac{1}{
\omega-\omega_{\mu}}
\int\frac{dj_1dj_2}{(2\pi i)^2(j+1-j_1-j_2)}\]\beq
\sum_{\mu_1,\mu_2}V_{\mu,\mu_1\mu_2}
(n_1^{\mu_1}(\rho_1;j_1)
n_{2}^{\mu_2}(\rho_2,\rho_3;j_2)+\ ST)
\eeq
It is instructive to study explicitely this term inserting the expressions for
$n_1^{\mu_1}$ and $n_2^{\mu_2}$ obtained earlier. Then we get,
after integrations over $j_1$ and $j_2$:
\[
n_3^{\mu}(\rho_1,\rho_2,\rho_3;j)= \frac{1}{
\omega-\omega_{\mu}}
\sum_{\mu_1,\mu_2,\mu_3,\mu_4}V_{\mu,\mu_1\mu_2}V_{\mu_2\mu_3\mu_4}\]\beq
\frac{1}{(\omega-\omega_{\mu_1}-\omega_{\mu_2})
(\omega-\omega_{\mu_1}-\omega_{\mu_3}-\omega_{\mu_4})}
E^*_{\mu_1}(\rho_1)E^*_{\mu_3}(\rho_2)E^*_{\mu_4}(\rho_3)
\frac{1}{\rho_1^4\rho_2^4\rho_3^4}
\label{n3eqj}
\eeq
To this term we have to add terms which symmetrize in the three dipoles.

Studying (\ref{n3eqj}) we see that it corresponds to the picture when first the
initial pomeron splits into two pomerons, 1 and 2, and afterwards the pomeron 2
splits into pomerons 3 and 4. One does not find here a local  vertex
for the transition of the initial pomeron into three final ones. It is not
difficult to see under which condition one would get such a local vertex.
If we forget about the dependence of the second denominator on $\mu_2$,
then one can sum over  $\mu_2$. Using  the completeness relation (\ref{comple})
one obtains
\beq
\sum_{\mu_2}V_{\mu\mu_1\mu_2}V_{\mu_2\mu_3\mu_4}=\left(\frac{g^2N_c}{8\pi^3}
\right)^2
\int\frac{d^2r_1d^2r_2d^2r_3d^2r_0}{r_{12}^2r_{10}^2r_{23}^2r_{30}^2}
E^*_{\mu}(r_1,r_0)E_{\mu_1}(r_1,r_2)E_{\mu_3}(r_2,r_3)E_{\mu_4}(r_3,r_0)
\label{vertrel}
\eeq 
which is just the vertex from one to three pomerons introduced by Peschanski.
However, the described summation is not possible due to the second
denominator. It implies that the pomeron 2 has to evolve in $y$ from
the point of its formation from the initial pomeron up to the
point of its splitting into the final pomerons 3 and 4.

Thus our conclusion is that the vertex for transition from 1 to $k$ pomerons
introduced by Peschanski, in fact, does not appear in the solution of
the Mueller equation (\ref{genkfold}) for the $k$-fold density, which rather 
corresponds to a set of all fan diagrams with only the triple pomeron 
coupling. 
Absence of higher-order couplings can be directly traced to the structure of 
the equation (\ref{genDeq}) for the generating functional $D$, 
quadratic in $D$, which in fact means that the basic vertex changes one dipole
into two.

The general structure of higher order vertices introduced by Peschanski 
hence refers to an extension, which he proposes, to the dipole model.
In his work the possibility of a splitting of one dipole into many dipoles
is considered, going therefore beyond the standard dipole model. 
His idea is related to the fact that in general a contribution to the 
transition from $1$ to $p$ pomeron given by the fan diagrams would require
a large rapidity interval at each splitting step in order to ensure the 
validity of the Regge approximation.
The derived effective field theory is therefore more complex but it
should be characterized by a strong symmetry since the vertices are
related each other by integral relations, as for example the one in eq.
(\ref{vertrel}).

A general analysis of the $n$-reggeon amplitudes approach, based on 
$s$-channel unitarity, is not known yet, even if some work in this direction 
is in progress. Any advance in this direction  will be of great interest
to understand the Regge limit effective behaviour of QCD.

To conclude this section we shall try to analyze in more details the picture
one derives from the triple pomeron interaction structure so far verified. 
It has been shown that at asymptotic energies the higher-order densities in 
the dipole approach correspond to the standard Regge-Gribov picture, in the
tree approximation (fan diagrams), with only the triple pomeron interaction,
which however has a highly complicated non-local form.
Indeed, the triple pomeron interaction present in (\ref{n3eqj}) corresponds 
to a structure
\beq
T=\frac{g^2N_c}{8\pi^3}
\int\frac{d^2r_1d^2r_2d^2r_3}{r_{12}^2r_{23}^2r_{31}^2}
\tilde{G}_3(r_1,r_2;r_1',r'_2)\tilde{G}_1(r_2,r_3;r'_2,r'_3)
\tilde{G}_2(r_3,r_1;r'_3,r'_1)
\label{triplestru}
\eeq
where $\tilde{G}_i,\ i=1,2,3$ are Green functions of the interacting
pomerons defined as
\beq
\tilde{G}(r_1,r_2;r'_1,r'_2)=
\sum_{\mu}\frac{E_{\mu}(r_1,r_2)E^*_{\mu}(r'_1,r'_2)}
{\omega-\omega_{\mu}}
\label{bfklGtilde}
\eeq
They are not the physical BFKL Green functions. The latter  include an extra
factor depending on $\mu$:
\beq
G(r_1,r_2;r'_1,r'_2)=\frac{1}{4\pi^8}
\sum_{\mu}a_{n+1,\nu}a_{n-1,\nu}\frac{E_{\mu}(r_1,r_2)E^*_{\mu}(r'_1,r'_2)}
{\omega-\omega_{\mu}}
\label{bfklG}
\eeq
However
in the limit $s\rightarrow\infty$ only the lowest conformal
weights contribute $n=\nu=0$ for which $a_{n\pm 1,\nu}=2\pi^4$
and (\ref{bfklGtilde}) and (\ref{bfklG}) coincide. 
Then we can forget about tildes in (\ref{triplestru}).

We transform the Green functions to  given total momenta of the pomerons
presenting
\beq
G_3(r_1,r_2;r'_1,r'_2)=\int \frac{d^2l_3}{(2\pi)^2}e^{il_3(R_3-R'_3)}
G_{l_3}(r_{12},r'_{12})
\eeq
where $R_3=(1/2)(r_1+r_2)$ and similarly for the two other Green functions.
Introducing $R=r_1+r_2+r_3$ we transform the integration over the coordinates
as follows
\beq
\int d^2r_1d^2r_2d^2r_3=
\int d^2Rd^2r_{12}d^2r_{23}d^2r_{31}
\delta^2(r_{12}+r_{23}+r_{31})
\label{measure3}
\eeq
The coordinates themselves are 
$r_1=(1/3)(R-r_{21}-r_{31})$ etc., where from we find
\beq
R_1=(1/6)(2R-r_{12}-r_{13}),\ R_2=(1/6)(2R-r_{21}-r_{23}),\ 
R_3=(1/6)(2R-r_{31}-r_{32})
\eeq
and
\beq
 i\sum_{j=1}^3l_jR_j=i(1/3)R\sum_{j=1}^3l_j-
i(1/6)(r_{12}l_{12}+r_{23}l_{23}+r_{31}l_{31})
\eeq
where we denoted $l_{12}=l_1-l_2$ etc.
The integral over $R$ gives $9(2\pi)^2\delta^2(l_1+l_2+l_3)$.
 Presenting  the remaining
$\delta$-function in (\ref{measure3}) as an integral over an auxiliary 
momentum $q$ we find an expression (for fixed $l_1,l_2$ and $l_3$)
\[\frac{g^2N_c}{8\pi^3}
\int\frac{d^2q}{(2\pi)^2}\frac{d^2r_{12}d^2r_{23}d^2r_{31}}
{r^2_{12}r^2_{23}r^2_{31}}\exp \left(ir_{12}(q-\frac{1}{6}l_{12})+
ir_{23}(q-\frac{1}{6}l_{23})+ir_{31}(q-\frac{1}{6}l_{31}\right)\]\beq
G_{l_3}(r_{12},r'_{12})G_{l_1}(r_{23},r'_{23})G_{l_2}(r_{31},r'_{31})
\eeq

At this point we recall the expression for the BFKL Green function with
a fixed total momentum:
\beq
G_l(r,r')=\frac{1}{(2\pi)^4}\int \frac{\nu^2 d\nu}{(\nu^2+1/4)^2}
s^{\omega(\nu)} E^{(l)}_\nu(r)E^{(l)}_\nu(r')
\label{green}
\eeq
where
\beq
E^{(l)}_\nu(r)=\int d^2R\exp (ilR)
\left(\frac{r}{|R+r/2||R-r/2|}\right)^{1+2i\nu}
\label{Elnu}
\eeq
and where we retained only the dominant isotropic term. At $s\rightarrow\infty$
the vicinity of $\nu=0$ gives the dominant contribution. If $l\neq 0$ then the
integral in (\ref{Elnu}) converges at large $R$ and we can take the 
functions $E$ out of the integral over $\nu$ at $\nu=0$. 
Taking then the asymptotics of the remaining integral, we find
\beq
G_l(r,r')\simeq\frac{1}{2\pi^4}s^{\Delta}\frac{\sqrt{\pi}}{(a\ln s)^{3/2}}
E^{(l)}_0(r)E^{(l)}_0(r')
\eeq
where $\Delta=\omega_{n=0,\nu=0}$ is the BFKL intercept and
$a=7g^2N_c\zeta(3)/(2\pi^2)$.
As we see, the Green function asymptotically factorizes in $r$ and $r'$.
This means that we obtain a quantum field theory of pomerons with a 
propagator
\beq
P(y,l)=\frac{2}{\pi^2}e^{y\Delta}\frac{\sqrt{\pi}}{(ay)^{3/2}}
\eeq
(not really depending on the momentum $l$) and an interaction vertex
\[
V(l_1,l_2,l_3)= \frac{9g^2N_c}{8\pi^3}
\int\frac{d^2q}{(2\pi)^8}\frac{d^2r_{12}d^2r_{23}d^2r_{31}}
{r^2_{12}r^2_{23}r^2_{31}}\]\beq
\exp \left(ir_{12}(q-\frac{1}{6}l_{12})+
ir_{23}(q-\frac{1}{6}l_{23})+ir_{31}(q-\frac{1}{6}l_{31}\right)
E^{(l_3)}_0(r_{12})E^{(l_1)}_0(r_{23})E^{(l_2)}_0(r_{31})
\eeq

The vertex factorizes under the sign of the integration over $q$:
\beq
V(l_1,l_2,l_3)=\frac{9g^2N_c}{8\pi^3}
\int\frac{d^2q}{(2\pi)^2}
J(l_3,q-\frac{1}{6}l_{12})J(l_1,q-\frac{1}{6}l_{23})
J(l_2,q-\frac{1}{6}l_{31})
\label{vertex}
\eeq
where
\beq
J(l,q)=\int\frac{d^2r}{(2\pi)^2 r^2}e^{iqr}E^{(l)}_0(r)
=\int\frac{d^2p}{2\pi p}\frac{1}{|p-q+l/2||p-q-l/2|}
\eeq

Note that for $l=0$ this derivation is incorrect. Calculations show
that in this case eq. (\ref{vertex}) for the vertex remains valid with
\beq
J(0,q)=\frac{1}{9q}
\eeq
However the Green function (\ref{green}) at $l=0$ has an asymptotics
\beq G_0(r,r')\simeq\frac{1}{2\pi^2}s^{\Delta}\sqrt{\frac{\pi}{a\ln s}}
rr'\exp\left(-\frac{\ln^2(r/r')}{a\ln s}\right)
\eeq
which is evidently no more factorizable in $r$ and $r'$.

% conclusions

\chapter{Conclusions}
As stated in the introduction, the thesis has been devoted to the study of
some issues relevant for the hard pomeron model. Both theoretical and
phenomenological aspects have been considered.

From a phenomenological point of view in the second chapter an extension
to include the running coupling for the hard pomeron is studied.
The leading singularity in this way is softened and once the infrared sector
is safely included in the model one can avoid also the problems due to the
diffusion mechanism which spoil out the validity of the pure BFKL pomeron.
The model, which preserves the gluon reggeization property by means of the 
bootstrap condition and the right gluon distribution behaviour in the DLLA,
has been studied both numerically and in some aspects analytically.
Numerical calculations were addressed to the computation of the spectrum
and eigenfunctions of the Schr\"odinger-like equation for the pomeron.
The spectrum has shown a discrete part and for the wide parameter range
two negative ``energies'' states corresponding to two supercritical
pomeron states.
The slope of each singularity has been calculated with a perturbative
approach. The sensitivity of the model is mainly with respect to only one
infrared parameter, a mass which we chose to fix forcing a value for the
slope of the leading singularity. 
We have applied such a model to the process of two virtual photon scattering
and to see some unitarization effects we have considered in the large $N_c$
limit a multi-scattering (eikonal) formulation which have shown small
unitarity corrections of such a kind to the cross sections. Triple pomeron
interactions, required by unitarity, are not taken into account in these
calculations and larger unitarity effects are therefore not excluded.
Similar considerations were applied to a toy model, with a Gaussian colour
dipole distribution, for the proton.

An application of this model to the inclusive jet production has been
considered.
A correct asymptotic behaviour of the differential cross section for large
jet transverse momenta has been found. In this case one must note that in
the BFKL approach already some unitarity correction, which have lead 
to modify the BFKL equation with a non-linear contribution generating fan 
diagrams, as well as coherence effects were, for example, invoked.
Since the leading singularity is a pole, the $y$ dependence in the
cross section depends on the inclusion of the non-leading states of the
spectrum. Here we calculated the contribution due to the two supercritical
pomerons and for a better estimate a much larger part of the spectrum
depending on the energy, should be considered.

The contribution from the whole spectrum have been included
in the analysis of the evolution of the gluon density and of the proton
singlet structure function.
The initial proton gluon distribution in rapidity has been found by matching
the related singlet structure function. The main problem has been to find a
way to couple the gluon to the photon taking in account the introduced
running of the coupling. A phenomenological approach has been chosen, based
on the requirement that for the structure function in the DLLA the analytical
asymptotic behaviour should match the one calculated using the DGLAP quark
distributions.
The numerical calculations have still shown a steep increase with decreasing
$x$ of the gluon distribution, but with a better behaviour than for a BFKL
pomeron.

In the theoretical part of the thesis we deal with the unitarity problem.
At the basis of this contribution are the works of J.Bartels et al.
where a program to ensure unitarity requirements has been developed and 
the analysis of the lower order corrections, up to 4 gluons in the
$t$-channel, started.
The idea of studying these corrections in the large $N_c$ limit,
a program initiated by M.Braun, is pursued in order to avoid the
otherwise complicated colour structure. The nice property of this limit is that
one does not loose many important features of the underlying interactions and
it is also possible to directly compare the obtained results with some
recent developments derived in the colour dipole picture.
In particular the system of four gluons has been studied for a non-zero
momentum flow in the $t$ channel in the leading and next-to-leading order in
$1/N_c$. The solution has been shown to have in general double pomeron exchange
and triple pomeron interaction contributions.
The latter is written in terms of a three point function which is
by itself infrared stable and conformal invariant.
Following the idea of Bartels and Wuestoff it is shown the
possibility of eliminating the DPE terms in the solution so that the unitarity
correction to this order may depend only on a particular three pomeron vertex,
plus some single pomeron terms. There is complete agreement with the Bartels
et al. works once the large $N_c$ limit is taken into account.
After coupling this vertex to pomerons one can easily compare the obtained two
pomeron amplitude with the double dipole colour density analyzed by
R.Peschanski in the standard dipole picture .
One finds a substantial agreement for the triple pomeron interaction form, apart
from an asymmetry factor and the previous mentioned single pomeron terms.
The properties of higher order dipole densities, as defined in the A.H. Mueller
formulation of the colour dipole model, are subsequently studied.
A picture is found where in the rapidity evolution a pomeron may split into
two pomerons, thus leading to a fan diagrams description with only three
pomeron vertices involved. The resulting pomeron effective field theory
would hence have a strong signature.
Therefore a corresponding analysis is needed in the reggeon approach based on
$s$-channel unitarity. The analysis starting from a system of $5$ and $6$
reggeized gluons and in general of higher order distributions will show if this
picture is correct or one has to deal with higher order effective pomeron
vertices. In such a case one should also consider extension in the colour
dipole picture, as for example proposed by R.Peschanski.
In any case the structure of a possible effective theory of QCD in the Regge
limit will then be strongly restricted. 

%%%%%%%%%%%%%%%%%%%%%%%%%%%%%%%%%%%%%%%%%%%%%%%%%%%%%%%%%%%%%%%%%%%%%%%%%%%%%
\newpage
\noindent
{\huge {\bf Acknowledgements}}\\

I would like to express my deep gratitude to my supervisor Professor Giovanni 
Venturi for his support, interest and encouragement and for countless 
stimulating discussions in different fields of Physics.

I am sincerely and infinitely grateful also to Professor Mikhail Braun,
for his guidance and help in the last years. His ideas have had a 
strong influence and most of the work present in the thesis is inspired
and arises from his collaboration.

It was a great experience to visit the University of Santiago de
Compostela and I would like to thank Profs. C. Pajares and L. Miramontes
for their hospitality.

Furthermore I thank Prof. R. Peschanski for pleasant and useful discussions.

Finally a special thank to Fabio Finelli for many discussions and a 
fruitful collaboration.
% appendix

\appendix
%\addcontentsline{toc}{chapter}{Appendix}
\chapter{}

%%%%%%%%%%%%%%%%%%%%%%%%%%%%%%%%%%%%%%%%%%%%%%%%%%%%%%%%%%%%%%%%%%%%%%%%%
\section{Asymptotic behaviour of the wave functions.}
%%%%%%%%%%%%%%%%%%%%%%%%%%%%%%%%%%%%%%%%%%%%%%%%%%%%%%%%%%%%%%%%%%%%%%%%%
\subsection {The momentum space behaviour}
Let us consider the asymptotic expression for the gluon Regge trajectory
given in (\ref{tra_int})
\beq
\omega(q) \mathop{\sim}_{q^2>>m^2} -\frac{N_c}{2b} 
\ln{ \frac{ \ln q^2}{\ln m^2}}
\eeq
Hence, for $N_c=3$, one can write the asymptotic form of the pomeron
non-homogeneous equation at high momentum for the semi-amputated wave 
function $\psi$ 
\beq
\ln\frac{\ln q^{2}}{\ln m^{2}} \psi(q)+\frac{am^{2}\ln m^{2}}{q^{2}\ln
q^{2}}={\tilde\epsilon}\psi(q)+\frac{1}{\pi}\int\frac{d^{2}q' \psi (q')}
{[(q-q')^{2}+m^{2}]\ln[(q-q')^{2}+m^{2}]}
\lb{eigen1}
\eeq
where we have put $E=(3/b)\tilde\epsilon$
(we consider the case $m=m_1$ for simplicity) 
Let us study the last (integral) term, which we denote by $D$. Changing the
variable $q'=|q|\kappa$ we present it in the form
\[
D=
\frac{1}{\pi}\int\frac{d^{2}\kappa \psi (|q|\kappa)}
{[(n-\kappa)^{2}+m^{2}/q^{2}]\ln[q^{2}(n-\kappa)^{2}+m^{2}]}\]\beq
=\frac{1}{\pi}\int\frac{d^{2}\kappa \psi (|q|(n+\kappa))}
{(\kappa^{2}+m^{2}/q^{2})\ln(q^{2}\kappa^{2}+m^{2})}
\eeq
with $n^{2}=1$.
In the last form it is evident that the leading terms at
$q\rightarrow\infty$
come from the integration region of small $\kappa$. Then we split the
total $\kappa$ space into two parts: $\kappa>\kappa_{0}$ and
$\kappa<\kappa_{0}$ where $\kappa_{0}$ is a small number
\beq
\kappa_{0}<<1
\lb{cond1}
\eeq
The contributions from these two parts we denote as $D_{1}$ and $D_{2}$,
respectively. Our first task is to show that $D_{2}$ cancels the kinetic
term in (\ref{eigen1}) irrespective of the asymptotics of $\psi(q)$.

With small enough $\kappa_{0}$ and for a "good enough" $\psi$ 
\beq
\psi (|q|(n+\kappa))\simeq\psi (|q|n)=\psi(q)
\lb{approx1}
\eeq
so that we can take it out of the integral over $\kappa$ in $D_{2}$.
Actually (\ref{approx1}) is our definition of a "good" function, so that we shall have
to check if this condition is indeed satisfied for the found asymptotic
$\psi(q)$. 

With (\ref{approx1}) $D_{2}$ simplifies to
\beq
D_{2}=\frac{\psi(q)}{\pi}\int_{0}^{\kappa_{0}^{2}q^{2}}\frac{d^{2}q' }
{({q'}^{2}+m^{2})\ln({q'}^{2}+m^{2})}=
\psi(q)\ln\frac{\ln(\kappa_{0}^{2}q^{2}+m^{2})}{\ln m^{2}}
\lb{d2_1}
\eeq
As $q\rightarrow\infty$ we assume that also
\beq
\kappa_{0}q\rightarrow\infty
\lb{limk0q}
\eeq
Evidently this condition fixes the manner in which $\kappa_{0}$ goes to zero
as $q$ turns large. This should be taken into account when verifying
condition (\ref{cond1}). Then (\ref{d2_1}) gives
\beq
D_{2}=\psi(q)\ln\frac{\ln\kappa_{0}^{2}q^{2}}{\ln m^{2}}
\eeq
We have also
\[\ln\ln\kappa_{0}^{2}q^{2}=\ln(\ln q^{2}+\ln\kappa_{0}^{2})\]
But according to (\ref{limk0q})
\[\ln q^{2}>>|\ln\kappa_{0}^{2}|\]
so that we have
\[\ln\ln\kappa_{0}^{2}q^{2}=\ln\ln q^{2}+\frac{\ln\kappa_{0}^{2}}{\ln q^{2}}
+ \cdots\]
Then our final result is
\beq
D_{2}=\psi(q)\left(\ln\frac{\ln q^{2}}{\ln m^{2}}+O(1/\ln q^{2})\right)
\lb{d2_2}
\eeq
The first term  exactly cancels the kinetic energy ( first) term
in the pomeron equation. The correction term in (\ref{d2_2}) evidently is much
smaller than ${\tilde\epsilon}\psi(q)$, since the factor which multiplies
the function $\psi$ goes to zero in (\ref{d2_2}). So we can safely neglect it.

As a result, the asymptotic equation becomes
\beq
\frac{am^{2}\ln m^{2}}{q^{2}\ln
q^{2}}={\tilde\epsilon}\psi(q)+D_{1}
\lb{eigen2}
\eeq
with the term $D_{1}$ given by
\beq
D_{1}=
\frac{1}{\pi}\int\frac{d^{2}q' \psi
(q')\theta((q-q')^{2}-\kappa_{0}^{2}q^{2})}
{(q-q')^{2}\ln (q-q')^{2}} 
\lb{d1}
\eeq
where we have omitted the $m^{2}$ terms in the denominator because of 
(\ref{limk0q}).

We consider the function $\chi(q)$ defined by
\beq  \chi(q)=\psi(q)q^{2}\ln q^{2}\eeq
Multiplying (\ref{eigen2}) by $q^{2}\ln q^{2}$ we find an equation for $\chi$
\beq
am^{2}\ln m^{2}={\tilde\epsilon}\chi(q)+
\frac{1}{\pi}\int\frac{d^{2}q' \chi
(q')\theta((q-q')^{2}-\kappa_{0}^{2}q^{2})}{{q'}^{2}\ln {q'}^{2}}
\frac{q^{2}\ln q^{2}}{(q-q')^{2}\ln (q-q')^{2}}
\lb{eigen3}
\eeq

At this point we make a second assumption. Namely we assume that in the
integral term of (\ref{eigen3}) values $m<<q'<<q$ give the dominant 
contribution (as
typical for logarithmic integrals). Of course, this assumption is also to be
checked for the final asymptotics.
With this assumption, we can forget about the $\theta$ function and also
put the last factor equal to unity in the integral term of (\ref{eigen3}).
We obtain
\beq
am^{2}\ln m^{2}={\tilde\epsilon}\chi(q)+
\int_{0}^{q^{2}}\frac{d{q'}^{2} \chi
(q')}{{q'}^{2}\ln {q'}^{2}}
 \eeq
Differentiating with respect to $q^{2}$
\beq
{\tilde\epsilon}\frac{d\chi}{dq^{2}}=-\frac{\chi}{q^{2}\ln q^{2}}
\eeq
or
\beq
{\tilde\epsilon}\frac{d\chi}{d\ln\ln q^{2}}=-\chi
\eeq
with a solution
\beq
\chi(q)=A\exp\left(-\frac{\ln\ln q^{2}}{\tilde\epsilon}\right)
=A(\ln q^{2})^{-1/
\tilde\epsilon}\eeq
The initial function $\psi(q)$ has the asymptotics
\beq
\psi(q)=\frac{A}{q^{2}}(\ln q^{2})^{\beta}
\lb{asymptmom}
\eeq
where
\beq
\beta=-1-\frac{1}{\tilde\epsilon}=-1-\frac{3}{bE}
\eeq

Now we must check that our assumptions are indeed fulfilled for the
found asymptotics. 

Let us begin with the second assumption that the values $m<<q'<<q$ give
the bulk of the contribution to the integral in (\ref{eigen3}). 
Evidently, in order
that the integral be dominated by large values of $q'$, it
 should diverge as $q\rightarrow\infty$. This
leads to the condition
\beq
E<0,\ \beta>-1
\eeq
So our asymptotics can only be valid for negative energies, that is, for
bound states.

Now for the second part of this assumption. To prove that values $q'<<q$
dominate we shall calculate the contribution from the region $q'>>q$ and
show that it is smaller. The corresponding integral is
\beq
I=Aq^{2}\ln q^{2}\int_{q^{2}}^{\infty}d{q'}^{2}(\ln {q'}^{2})^
{\beta-1}/{q'}^{4}
\eeq
 In terms of $x=\ln {q'}^{2}$
\beq
I=q^{2}\ln q^{2}\int_{\ln q^{2}}^{\infty}dxx^{\beta-1}\exp (-x)
\eeq
The integral over $x$ can be developed in an asymptotic series in $1/\ln
q^{2}$:
\beq
\int_{\ln q^{2}}^{\infty}dxx^{\beta-1}\exp (-x)=
-\int_{\ln q^{2}}^{\infty}x^{\beta-1}d\exp (-x)=
\frac{(\ln q^{2})^{\beta-1}}{q^{2}}+
(\beta-1)\int_{\ln q^{2}}^{\infty}dxx^{\beta-2}\exp (-x)=...
\eeq
From this we conclude that the integral $I$ has the asymptotics
\beq
I=A(\ln q^{2})^{\beta}
\eeq
to be compared to the contribution from the region $q'<<q$ which behaves as
$(\ln q^{2})^{\beta+1}$.   We see that we have lost one power of
$\ln q^{2}$, so that the region $q'>>q$ indeed can be neglected.

Now to the assumption (\ref{approx1}). We have explicitly
\[ \psi(|q|(n+\kappa))=(A/q^{2})(n+\kappa)^{-2}(\ln q^{2}(n+\kappa)^{2})^
{\beta}=\]
\beq
(A/q^{2})(1-2n\kappa-\kappa^{2}+4(n\kappa)^{2})
(\ln q^{2})^{\beta}(1+(\beta (2n\kappa+\kappa^{2})/\ln q^{2})
\lb{approx1_2}
\eeq
and it is evident that (\ref{d1}) is satisfied with the behaviour of $\kappa$ as
indicated in (\ref{limk0q}).
Indeed take $\kappa\sim q^{-\delta}$ with $\delta<1$.
Then $q\kappa\sim q^{1-\delta}\rightarrow\infty$ and all the correcting
terms in (\ref{approx1_2}) have the order $q^{-\delta}$.

Note that this does not mean that (\ref{approx1}) is quite obvious.  It is not
valid for, say, the exponential function. In fact we have
\[\exp(aq^{2}(n+\kappa)^{2})=\exp(aq^{2})\exp
(aq^{2}(2n\kappa+\kappa^{2}))\]
and since $q\kappa$ is large the second factor cannot be neglected.

In conclusion, we have verified that our assumptions are fulfilled and
therefore the asymptotics (\ref{asymptmom}) is correct for negative 
``energies''.

\subsection{The coordinate space  behaviour}

We are also interested in the coordinate space behaviour of the semi-amputated
wave function
\beq
\psi(r)= \int_0^{\infty} \frac{q dq}{2 \pi} J_0(qr) \psi(q)
\eeq

To estimate $\psi(r)$ for $r \to 0$ we make use of the asymptotic momentum
behaviour found previously (\ref{asymptmom}), so we can write
\beq
\psi(r) \approx \int_0^{q_0} \frac{q dq}{2 \pi} J_0(qr) \psi(q) +
         \int_{q_0}^{\infty} \frac{q dq}{2 \pi} J_0(qr) 
         \frac{ln(q^2)^{\beta}}{q^2}
\eeq         

The first integral for $r \to 0$ is finite; 
on the other hand the second
integral results to be not bounded. Infact using the integration variable
$y=qr$ we split the $y$ integration region in two parts, $q_0 r < y < y_0 \ll 1$
and $y \ge y_0$; defining the two contribution $I_1$ and $I_2$ respectively,
we have
\beq
I_1 \sim \int_{q_0 r}^{y_0} \frac{dy}{y} \bigl(\ln \frac{y}{r} 
    \bigr)^\beta
    \mathop{\sim}_{r \to 0} \bigl( \ln \frac{1}{r} \bigr) ^{\beta+1}
\eeq
and
\beq
I_2 \sim \int_{y_0}^{\infty} \frac{dy}{y} J_0(y) \bigl( \ln y + \ln 
    \frac{1}{r} \bigr)^{\beta}  \mathop{\sim}_{r \to 0}
     \bigl( \ln \frac{1}{r} \bigr) ^{\beta}
\eeq
So the asymptotic small $r$ behaviour will be divergent, in particular
\beq
\psi(r)  \mathop{\sim}_{r \to 0} \bigl( \ln \frac{1}{r} \bigr) ^{\beta+1}
\eeq

%%%%%%%%%%%%%%%%%%%%%%%%%%%%%%%%%%%%%%%%%%%%%%%%%%%%%%%%%%%%%%%%
\newpage
\section{Two gluons coupling to quark loop for non-zero transfer momentum}
Consider a $q\bar q$ loop for the scattering of a virtual photon
$\gamma^*(q)+...\rightarrow\gamma^*(q+l)+...$, $q^2=-Q^2$. The momentum
transfer $l$ is taken to be purely transversal. 
We are interested in a multi-discontinuity amplitude so that at the lowest
order one retains at each discontinuity a simple $q\bar{q}$ state with
the quark and antiquark on shell (due to unitarity) and then with their 
propagators substituted by the corresponding delta distributions.
The calculations are quite straightforward.

For example let us consider the most involved case of a transverse polarized 
photon and the gluons (with colours $a_1$ and $a_2$) coupled one 
(with momentum $q_1$) to the quark 
(with momentum $k_1$) and the other to the anti-quark (momenta $q_2$ and $k_2$ 
respectively) so that $q=k_1+k_2$. In this case the propagators for the
quark with momentum $k_1+q_1$ and the anti-quark with momentum $k_2$ have to
be replaced by the mass-shell conditions.

In the high energy limit only the longitudinal component of the gluons 
polarization give a contribution. It is convenient to work in the light cone
coordinates and therefore one has for the trace present in the amplitude the
following expression
\beq
N_\perp=\frac{\delta_{a_1a_2}}{2}g_{\alpha\beta}Tr\Bigl[
\gamma_\perp^\alpha (m_f+ \gamma \cdot k_1) \gamma_- (m_f+\gamma \cdot 
(k_1+q_1)) \gamma_\perp^\beta (m_f-\gamma \cdot (k_2+q_2)) \gamma_- 
(m_f-\gamma \cdot k_2)
\Bigr]
\eeq
After some algebra and performing the trace one obtains
\beq
N_\perp=-8 \delta_{a_1a_2} q_-^2 \Bigl\{ m_f^2 +[\alpha^2+(1-\alpha)^2]
k_{1\perp} \cdot (k_1+q_1)_\perp -\alpha^2 k_{1\perp} \cdot l_\perp \Bigr\}
\eeq
where $\alpha$ is defined such that $k_1=\alpha q_-$.
The propagators contain in the denominator two terms which can be written,
defining $Q^2=-q^2$, as
\bea
D_1=(k_1^2-m_f^2)&=& -\frac{1}{1-\alpha} 
\Bigl[ Q^2\alpha (1-\alpha) +(m_f^2+k_{1\perp}^2 )\Bigr] \nonumber \\
D_2=[(k_2+q_2)^2-m_f^2] &=& -\frac{1}{\alpha} \Bigl\{
\alpha (1-\alpha) (Q^2+l^2)+m_f^2+
\Bigl[ (k_1+q_1)_\perp -\alpha l_\perp\Bigr]^2 \Bigr\}
\eea
Taking also into account the contribution of the $\delta$ distributions,
 $1/4 q_-\alpha(1-\alpha)$, one easily obtains the full expression.
Similar expressions can be obtained for different photon polarizations and 
configurations of gluons.

Thus the calculations give for the function $f(1,2)$ corresponding to the loop
with gluon 1 attached to $q$ and gluon 2 attached to $\bar q$ the following
expression
\beq
f(k_1,k_2)=e_f^2\int_0^1d\alpha\int\frac{d^2k}{(2\pi)^3}\frac{N}{D}
\label{firstquark}
\eeq
where now we use ${\bf k_i}$ to indicate the transverse momentum vectors of 
the gluons coupled to the photon.
\beq
D=(\epsilon^2+{\bf k}^2)(\epsilon_1^2+({\bf k+k_1}-\alpha {\bf l})^2)
\eeq
\beq
\epsilon^2=Q^2\alpha(1-\alpha)+m_f^2,\ \
\epsilon^2_1=(Q^2+{\bf l}^2)\alpha(1-\alpha)+m_f^2
\eeq
$e_f$ and $m_f$ are the quark electric charge and mass and the numerator
for a transversal photon is
\beq
N^{\perp}=m_f^2 +(\alpha^2+(1-\alpha)^2){\bf k}({\bf k+k_1})-\alpha^2{\bf kl}
\eeq
and for a longitudinal photon is
\beq
N^{L}=4Q^2\alpha^2(1-\alpha)^2
\label{lastquark}
\eeq

This expression can be conveniently represented as an integral over the
colour dipole density $\rho$ created by the $q\bar q$ pair at a given
distance in the transverse space:
\beq
f(k_1,k_2)=\int d^2r\rho_l(r)e^{ik_1r}
\eeq
From (\ref{firstquark}) - (\ref{lastquark}) one finds for the transverse 
and longitudinal photons:
\bea
\rho^\perp_l(r)&=&\frac{e_f^2}{(2\pi)^3}e^{-i\alpha lr}\int_0^1d\alpha
\Bigl[ m_f^2K_0(\epsilon r)K_0(\epsilon_1 r)+
(\alpha^2+(1-\alpha)^2)\epsilon\epsilon_1 K_1(\epsilon r)K_1(\epsilon_1 r)-
\nonumber \\
&&\alpha(1-\alpha)(1-2\alpha)\frac{i\epsilon{\bf lr}}{r}K_0(\epsilon_1 r)
K_1(\epsilon r) \Bigr] \nonumber \\
\rho^L_l(r)&=&\frac{4e_f^2Q^2}{(2\pi)^3}e^{-i\alpha lr}\int_0^1d\alpha
\alpha^2(1-\alpha)^2K_0(\epsilon r)K_0(\epsilon_1 r)
\eea

%%%%%%%%%%%%%%%%%%%%%%%%%%%%%%%%%%%%%%%%%%%%%%%%%%%%%%%%%%%%%%%%%%%%
\newpage
\section{Coordinate representation of the non-forward $G$ function}

For the non-forward direction a function $G(k_1,k_2,k_3)$ can be defined
by a natural generalization of the infrared stable function introduced by
J.Bartels \cite{bartwue}:
\[
G(k_1,k_2,k_3)=-g^2 N_c\, W_2(k_1,k_2,k_3)\]\beq-
D(k_1,k_2+k_3)(\omega(k_2)-\omega(k_2+k_3))-
D(k_1+k_2,k_3)(\omega(k_2)-\omega(k_1+k_2))
\eeq
where $D(k_1,k_2)$ is the amputated BFKL function with the gluon momenta
$k_1$ and $k_2$; $\omega(k)$ is the gluon trajectory and $W_2(k_1,k_2,k_3)$
is the $2\rightarrow 3$ vertex integrated with the function $D$. 
A graphical representation of the function $G$, following the notation 
introduced by Bartels, has been given in Fig. \ref{figG}.
$G$ can be presented as a sum of proper (integral) $G_1$ and improper 
(terms with $\omega$) $G_2$ parts which, by definition, are given by
\beq
G_2(k_1,k_2,k_3)=-D(k_1,k_2+k_3)(\omega(k_2)-\omega(k_2+k_3))-
D(k_1+k_2,k_3)(\omega(k_2)-\omega(k_1+k_2))
\label{improper}
\eeq
and
\[
G_1(k_1,k_2,k_3)=
g^2N_c\,\int \frac{d^2q_1d^2q_3}{(2\pi)^3}
\delta^2 (q_1+q_3-k_1-k_2-k_3)D(q_1,q_3)\]\beq
\left(\frac{(k_2+k_3)^2}{(q_1-k_1)^2q_3^2}
+\frac{(k_1+k_2)^2}{q_1^2(q_3-k_3)^2}-\frac{k_2^2}{(q_1-k_1)^2
(q_3-k_3)^2}-\frac{(k_1+k_2+k_3)^2}{q_1^2q_3^2}\right)
\label{proper}
\eeq

For the following we recall that the gluon trajectory is
\beq
\omega(k)=-(1/2)g^2N_c\,\int \frac{d^2q}{(2\pi)^3}
\frac{k^2}{q^2(q-k)^2}=-\frac{g^2N_c}{8\pi^2}\ln\frac{k^2}{m^2}
\eeq
where the gluon mass $m$ is a regularization parameter 
(to be set to zero in the infrared finite expressions).

We shall consider the Fourier transform
\beq
G(r_1,r_2,r_3)=\frac{1}{(2\pi)^6}\int
d^2k_1d^2k_2d^2k_3\exp (i\sum_{j=1}^{3}k_jr_j)G(k_1,k_2,k_3)
\eeq
To simplify the notation we denote
\beq
d\tau(k)=\frac{1}{(2\pi)^6}
d^2k_1d^2k_2d^2k_3 \quad , \quad
d\tau(q)=\frac{d^2q_1d^2q_3}{(2\pi)^3}
\delta^2 (q_1+q_3-k_1-k_2-k_3)
\eeq
and
\beq
d\tau(\rho)=d^2\rho_1 d^2\rho_3
\eeq

\subsection{Proper part}
Let us start by considering the terms (four) which appear in the proper part 
(\ref{proper}) and which will be denoted as $G_{11},\ G_{12}, ...$ etc. 
For the first term one has
\beq
G_{11}=g^2 N_c\int d\tau(k)d\tau(q)d\tau(\rho)D(\rho_1,\rho_3)
\frac{(k_2+k_3)^2}{(q_1-k_1)^2q_3^2}\exp (i\sum_{j=1}^3 k_jr_j-
iq_1\rho_1-iq_3\rho_3)
\eeq
After the substitution $(k_2+k_3)^2\rightarrow -(\partial_2+\partial_3)^2$,
 $1/q_3^2\rightarrow -\partial_{\rho_3}^{-2}$ and $q_1-k_1\rightarrow k_2$
(all operators $\partial$ refer to coordinate space),
choosing $q_3$ as independent variable one obtains
\[
G_{11}=g^2 N_c(\partial_2+\partial_3)^2
\int d\tau(k)d\tau(q)d\tau(\rho)\frac{1}{k_2^2}\]\beq
\exp (i k_1r_1+ik_3r_3+i(q_3-k_3+k_2)r_2
-i(k_1+k_2)\rho_1-iq_3\rho_3) \partial_{\rho_3}^{-2}D(\rho_1,\rho_3)
\eeq
Integrations over $k_1$, $k_3$ and $q_3$ give
$(2\pi)^2\delta^2(r_1-\rho_1)$, $(2\pi)^2\delta^2(r_2-r_3)$ and
$(2\pi)^2\delta^2(r_2-\rho_3)$ respectively and
the remaining integral over $k_2$ has to be regularized:
\beq
\int\frac{d^2k_2}{k_2^2+m^2}\exp ik_2(r_2-\rho_1)=
2\pi(c-\ln |r_2-\rho_1|),\ \ m\rightarrow 0,
\label{regpropa}
\eeq
where
\beq
c=\ln(2/m)+\psi(1)
\eeq
Doing the integrations by means of $\delta$-functions, we obtain
\beq
G_{11}=\frac{g^2N_c}{4\pi^2}(\partial_2+\partial_3)^2\delta^2(r_{23})
(c-\ln r_{12})\partial_3^{-2}D(r_1,r_3)
\eeq
Here and in the following $r_{ij}=r_i-r_j$.
Note that the operator $\partial_2+\partial_3$ does not act on the
$\delta$-function. Its action on the $\ln r_{12}$ and $D$ is evidently
reduced to $\partial_2$ and $\partial_3$ respectively. However in the end
one has to put $r_2=r_3$ so that we obtain finally
\beq
G_{11}=\frac{g^2N_c}{4\pi^2}\delta^2(r_{23})
\partial_3^2(c-\ln r_{13})\partial_3^{-2}D(r_1,r_3)
\eeq

The Fourier transform of the second term in (\ref{proper}) is obtained from
 $G_{11}$ by interchanging $1\leftrightarrow 3$:
\beq
G_{12}=\frac{g^2N_c}{4\pi^2}\delta^2(r_{12})
\partial_1^2(c-\ln r_{13})\partial_1^{-2}D(r_1,r_3)
\eeq

Now we consider to the Fourier transform of the
third term in (\ref{proper}). Its explicit form is
\beq
G_{13}=-g^2 N_c \int d\tau(k)d\tau(q)d\tau(\rho)D(\rho_1,\rho_3)
\frac{k_2^2}{(q_1-k_1)^2(q_3-k_3)^2}\exp (i\sum_{j=1}^3 k_jr_j-
iq_1\rho_1-iq_3\rho_3)
\eeq
As usual we substitute $k_2^2\rightarrow -\partial_2^2$, 
$k_1\rightarrow k_1+q_1$,
$k_3\rightarrow k_3+q_3$ and choose $k_2$ as a dependent variable,
obtaining
\[
G_{13}=g^2 N_c\partial_2^2\int d\tau(k)d\tau(q)d\tau(\rho)D(\rho_1,\rho_3)\]
\beq
\frac{1}{k_1^2k_3^2}\exp (i (k_1+q_1)r_1+i(k_3+q_3)r_3-i(k_1+k_3)r_2-
iq_1\rho_1-iq_3\rho_3)
\eeq
The integrals over $q_1$ and $q_3$ give $(2\pi)^2\delta^2(r_1-\rho_1)$
and $(2\pi)^2\delta^2(r_3-\rho_3)$ respectively. The two integrals over
$k_1$ and $k_3$  give $2\pi(c-\ln r_{12})$ and $2\pi (c-\ln r_{23})$,
so that one has
\beq
G_{13}=\frac{g^2 N_c}{8\pi^3}\partial_2^2
(c-\ln r_{12})(c-\ln r_{23})D(r_1,r_3)
\eeq
and, after performing the derivatives in $r_2$, 
\beq
G_{13}=-\frac{g^2 N_c}{8\pi^3}\left(2\frac{{\bf r}_{12}{\bf r}_{23}}
{r_{12}^2
r_{23}^2}+2\pi (c-\ln r_{13})(\delta^2 (r_{23})+\delta^2(r_{12}))\right)
D(r_1,r_3)
\eeq

We are left with the last proper term
\beq
G_{14}=-g^2 N_c \int d\tau(k)d\tau(q)d\tau(\rho)D(\rho_1,\rho_3)
\frac{(k_1+k_2+k_3)^2}{q_1^2q_3^2}\exp (i\sum_{j=1}^3 k_jr_j-
iq_1\rho_1-iq_3\rho_3)
\eeq
Substituting all $k$'s by the derivatives w.r.t. $r$'s and $q$'s by the
derivatives w.r.t. $\rho$'s we can do all the integrals to obtaining
\beq
G_{14}=\frac{g^2 N_c}{2\pi}(\partial_1+\partial_2+\partial_3)^2
\delta^2(r_{12})\delta^2(r_{23})\partial_1^{-2}\partial_3^{-2}D(r_1,r_3)
\eeq
Since the sum $\sum_{i=1}^{3}\partial_i$ does not act on the two $\delta$-
functions, we can transfer the first operator to the right and note that,
 acting on $D$, $\partial_2$ gives zero. So finally one has
\beq
G_{14}=\frac{g^2 N_c}{2\pi}
\delta^2(r_{12})\delta^2(r_{23})(\partial_1+\partial_3)^2
\partial_1^{-2}\partial_3^{-2}D(r_1,r_3)
\eeq

\subsection{Improper part}
A similar notation is used for the improper part: $G_{21},
G_{22},...$ correspond to the 4 terms in (\ref{improper}).
Take the first term
\beq
G_{21}=-\int d\tau(k)d\tau(\rho)\omega(k_2)D(\rho_1,\rho_3)
\exp (i\sum_{j=1}^{3}k_jr_j-ik_1\rho_1-i(k_2+k_3)\rho_3)
\label{G21first}
\eeq
We transform this expression in two possible ways. One is evident:
we just take $\omega$ out of the integral as an operator to obtain
immediately
\beq
G_{21}=-\omega(-i\partial_2)\delta^2(r_{23})D(r_1,r_3)
\eeq
In this form it is clear that the contribution can be split into a sum
of two independent part, a holomorphic one acting on $r$'s and an
antiholomorphic one acting on $r^*$'s. However this expression is
difficult to analyze as to its conformal properties, due to the
fact that the operator $\omega$ is applied to the $\delta$-function,
that is, to both $D$ and a function which is to be integrated with $G$

A convenient alternative form for $G_{21}$ is obtained
by means of integration over $k_1$ and $k_3$ in (\ref{G21first}) 
\beq
G_{21}=-D(r_1,r_3)\int\frac{d^2k_2}{(2\pi)^2}e^{ik_2r_{23}}\omega(k_2)=
-\tilde \omega(r_{23})D(r_1,r_3)
\eeq
where $\tilde \omega(r)$ is just the Fourier transform of $\omega(k)$.

A useful relation, similar to (\ref{regpropa}),
\beq
\int \frac{d^2r}{r^2+\epsilon^2}e^{ikr}=2\pi(c_1-\ln k)
\quad , \quad
c_1=\ln(2/\epsilon)+\psi(1)
\eeq
allows us to write
\beq
\omega(k)=-\frac{g^2 N_c}{8\pi^2}\ln\frac{k^2}{m^2}=
\frac{g^2 N_c}{8\pi^3}\Bigl[\int \frac{d^2r}{r^2+\epsilon^2}e^{ikr} -
2\pi(c_1-\ln m)\Bigr]
\label{omegak}
\eeq
On noting that
\beq
c_1-\ln m=c-\ln\epsilon
\eeq
from eq. (\ref{omegak}) we conclude that
\beq
\tilde \omega(r)=
\frac{g^2 N_c}{8\pi^3}\Bigl[
\frac{1}{r^2+\epsilon^2} +2\pi(\ln \epsilon-c)\delta^2(r)\Bigr]
\label{omegar}
\eeq
It also follows that the right-hand side of (\ref{omegar}) does not, in fact, 
depend on $\epsilon$. 
Evidently the term with $\ln\epsilon$ serves to regularize
the singularity of the first term at $r=0$. In the following this 
regularization will always be implied, so that
\beq
\frac{1}{r^2}\equiv \frac{1}{r^2+\epsilon^2}+2\pi\ln\epsilon\delta^2(r),\ \ 
\epsilon\rightarrow 0,
\label{regular}
\eeq
With this convention we finally obtain
\beq
\tilde \omega(r)=
\frac{g^2 N_c}{8\pi^3}(\frac{1}{r^2} -2\pi c\delta^2(r))
\eeq
The same result could be obtained directly by Fourier transforming the
integral representation of $\omega(k)$ but in this way we obtain
(\ref{regular}) as a byproduct.
Using this we get our final expression
\beq
G_{21}=-\frac{g^2 N_c}{8\pi^3}(\frac{1}{r_{23}^2} -2\pi c\delta^2(r_{23}))
D(r_1,r_3)
\eeq

The term $G_{23}$ is obtained from this by the permutation 1 and 3 in the
operator acting on $D$:
\beq
G_{23}=-\frac{g^2 N_c}{8\pi^3}(\frac{1}{r_{12}^2} -2\pi c\delta^2(r_{12}))
D(r_1,r_3)
\eeq

Improper terms 2 and 4 are calculated in a simpler manner. In fact we have
\beq
G_{22}=-\int d\tau(k)d\tau(\rho)\omega(k_2+k_3)D(\rho_1,\rho_3)
\exp (i\sum_{j=1}^{3}k_jr_j-ik_1\rho_1-i(k_2+k_3)\rho_3)
\eeq
In this case we can substitute $\omega(k_2+k_3)$ by an operator
$\omega(-i\partial_{\rho_3})$ acting on $D$. After that all integrations are
trivially done and we obtain
\beq
G_{22}=\delta^2(r_{23})\omega(-i\partial_3)D(r_1,r_3)
\eeq
and in a similar manner
\beq
G_{24}=\delta^2(r_{12})\omega(-i\partial_1)D(r_1,r_3)
\eeq

\subsection{Infrared stable pieces}

Let us see how to combine different proper and improper terms in order to 
obtain expressions which do not depend on $m$ and so are infrared finite. 
There are three infrared finite combinations.

Summing $G_{11}$ and $G_{22}$ we find
\beq
G^{(1)}= G_{11}+G_{22}=\delta^2(r_{23})\frac{g^2N_c}{8\pi^2}
\Bigl[2\ln2+2\psi(1)-k_3^2\ln r_{13}^2k_3^{-2}-\ln k_3^2\Bigr]D(r_1,r_3)
\eeq
Here we have used the explicit expression for $\omega$ in the momentum space
and also used the notation $k_3$ for the operator $-i\partial_3$
acting in coordinate space. The important point is that the terms
with $\ln m$ have cancelled in the sum, so that the expression is infrared 
finite. Note that the expression in the brackets is a half of the Hamiltonian
for the two gluons (with a minus sign). It represents the Hamiltonian of the
gluon 3 in interaction with the gluon 1 considered as an external source.

The other half of the Hamiltonian appears in the sum
 $G_{12}$ and $G_{24}$, for which we also find an infrared finite
 expression
\beq
G^{(2)}= G_{12} + G_{24}=\delta^2(r_{12})\frac{g^2N_c}{8\pi^2}
\Bigl[2\ln2+2\psi(1)-k_1^2\ln r_{13}^2k_1^{-2}-\ln k_1^2\Bigr]D(r_1,r_3)
\eeq

The rest of the terms give an expression
\[
G^{(3)}=G_{13}+G_{14}+G_{21}+G_{23}\]\beq=
\frac{g^2 N_c}{8\pi^3} \Bigl[-4\pi^2\delta^2(r_{12})\delta^2(r_{23})(k_1+k_3)^2
k_1^{-2}k_3^{-2}+2\pi(\delta^2(r_{23})+\delta^2(r_{12})\ln r_{13}-
\frac{r_{13}^2}{r_{12}^2r_{23}^2} \Bigr] D(r_1,r_3)
\eeq
It is also infrared finite since all the $\ln m$ terms cancel. 
One has only to remember that the singularities
of the last term at $r_{12}=0$ and $r_{23}=0$ have to be regularized, if 
necessary, according to the prescription (\ref{regular}).

%%%%%%%%%%%%%%%%%%%%%%%%%%%%%%%%%%%%%%%%%%%%%%%%%%%%%%%%%%%%%%%%%%%%%%%
\newpage
\section{Some useful relations in graphical notation.}
\begin{figure}
\centering
\includegraphics[width=4.0in]{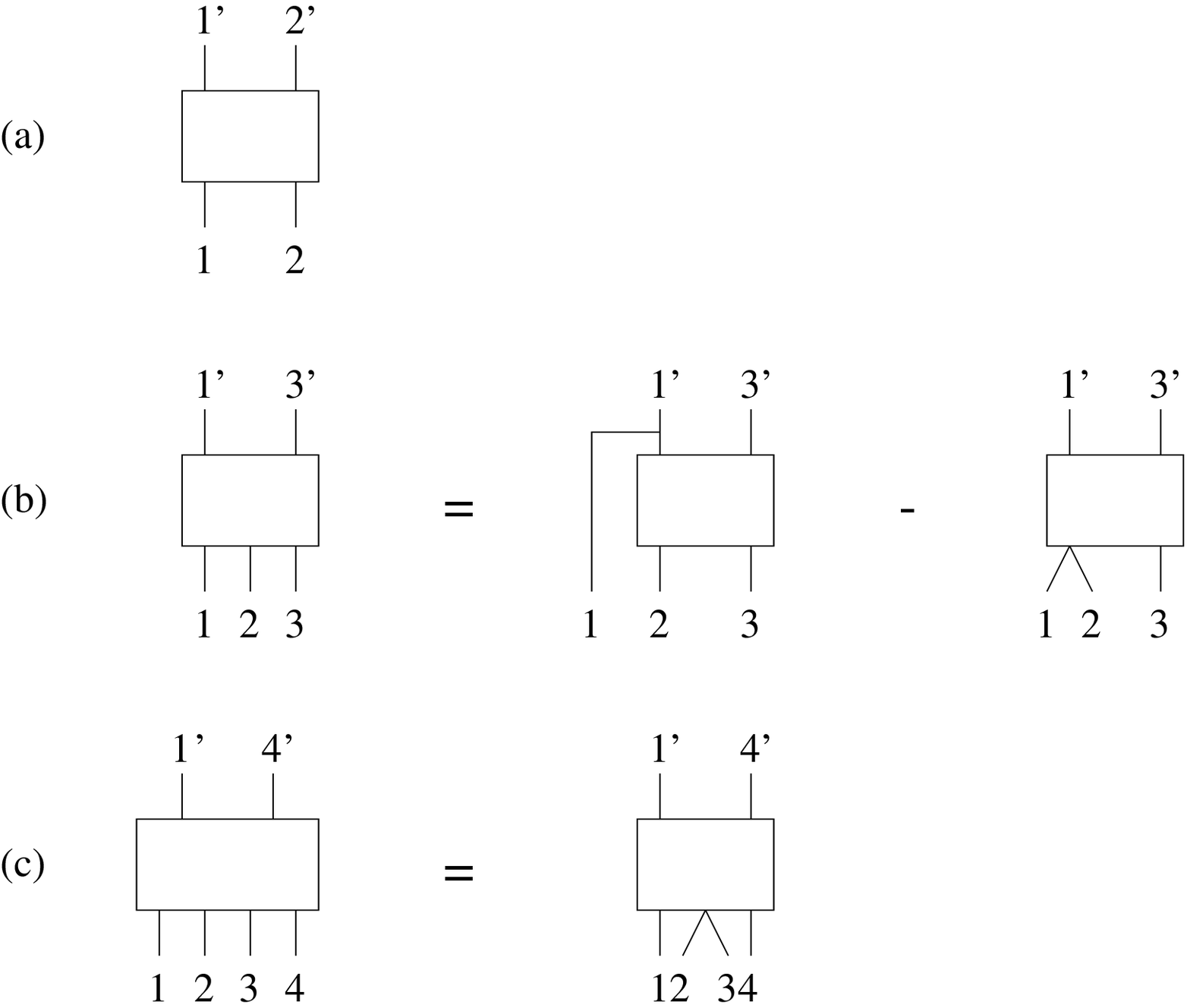}
\captionm{Graphical representation of the elementary vertices.}
\label{figvert}
\end{figure}

\begin{figure}
\centering
\includegraphics[width=4.0in]{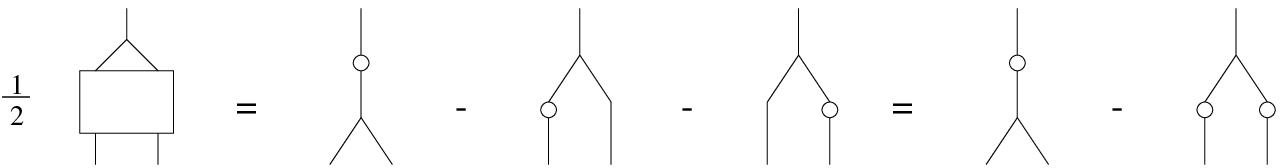}
\captionm{Graphical representation of the bootstrap condition.}
\label{figbo}
\end{figure}

We introduce in the following a graphical notation in order
to easily derive some relations involving the elementary reggeized gluons
(see also in \cite{braun4} for the analytical counterpart).
A line will denote a reggeized gluon momentum, a line split a dependence 
in the sum of the momenta corresponding to the lines, the squares the 
interaction vertices defined in (\ref{vert22})-(\ref{vertices}) and
a small circle the contribution of a Regge trajectory (if more than one is
present a sum of all the possible one-circle graphs is meant).
In Fig. \ref{figvert} we show the vertices with their explicit definition, 
related to the $2\to2$, $2\to3$ and $2\to4$ transitions in $(a)$, $(b)$ and 
$(c)$ respectively, and in Fig. \ref{figbo} a possible representation of the 
bootstrap condition.

\begin{figure}
\centering
\includegraphics[width=6.0in]{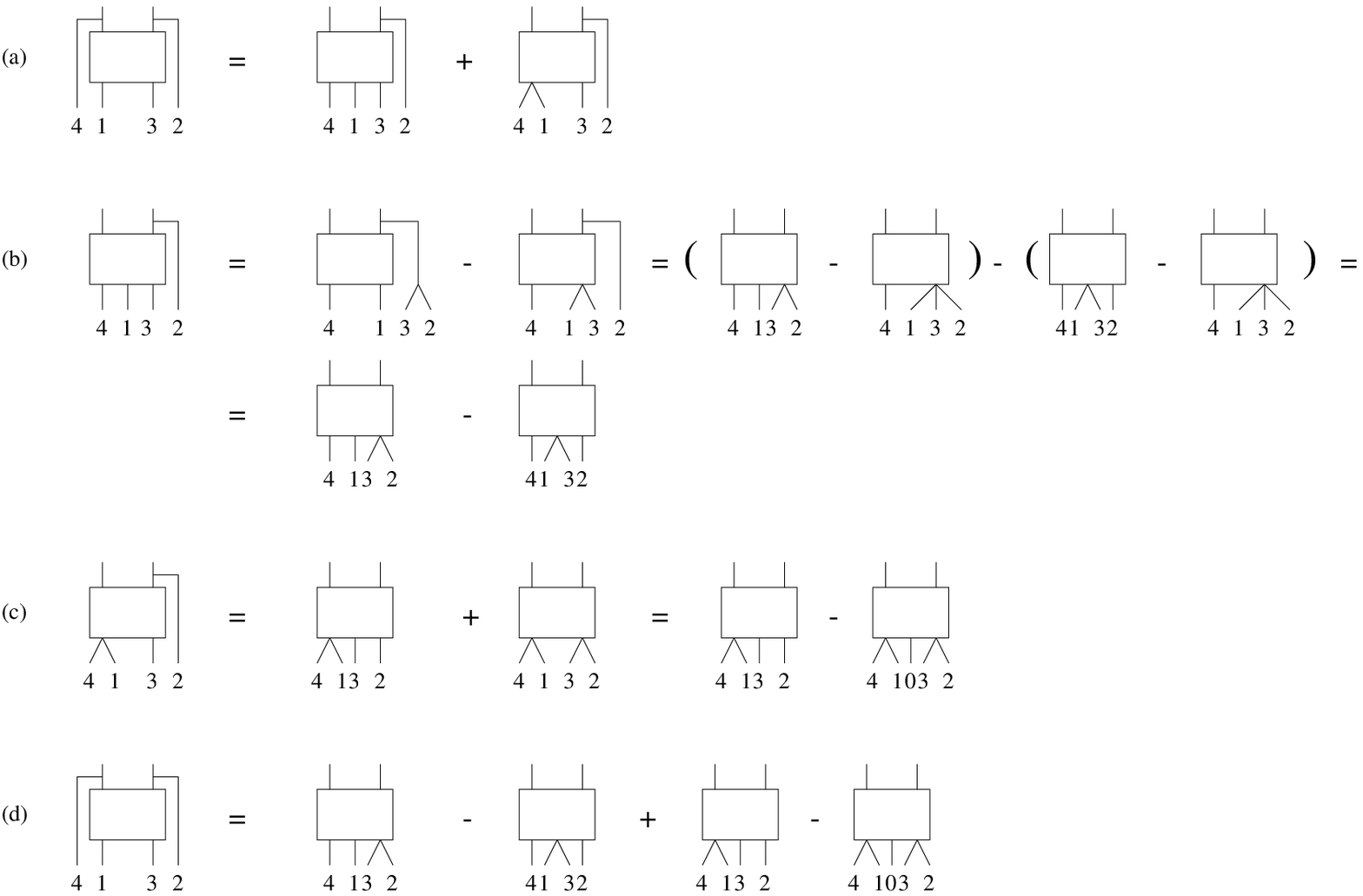}
\captionm{Proof of the relation (\ref{relaz1}).}
\label{figre1}
\end{figure}

It is simple to check the validity of the relation (\ref{relaz1}) which was
needed to calculate the triple pomeron vertex $Z$.
The steps are shown in Fig. \ref{figre1} and depend exclusively on the
repeated application of the relation $(b)$ of Fig. \ref{figvert}.
The relation in line $(a)$ is directly derived by applying on $W(4,1,3)$
(first term in the rhs) this relation with the indices $(123)\to(413)$.
In $(b)$ and $(c)$ in a similar manner other useful relations are derived 
and, applying them to $(a)$, one finds in $(d)$ the identity (\ref{relaz1}). 

\begin{figure}
\centering
\includegraphics[width=6.0in]{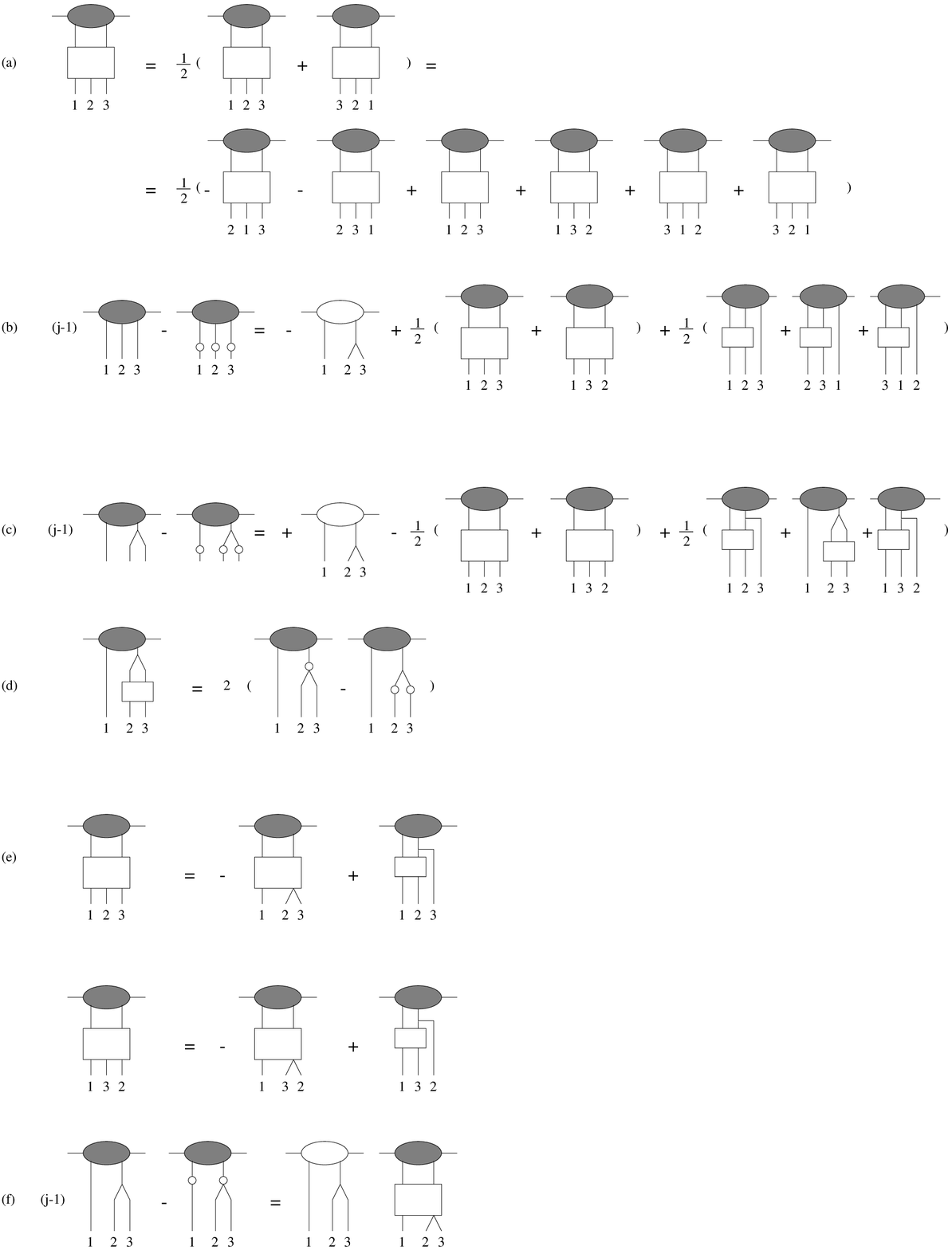}
\captionm{Proof of the relation (\ref{sol3glu}).}
\label{figre2}
\end{figure}

Let us show in this graphical representation how to apply the bootstrap 
condition, giving the proof that the solution of eq. 
(\ref{eq3glu}) is in fact the expression (\ref{sol3glu}).
The proof is based on the reduction, by means of the bootstrap, of the equation
for the 3 gluon system to the equation for two gluon in the singlet state,
i.e. a BFKL non-homogeneous equation.
Precisely one is able to show this for an equation with an inhomogeneous term
which is a part of the one present in (\ref{eq3glu}) so one completes the 
solution by applying the linearity of the equation. 
Therefore we decompose the second inhomogeneous term $D_{2\to3}$ in a sum of
three terms as in $D_{30}$ and study one of the three similar equations,
corresponding to the case with $D_{20}(1)$ and the third and fourth terms in 
the decomposition of $D_{2\to3}$ given in Fig. \ref{figre2}.$a$.

We can now easily check that the solution of the resulting equation 
(Fig. \ref{figre2}.$b$) is given in fact by $D_3^{(1)}=-D_2(1,2+3)$.
With this substitution the equation is represented in Fig. \ref{figre2}.$c$ 
for the $D_2$. On applying the bootstrap condition one has the relation in 
Fig. \ref{figre2}.$d$ and in Fig. \ref{figre2}.$e$ two relations are derived 
from Fig. \ref{figvert}.$b$. 
Substituting these identities in the relation of Fig. \ref{figre2}.$c$ one 
finally obtains for $D_2$ the equation in Fig. \ref{figre2}.$f$ which is the 
expected BFKL equation.

\newpage

% references

\end{spacing}
\end{document}